%% file: main.tex
\documentclass{aa}    %\documentclass[referee]{aa} 

\usepackage{graphicx}%,url,twoopt,natbib, nth
\usepackage[varg]{txfonts}
\usepackage[detect-all]{siunitx}
\usepackage{arydshln}
\usepackage{color,hyperref}
\hypersetup{colorlinks=true, linkcolor=blue,
citecolor=blue, filecolor=blue, urlcolor=blue}

\bibpunct[, ]{(}{)}{,}{a}{}{,}
\newcommand{\farc}{\hbox{$\text{ }\!\!^{\prime\prime}$}}

\begin{document}

\title{Dust reddening and extinction curves toward \\
gamma-ray bursts at z > 4}

\titlerunning{Dust at $z > 4$}

\author{J.~Bolmer \inst{1,2,3} 
\and J. Greiner \inst{2}
\and T. Kr\"uhler \inst{2}
\and P. Schady \inst{2}
\and C. Ledoux \inst{1}
\and N. R. Tanvir \inst{4}
\and A. J. Levan \inst{5}
}

\institute{European Southern Observatory, Alonso de C\'{o}rdova 3107, Vitacura, Casilla 19001, Santiago 19, Chile\\
\email{jbolmer@eso.org}
\and Max-Planck-Institut für extraterrestrische Physik, Giessenbachstraße, 85748 Garching, Germany
\and Technische Universit\"{a}t M\"{u}nchen, Boltzmannstraße 2, D - 85748 Garching, Germany
\and University of Leicester, Department of Physics and Astronomy and  Leicester Institute of Space \& Earth Observation, University Road, Leicester, LE1 7RH, UK,
\and Department of Physics, University of Warwick, Coventry, CV4 7AL, UK,
}

\date{Received/accepted}

\authorrunning{Bolmer et al.}

  \abstract
        {Dust is known to be produced in the envelopes of asymptotic giant branch (AGB) stars, the expanded
    shells of supernova (SN) remnants, and in situ grain growth within the interstellar medium (ISM),
    although the corresponding efficiency of each of these dust formation mechanisms at different
    redshifts remains a topic of debate. During the first Gyr after the Big Bang, it is widely believed
    that there was not enough time to form AGB stars in high numbers, hence the dust at this epoch is
    expected to be purely from SNe or subsequent grain growth in the ISM. 
    The time period corresponding to $z\sim\numrange{5}{6}$ is thus expected to display the
    transition from SN-only dust to a mixture of both formation channels as is generally recognized at present.}
    {Here we aim to use afterglow observations of gamma-ray bursts (GRBs) at redshifts larger
    than $z > 4$ to derive host galaxy dust column densities along their line of sight
    and to test if a SN-type dust extinction curve is required
    for some of the bursts.}
    {We performed GRB afterglow observations with the seven-channel Gamma-Ray Optical and
    Near-infrared Detector (GROND) at the 2.2 m MPI telescope in La Silla, Chile (ESO), and
    we combined these observations with quasi-simultaneous data gathered with the XRT telescope on board the
    \textit{Swift} satellite.} 
    {We increase the number of measured $A_V$ values for GRBs at $z > 4$ by a factor
    of $\sim$\numrange{2}{3} and find that, in contrast to samples at mostly lower redshift,
    all of the GRB afterglows have a visual extinction of $A_V < 0.5$ mag. Analysis of the GROND
    detection thresholds and results from a Monte Carlo simulation show that although we partly
    suffer from an observational bias against highly extinguished sight-lines, GRB host galaxies
    at $4<z<6$ seem to contain on average less dust than at $z\sim2$. Additionally, we find that
    all of the GRBs can be modeled with locally measured extinction curves and that the SN-like
    dust extinction curve, as previously found toward GRB 071025, provides a better fit for only
    two of the afterglow SEDs. However, because of the lack of highly extinguished sight lines
    and the limited wavelength coverage we cannot distinguish between the different scenarios.
    For the first time we also report a photometric redshift of $z_{\mathrm{phot}} =
    7.88^{+0.75}_{-0.94}$ for GRB 100905A, making it one of the most distant GRBs
    known to date.}{}

   \maketitle

% \onecolumn

\section{Introduction}

As a result of their high luminosities, gamma-ray bursts (GRBs) provide a powerful
and unique probe to study the interstellar medium (ISM) out to very high redshifts,
up to the epoch of reionization \citep[e.g.,][]{gehrels2009,
kumar2015}. Shining through their host galaxies, deviations
from their simple, smooth and featureless, intrinsic power-law spectra
caused by dust, metals, or gas, allow detailed studies of the illuminated regions along the
line of sight of their host galaxy and the intergalactic medium (IGM) \citep{galama2001,kann2006,
schady2007}. Broadband photometric and spectroscopic observations of the GRB afterglow are now
routinely used to measure metal, molecule, and dust column densities along with depletion patterns
or dust-to-metal ratios to high accuracy \citep[e.g.,][]{ledoux2009,
kruehler2013, decia2013, sparre2014, wiseman2016}. Likewise, interstellar extinction curves
have been tested out to very high redshifts \citep{zafar2011b, greiner2011, perley2011, schady2012},
including detections of the characteristic $2175$ \AA\ bump \citep{kruehler2008, zafar2012} as known
from the Milky Way (MW) and the Large Magellanic Cloud (LMC) or more unusual features
\citep{savaglio2004, perley2008gd, fynbo2014}, which might give new clues about dust
production and properties throughout the Universe. 

        \begin{figure}
                \centering
                \includegraphics[width=9cm]{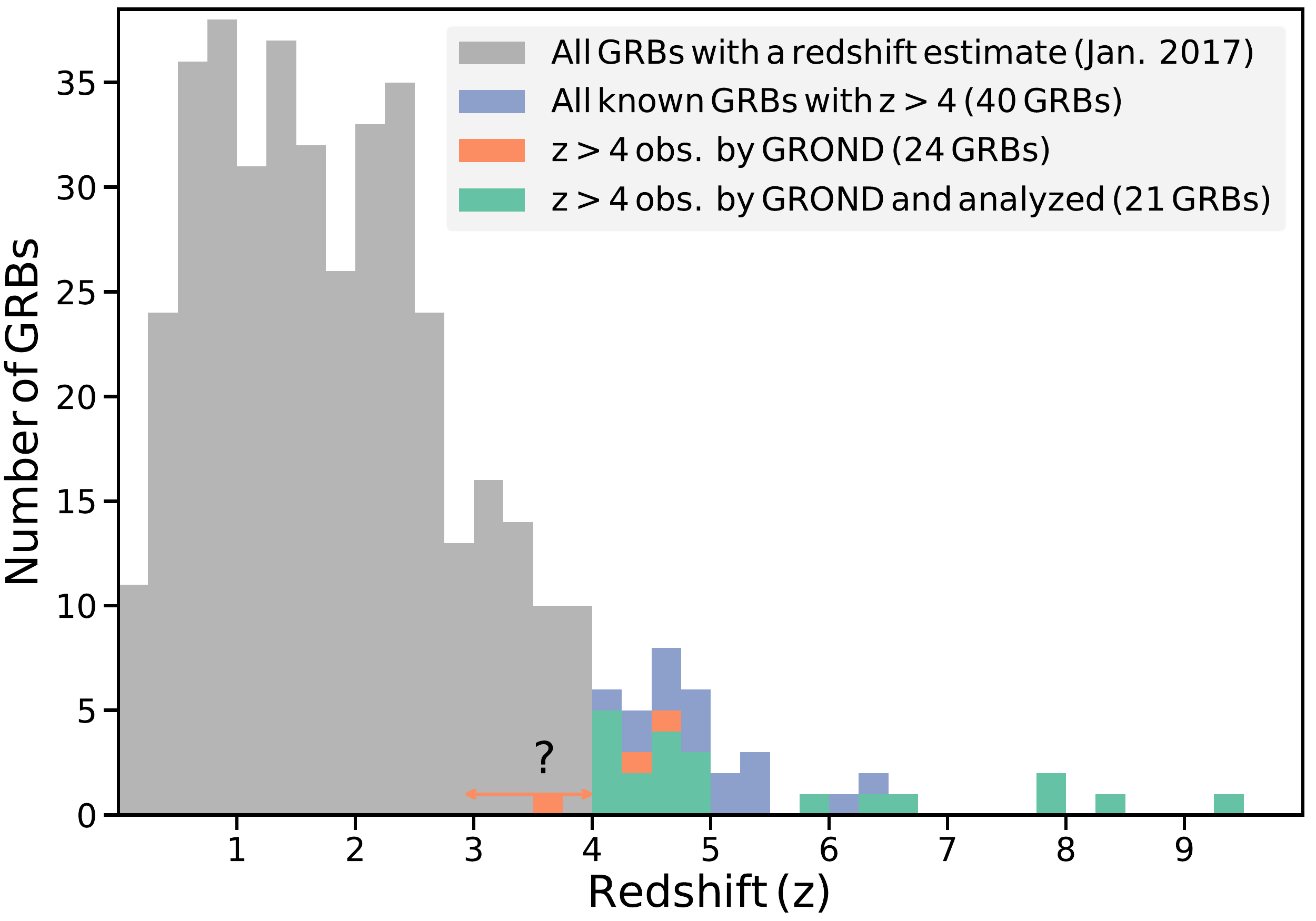}
                \caption{Distribution of GRBs per redshift bin ($\Delta z = 0.25$) for the
        events discussed in this paper compared to all LGRBs with a reported redshift estimate
        (indicated in gray; %taken from \url{http://www.mpe.mpg.de/~jcg/grbgen.html}
        including all events before January 2017). A total of 40 bursts (blue) have a well-derived
        redshift estimate of $z > 4$; 24 (orange) of these have been observed with GROND,
        from which 21 (green) are analyzed in this paper.}
                \label{fig:zdist}
        \end{figure}

Measuring dust at high redshift comes with substantial observational biases. Firstly,
while $\gtrsim 90$\% of all \emph{Swift}-detected GRBs ($> 1000$) are detected and
localized in X-rays with the \emph{Swift}/XRT, only about 30\% have a redshift estimate.
Secondly, at redshifts $z > 2$ dust reddening forms an increasing hindrance in detecting the
optical and near-infrared (NIR) afterglow.
For instance, for a GRB at a redshift of $z = 4$, depending on the extinction law,
a rest-frame $A_V = 1$ mag corresponds to an observer frame $A_V \sim\numrange{4}{5}$ mag, 
just due to redshifting the bandpass because the attenuation by dust
usually increases from red to blue wavelengths. Heavily obscured afterglows, the so-called
\emph{dark} GRBs, are generally found to occur in more massive and redder galaxies
\citep{kruehler2011b, rossi2012, perley2013, hunt2014}, and it was argued that these
bursts were more likely to be missed in follow-up campaigns.

Various approaches were therefore made to create
optically unbiased samples of GRBs that are representative of the whole population.
\cite{cenko2009paper} and \cite{greiner2011} for example chose only those GRBs that were
observed within a few hours after the \emph{Swift}/BAT trigger by instruments dedicated to
observe every GRB. These and similar approaches, such as the BAT6 sample \citep{salv2012, covino2013}
or the TOUGH \citep{hjorth2012} and SHOALS \citep{perley2016a, perley2016b} surveys, find
the percentage of \emph{dark} GRBs to be around $\sim 20-40$\% without necessarily considering
a potential evolution with redshift. Although much smaller ($< 100$), these samples then reach a
completeness in redshift of $> 90$\%. Furthermore, especially at even higher redshifts
($z > 3.5$), when optical and NIR SEDs and spectra are increasingly absorbed by the
Ly$\alpha$ forest, additional absorption by dust can theoretically make a detection of the
afterglow nearly impossible, even for 8 m class telescopes.
    
        \begin{figure}[ht]
                \centering
                \includegraphics[width=9cm]{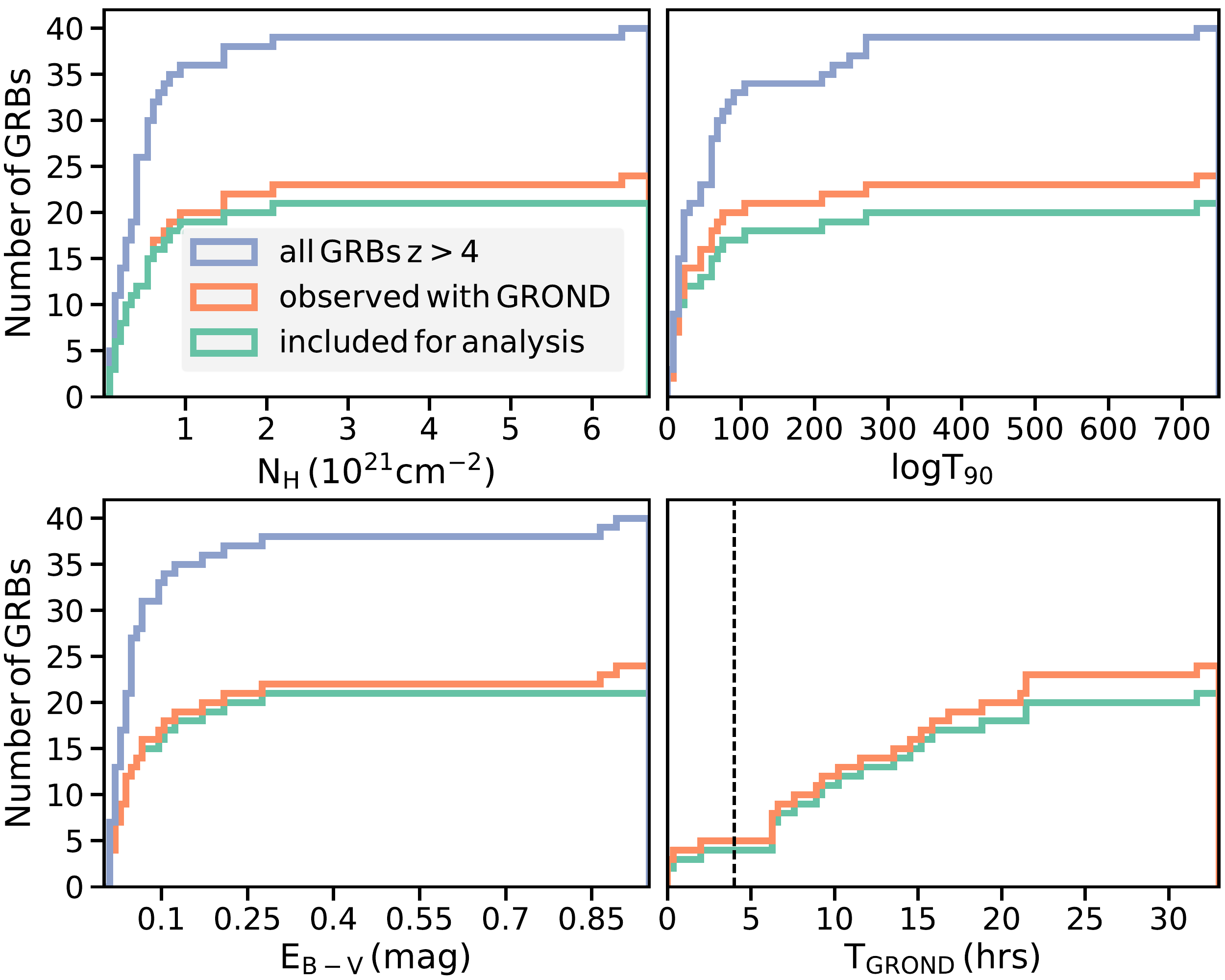}
                \caption{Cumulative distribution of the basic properties of the GRBs at $z>4$.
        Shown are the galactic foreground reddening $E_{B-V}$ and hydrogen column density
        $N_{\element[][]{H}}$, as well as the duration of the prompt emission $T_{90}$
        and the time of the first GROND observation $T_{\mathrm{GROND}}$. With two exceptions,
        all GRBs are behind modest galactic dust and hydrogen column densities (left panels). The value
        $T_{90}$ is between 2 and 300 s for all of the GRBs with the only outliner being
        GRB 140614A with 720 s (top right panel). The majority of the GRBs were observed
        by GROND between 4 to 18 hrs after the trigger; only 6 are part of the unbiased
        GROND 4h sample \citep{greiner2011} (bottom right panel). }
                \label{fig:stats}
        \end{figure}

At redshifts ($z > 4-6$), when the Universe is thought to have been
still too young to have formed AGB stars in high numbers, SNe are expected
to be the main source of dust.
However, it is still under debate how effectively dust is produced in the 
expanded shells of SN and how high the contribution from AGB stars might be
\citep{val2009, hir2014}. This is mainly because a high percentage
of the SN produced dust might be destroyed by the reverse shock
of the SN itself \citep{nozawa2007, schneider2012a}. It is therefore likely that a
significant initial production of dust in SN ejecta is required to explain
the increasing evidence of large dust
masses and high star formation rates found in high redshift galaxies
(e.g., through ALMA) \citep{mancini2015, watson2015dust, laporte2017}.
Dust production in SN ejecta is observed in some local
SNe remnants \citep{gomez2012, indebetouw2014, matsuura2015, delooze2017}
and predicted by some analytical models \citep{schneider2012b, silvia2012}, which
cover a broad range of possible dust survival rates or a significant contribution
from subsequent grain growth in the ISM \citep{nozawa2012, sarangi2013, nozawa2015,
dustmicha2015}. 

%Added by TeX Support
       \input{table/sample.tex}

        \begin{figure*}
                \centering
                \resizebox{\hsize}{!}{\includegraphics{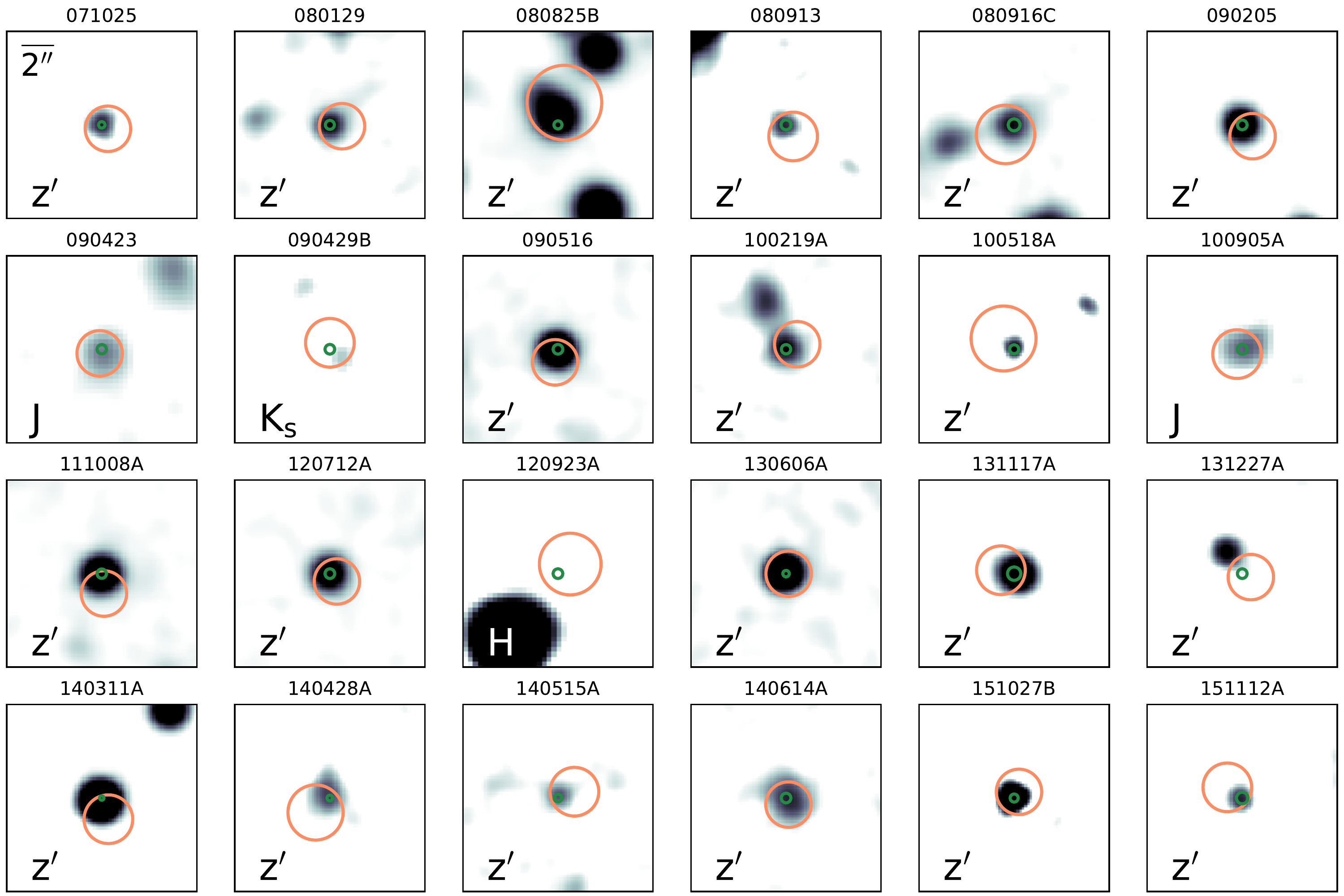}}
                \caption{Thumbnails for the 24 GRBs observed with GROND. The orange circles represent the
        \emph{Swift}/XRT 90\% error circles as taken from \url{www.swift.ac.uk/xrt_positions/}, and
        the green circles show the position of the NIR or optical afterglow as given in Tab. \ref{tab:sample}.
        A 2\farc\ scale bar is shown in the top left plot for GRB 071025 and the particular GROND
        filter is denoted in the bottom left corner of each finding chart. With the exception of
        GRBs 090429B, 120923A, and 131227A, we were able to detect the NIR or optical counterpart with GROND.}
                \label{fig:fcs}
        \end{figure*}

The theoretical model from \cite{todini2001} shows that SN-origin dust would produce
a characteristic extinction curve, which could be measured in absorption systems
toward background sources such as quasars and GRBs.
Indeed, \cite{maiolino2004}
reported evidence for the SN origin of dust in a quasar at redshift $z = 6.2$
\footnote{ \cite{hjorth2013} come to a different conclusion.}.
Similar evidence for extinction caused by SN synthesized dust was found in two GRB
afterglows at a redshift of $z > 4.8$. While two different authors reached the same
conclusion for GRB 071025 \citep{perley2010, jang2011}, the claim for SN-type dust
in GRB 050904 is more controversial \citep{stratta2007, zafar2010, stratta2011}.

The aim of this paper is to provide a detailed and consistent
study of the dust extinction properties in the afterglows of the most distant GRBs
to find out about a potential evolution with redshift and whether a SN-like dust extinction
curve is required for some of the bursts. The paper is arranged as follows: In Section
\ref{sec:sample} we describe the current sample of GRBs at $z > 4$. Section \ref{sec:redana}
presents our data analysis and reduction technique, and the main results are summarized
in Section \ref{sec:results}. Finally, we discuss the results and conclude in Section
\ref{sec:discuss}. Throughout the paper all magnitudes are given in the AB system and we
adopt the convention that the GRB flux density is described by $F_{\nu} (t) \propto
t^{-\alpha}\nu^{-\beta}$. Unless indicated otherwise, all errors are given at $1\sigma$
confidence.

\section{The sample}\label{sec:sample}

\begin{figure*}[th]
                \centering
                \resizebox{\hsize}{!}{\includegraphics{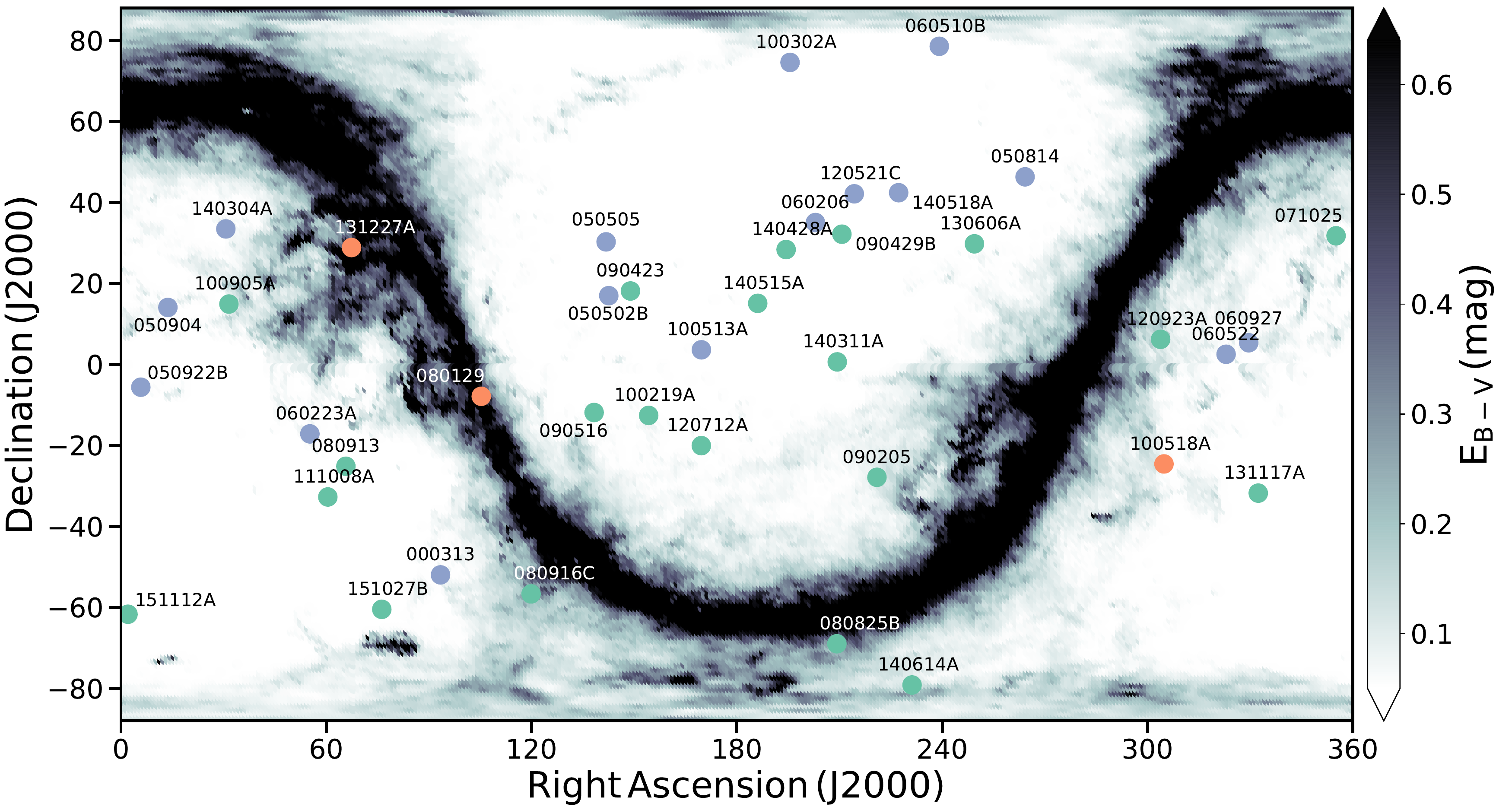}}
                \caption{All-sky map of the galactic dust reddening $E_{B-V}$ as given by
        \cite{sf2011}. The GRBs analyzed here are indicated in green. Those indicated
        in orange were excluded from the sample and those colored in blue are all other
        known GRBs with a redshift of $z > 4$. The sky above Dec $> 44^{\circ}$ cannot be
        observed from La Silla (Chile), where GROND is mounted at the 2.2 m MPI telescope.}
                \label{fig:map}
        \end{figure*}

The GRB afterglow sample presented here is based on selecting all 40 observed events
with a previously reported spectroscopic or photometric redshift of $z > 4$ (complete
up until 1 January 2017; see Fig. \ref{fig:zdist})\footnote{This was performed on the
basis of the public GRB table
maintained by one of the co-authors: \url{http://www.mpe.mpg.de/~jcg/grbgen.html}; bursts
classified as short and redshift values reported with a question mark were ignored.}.
This sample is presented in Tab. \ref{tab:sample}, which is divided into three parts. Of these
40 GRBs, we were able to observe 24 via GROND, and of these, 21 GRBs (top part) were
selected for our analysis and 3 GRBs (middle part) were excluded for reasons given below.
The remaining 16 GRBs at $z > 4$ are listed in the bottom portion of the table. These either
occurred before the GROND commissioning in 2007, were too far north to be observable from
Chile (see Fig. \ref{fig:map}), or were not observable with GROND because of bad weather (GRB 100513A).
For each GRB, we give the coordinates of the NIR or optical afterglow, redshift, duration of
the prompt emission ($T_{90}$), galactic foreground reddening $E_{B-V}$ and hydrogen column
density ($N_{\element[][]{H}}$), as well as the time after which we started observing the
afterglow with GROND ($T_{\mathrm{GROND}}$). These properties of the sample are also
visualized in Fig. \ref{fig:stats}.

Two of the GRBs observed by GROND lie close to the galactic plane and are behind high
galactic dust and hydrogen column densities, namely GRB 080129 and GRB 131227A,
and were thus excluded from the sample (see Fig. \ref{fig:map}). Additionally,
we excluded GRB 100518A because our analysis results in a photometric redshift
of only $z_{\mathrm{phot}} = 3.50^{+0.50}_{-0.62}$ (see Sec. \ref{sec:phot}). Also
only 4 out of the 21 analyzed GRBs are part of the unbiased GROND 4h sample, which
contains all GRBs that have been observed within at least 4 hrs post trigger \citep{greiner2011}.
However, a large percentage of the bursts could be observed between 4 and 18 hrs after their
detection, during the first night in La Silla (Chile) usually providing a high chance of
reliable detections in most of the seven GROND filter bands. 
Only three bursts (GRB 071025,080916C, and 090429B) could not be observed with GROND
during the first night of their trigger, mainly due bad weather in La Silla. Finally, although all 24 GRB
afterglows observed with GROND were also observed and detected with \emph{Swift}/XRT, the
prompt emission of three of these was initially detected by instruments on other satellites
(GRB 080825B: \emph{Agile}/GRID, GRB 080916C: \emph{Fermi}/GBM+LAT, and GRB 100518A:
\emph{Integral}/IBIS).

       \section{Data reduction and analysis}\label{sec:redana}

For the aim of this paper all GRBs with a redshift of $z > 4$ and observed by GROND and
\emph{Swift}/XRT, were analyzed to create broadband SEDs to measure dust
column densities and test extinction curves along their line of sight.

       \subsection{X-ray data analysis}

The XRT X-ray light curves and spectra of the afterglow were taken from the automated
data products provided by the public \emph{Swift}/XRT repository \citep{evans2009}.
To ensure that there were at least 20 counts per bin, we further regrouped the spectral data
from 0.3-10 keV in the chosen time interval  from the photon counting mode (PC) alone
with the \emph{grappha} task from the {\tt HEAsoft} package
and the response matrices from CALDB (Version {\tt v20120209}); this also
ensured that bad columns could be ignored. In cases where no X-ray data were
available simultaneously to the GROND observations, the XRT spectra were
additionally flux normalized to the mid-time of the chosen GROND exposure using
the temporal decay model, which best fit the XRT light curve. The common reference time was
generally chosen to be after any optical rise/late-time re-brightening, steep decay,
or plateau phase, at least in those cases where such a distinction was possible.
We also avoided selecting time intervals from periods of X-ray flares or spectral
evolution, i.e., time intervals of changing temporal decay.

       \subsection{Near-infrared and optical data analysis}
        
Image reduction and photometry of the GROND observations were carried out
with the standard \emph{Image Reduction and Analysis Facility} tasks (IRAF;
\cite{tody1993}), as described in \cite{kruehler2008}.
The absolute calibration of the GROND observations in $g'r'i'z'$
was carried out with stars observed in the Sloan Digital Sky Survey (SDSS).
In cases where the GRB was not in a field covered by the SDSS, an SDSS
field and the GRB field were observed consecutively during photometric
conditions to cross-calibrate the zero points. The absolute NIR
calibration in $JHK_s$ was performed using the Two Micron Sky Survey (2MASS;
\cite{skrutskie2006}) stars within the field of the GRB. This method results
in typical systematic errors of 0.03 mag for the $g'r'i'z'$ bands,
0.05 mag for the $JH$ bands, and 0.07 mag for the $K_s$ band. Finally,
before the fitting process, all magnitudes were corrected for the galactic foreground
reddening according to the values given by \cite{sf2011} and listed in 
Tab. \ref{tab:sample}. The complete set of GROND photometry for the GRBs
at $z > 4$ is given in the Appendix in Tab. \ref{tab:grondmags}.

\begin{figure}
                \resizebox{\hsize}{!}{\includegraphics{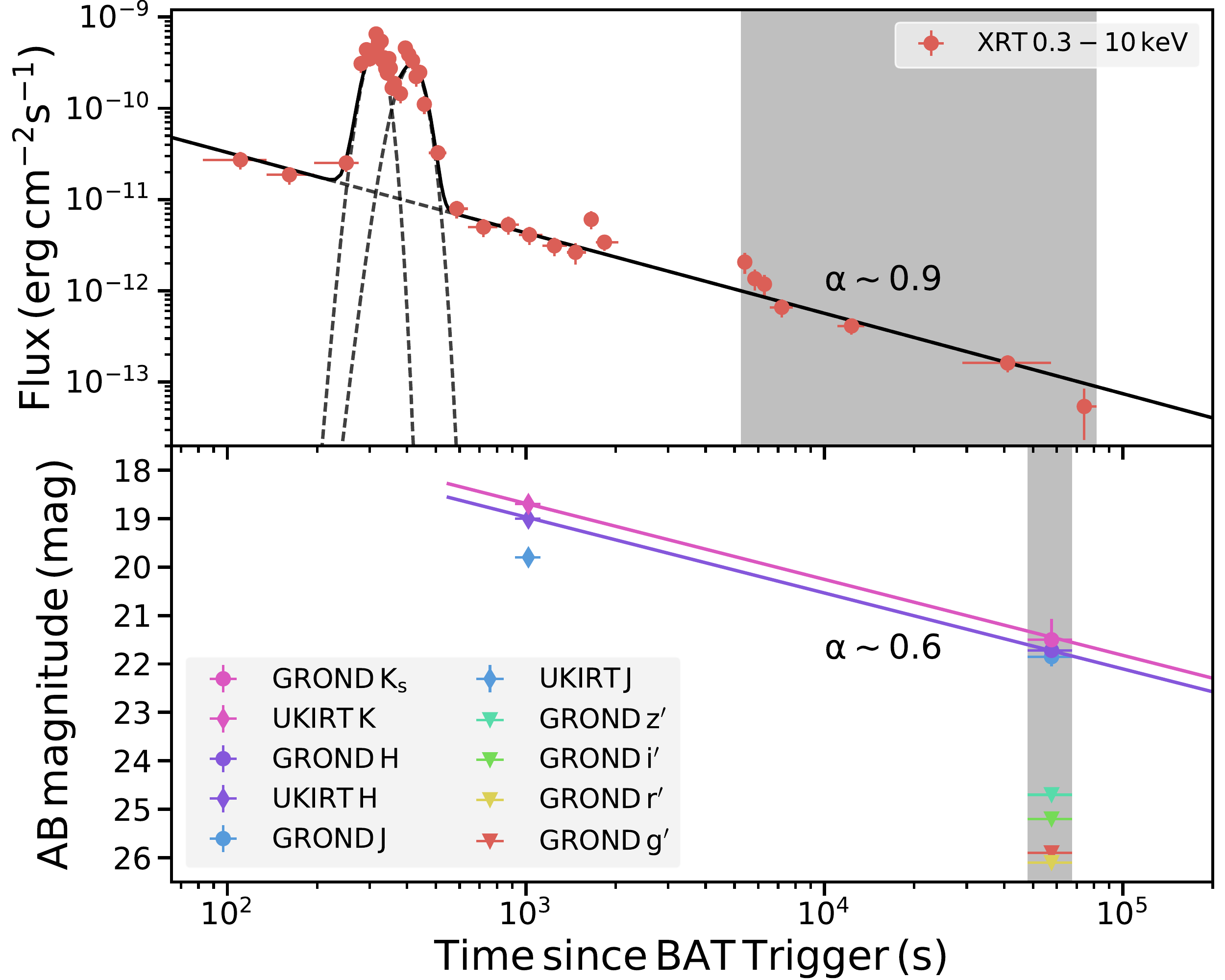}}
                \caption{Near-infrared and optical and XRT X-ray light curves of GRB 100905A. In
        addition to the GROND detections and upper limits we also plot
        UKIRT $J$-, $H-$ and $K$-band magnitudes as reported by \cite{im2010}.
        The XRT light curve shows some flaring activity between $T_0 + 300$
        and $T_0 + 600$ s and is otherwise best fit with a single power law
        and a temporal decay slope of $\alpha_\mathrm{X} \sim 0.9$. The GROND and UKIRT $K-$
        and $H$-band light curves indicate a temporal decay slope of $\alpha_\mathrm{o}
        \sim 0.6$. The time intervals used to created the quasi-simultaneous
        broadband SED are indicated in gray.}
                \label{fig:100905Alc}
        \end{figure}

As an example, the optical, NIR, and XRT X-ray light curves of GRB 100905 are
shown in Fig. \ref{fig:100905Alc}. The light curves of all other GRBs are moved
to the Appendix (Fig. \ref{fig:071025lc} to Fig. \ref{fig:151112Alc}), where 
we also describe the light-curve analysis in more detail and list the
best-fit models in Tab. \ref{tab:lcfit}.

\subsection{Position of the near-infrared and optical afterglow}

The position of the afterglow was determined by using the USNO or SDSS field
stars as astrometric reference. For the majority of the GRBs we used the GROND $z'$ band
observations and averaged over all detections. For GRB 090423 and 100905A, which were not
detected in $z'$, we used the 2MASS field stars in the GROND $J$ band. This method results
in a typical absolute error of 0\farcs{3} in each coordinate. For the bursts that were observed
but not detected by GROND, namely GRB 090429, 120923A, and 131227A, we collected coordinates
from the literature \citep{cucchiara2011, cucchiara20132,tanvir2017n}. For the rest of the
bursts, in Tab. \ref{tab:sample} we simply list the XRT positions taken from
\url{http://www.swift.ac.uk/xrt_positions/}, which are typically good to $\approx2$\arcsec 
; for GRB 000131 we use the position given by \cite{andersen2000}.

     \subsection{Photometric redshifts}\label{sec:phot}

From the 24 GRBs observed with GROND, 15 have a spectroscopically measured
redshift. All the other GRBs have photometric redshift measurements from the
afterglow, of which some are studied in detail in refereed publications and
others are only rough estimates that were published in GCNs. For the latter
we here provide new and more precise constraints based on carefully
analyzed and calibrated GROND data.

The photometric redshift for GRB 080825B of $z_{\mathrm{phot}}\sim 4.3$ was determined
by \cite{kruehler2011a} based on the Lyman-break technique, using multi-band photometry
from GROND and UVOT. We used that same, robust and reliable method to determine
and confirm photometric redshifts of $z_{\mathrm{phot}} > 4$ for GRB 071025, 080916C,
100905A, 140428A, and 151112A. For GRB 100518A, we cannot confirm the previously determined
redshift of $z_{\mathrm{phot}} > 4$. As already noted by \cite{kann2013}, for GRB 131227A a
detection with GROND is ambiguous, and finally, for GRB 090429B we used the photometric redshift
determined by \cite{cucchiara2011}. All spectroscopic and photometric redshifts and
the corresponding references are listed in Tab. \ref{tab:sample}.

Since this is the first time we publish the GROND data and a photometric redshift for
GRB 100905A, we briefly describe the burst in the following subsection. Details and plots
regarding the other five bursts are moved to the Appendix.

        \begin{figure}
                \resizebox{\hsize}{!}{\includegraphics{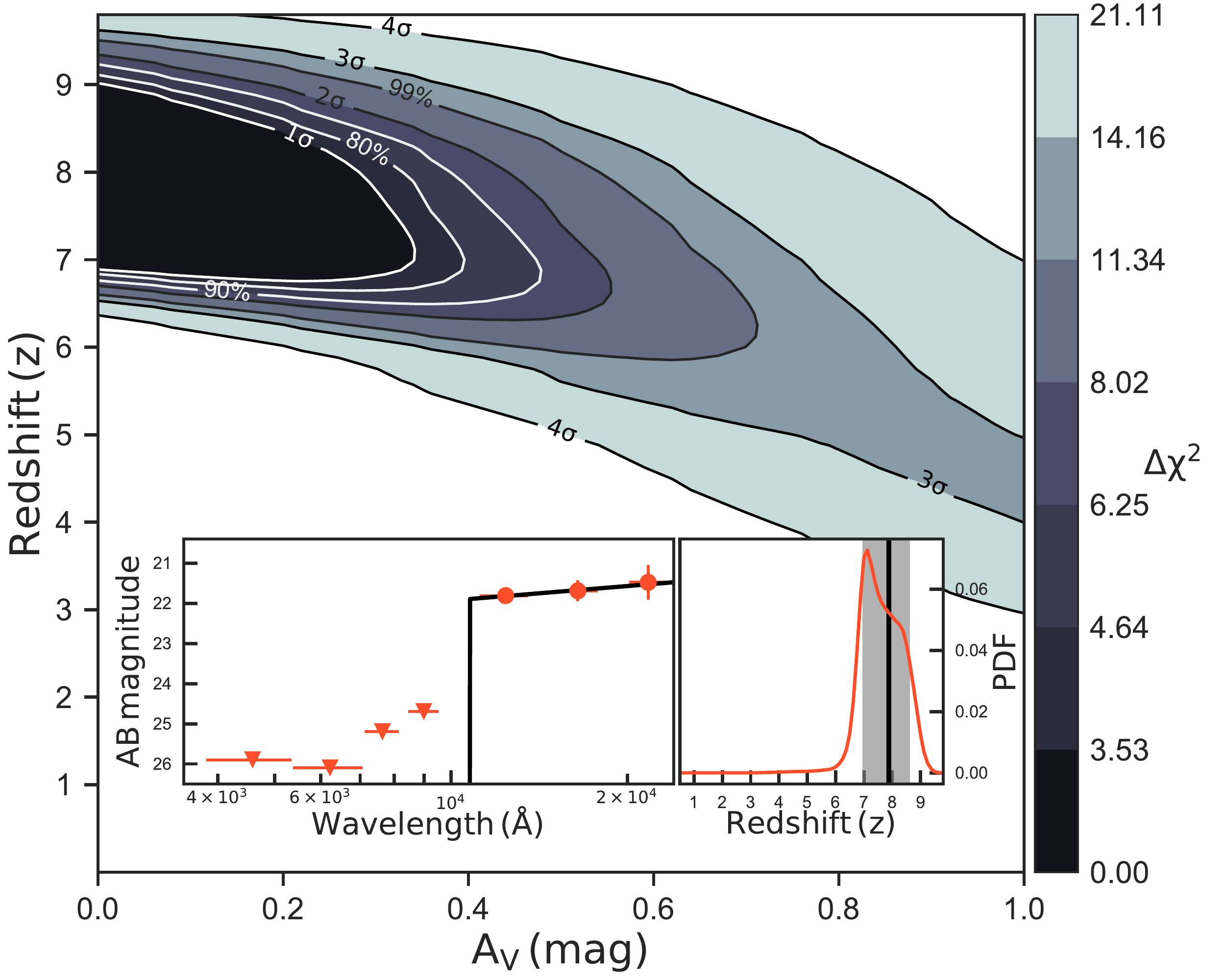}}
                \caption{Contour plot of the $\Delta\chi^2$ values for each of fitted host-intrinsic
        visual extinction $A_V$ and redshift (z) parameters for the best-fit power-law index
        of $\beta = 0.45$. For three degrees of freedom, the significance levels of $1\sigma$
        (68.27\%), $2\sigma$ (95.45\%), $3\sigma$ (99.73\%), and $4\sigma$ (99.99\%) correspond
        to $\Delta\chi^2 = 3.53$, 8.02, 11.35, and 21.11, respectively. As shown in the left
        inset, the GROND SED is best fit with the SMC extinction curve, a power-law slope of
        $\beta = 0.45$, no dust extinction ($A_V = 0.00$ mag), and a photometric redshift of
        $z_{\mathrm{phot}} = 7.88^{+0.75}_{-0.94}$. In the inset on the right we also show the
        corresponding redshift probability density function. The gray shaded area indicates
        the $1\sigma$ confidence interval.}
                \label{fig:100905Acontour}
        \end{figure}

        \begin{figure*}
    \sidecaption
        \includegraphics[width=12.0cm]{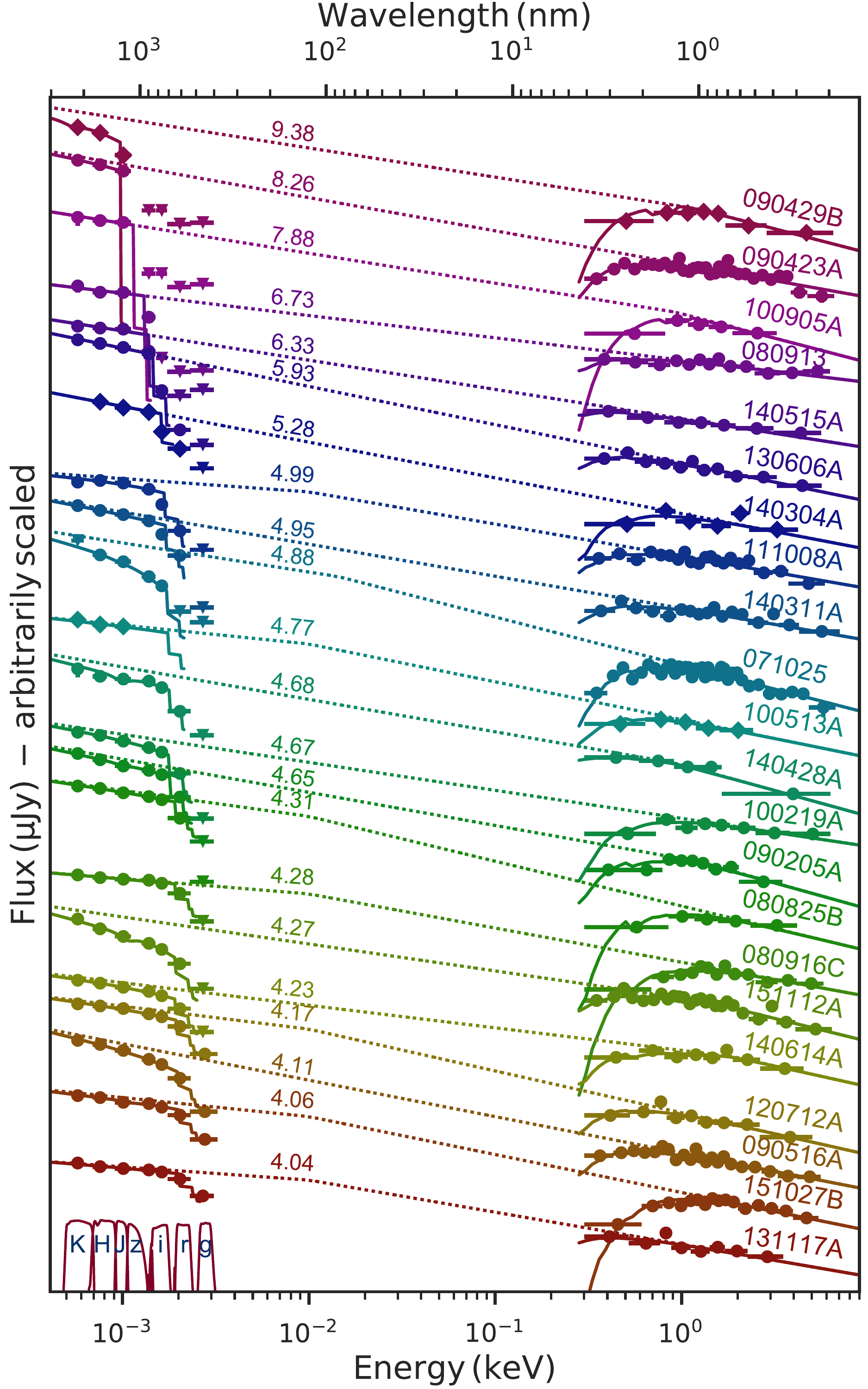}
        \caption{Spectral energy distribution for the 22 GRBs analyzed in this paper with
                increasing redshift from the bottom to top (as labeled). Data for GRBs
        detected with GROND are plotted with circles, data from other instruments with
        diamonds. The X-ray spectrum, if not available simultaneously to the optical and NIR
        data, was flux normalized to the mid-time of the chosen GROND exposure.
        Dashed lines indicate the unabsorbed best-fit models. Solid lines
        indicate the best-fit model including absorption: in the X-rays due to
        galactic plus host intrinsic absorption by medium weight metals; in the optical and NIR
        range due to host intrinsic absorption by dust (the data
        were corrected for galactic foreground reddening beforehand). The flux on
        the y-axis is completely arbitrary to get a better visualization.
        In the left corner at the bottom of the plot we also show the GROND filter curves.}
        \label{fig:allsed}
        \end{figure*}

      \input{table/av.tex}

      \subsection{GRB 100905A}

\emph{Swift}/BAT triggered on GRB 100905A on September 6, 2010 at
$T_0 =$ 15:08:14 UT (MJD $= 55444.63072$) with a duration of $T_{90}= 3.4 \pm 0.5$ s
\citep{barthelmy2010b, marshall2010}.
The XRT started observing the field around
100 s after the trigger. The UVOT observations only lead to upper limits \citep{siegel2010}.
We started observing the field with GROND at around 13 hrs after the BAT trigger and
detected the afterglow in $J$, $H$ and $K_s$ at a common position of RA, Dec
= +02:06:12.04, +14:55:45.80 with an absolute accuracy of 0\farcs31 in each coordinate
(as for all of the other GRBs, the magnitudes and upper limits are given in Tab.
\ref{tab:grondmags}). The afterglow was also detected in $J$, $H,$ and $K$ by \cite{im2010}
using the United Kingdom Infra-Red Telescope (UKIRT). The XRT X-ray, optical, and NIR
light curves are shown in Fig. \ref{fig:100905Alc}. Besides some flaring activity between
$T_0 + 300$ and $T_0 + 600$ s, the XRT light curve is best fit with a single power-law and
a temporal decay index of $\alpha_{\mathrm{x}} = 0.88 \pm 0.03$. When compared to the UKIRT observations,
our GROND magnitudes indicate a somewhat weaker fading of the optical and NIR afterglow
($\alpha_{\mathrm{o}} = 0.60 \pm 0.06$), when assuming a singe power-law decay.

Using the method presented in \cite{kruehler2011a}, we determine a photometric redshift
of $z_{\mathrm{phot}} = 7.88^{+0.75}_{-0.94}$, when fitting the GROND magnitudes with a
single power law, which is reddened by dust following the Small Magellanic Cloud (SMC)
extinction law. The big errors
are the result of the missing wavelength coverage between the $z'$ and $J$ band. The
corresponding $\Delta\chi^2$ contours, given the best-fit spectral slope of
$\beta = 0.45$, as well as the GROND SED and the redshift probability density function,
are shown in Fig. \ref{fig:100905Acontour}. Using only the GROND magnitudes we find no
evidence for absorption by dust ($A_V = 0.00^{+0.27}_{-0.00}$ mag).

So far, no robust redshift measurement for this event is reported in the literature, but,
if correct, it would make GRB 100905A one of the most distant GRBs known to date.
\footnote{\cite{starling2013} assumed a photometric redshift of $z = 7$, based on private
communication with Im. et al. and \cite{little2013} list $z \sim 7.25$ without giving a
reference.}

The very high redshift interpretation is consistent with the non-detection of a host galaxy
in deep NIR imaging from the Hubble Space Telescope. In a total of 10423~s of exposure in the
WFC3/F140W filter, no host is detected down to a 3$\sigma$ limiting magnitude of F140W >
28.5~mag$_{\mathrm{AB}}$. These faint magnitudes are characteristic for high-$z$ GRB hosts
in general \citep{2012ApJ...754...46T, 2016ApJ...825..135M} and would be somewhat unexpected
for a lower redshift ($z\sim2$) galaxy hosting a dust-extinguished GRB \citep{kruehler2011b}.

\subsection{Spectral energy distribution fitting}
\label{sec:34}

In theory, GRB afterglow spectra are featureless and non-thermal, synchrotron spectra
made up of a number of connected power laws. Observations in the X-ray, optical, and NIR regime,
typically sample the same portion of the synchrotron spectrum, i.e., a single power law describes
the X-ray to NIR SED; in some case, however, the synchrotron cooling frequency lies between
the X-ray and NIR spectral range, producing a change in spectral slope at higher energies.
Hence, after rescaling XRT and GROND data to a common reference time and correcting
the GROND magnitudes for the foreground reddening given in Tab. \ref{tab:sample},
the NIR  to X-ray broadband SEDs were fitted in XSPEC with
the combined model of {\tt PHABS$\cdot$ZPHABS$\cdot$ZDUST$\cdot$POW} and
{\tt PHABS$\cdot$ZPHABS$\cdot$ZDUST$\cdot$BKNPOW}. The redshift was fixed
to that obtained from spectroscopic measurements of the afterglow,
host, or photometric dropout, in this order of priority, either
from literature or determined in this work. The galactic foreground hydrogen column
densities were fixed to the values listed in Tab. \ref{tab:sample} and in the case of a broken
power law, the difference in slope between X-ray, optical and NIR wavelengths was
fixed to 0.5, according to the standard fireball model for
the slope difference around the cooling frequency and in line with the majority of
GROND-measured SEDs (\cite{greiner2011}. Filters blueward of Ly$\alpha$ at the GRB redshift were
ignored owing to additional absorption from the Ly$\alpha$ forest. All other parameters
were left free and $\chi^2$ was minimized when fitting each of the three extinction
curves featured in the {\tt ZDUST} model, namely those for the SMC and LMC as well as the
Milky Way (MW), with a single or a
broken power law. Additionally, for the burst showing evidence for dust extinction (i.e.,
$A_V > 0.1$ mag) we also fitted the model from \cite{todini2001} \& \cite{maiolino2004}
for an extinction curve caused by supernova synthesized dust. Its most characteristic
feature is the flattening between 3,000 \AA\ and 1,700 \AA\ followed by a steep decline
(see Fig. 2 in \cite{maiolino2004}). The reason we explored only
one out of various extinction curves proposed for SN-type dust is discussed in Sec. \ref{sec:discuss}.

        \begin{figure}
    \resizebox{\hsize}{!}{\includegraphics{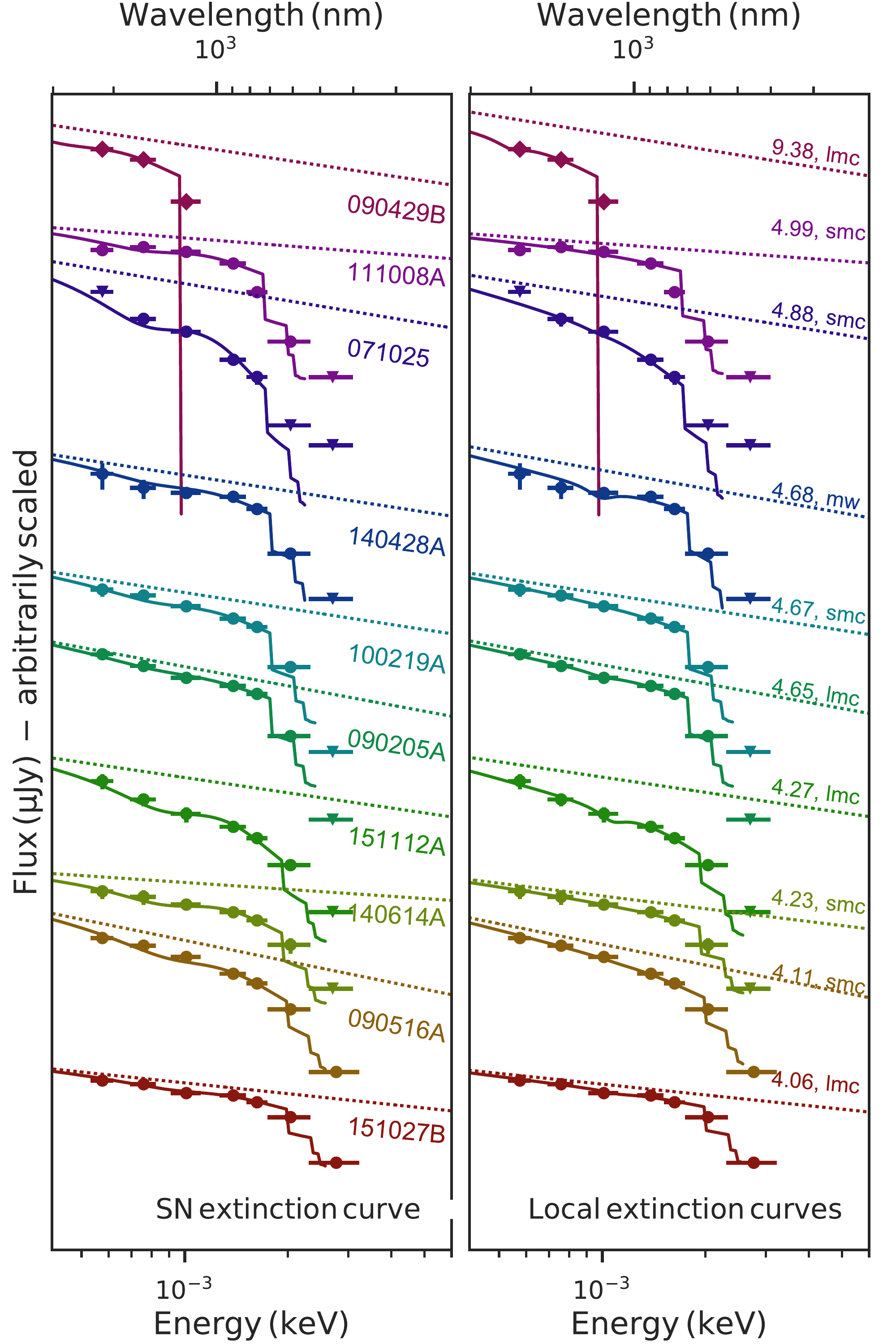}}
        \caption{Similar to Fig. \ref{fig:allsed}, but zoomed into the optical and NIR regime
    (the full plot is moved to the Appendix: Fig. \ref{fig:allsed_sncompl}).
    In the left inset we show the SEDs for the 10 GRBs that show evidence for a medium
    amount of dust ($A_V > 0.1$ mag) fitted with the SN extinction curve. For comparison,
    in the right inset we show again the best result when performing the fit with local
    extinction curves. The goodness of fit is comparable in both scenarios (see Tab.
    \ref{tab:avs} and Section \ref{sec:results}).}
        \label{fig:allsed_sn}
        \end{figure}

Especially for GRBs at high redshift, when more and more of the GROND bands
are affected by absorption caused by the Ly$\alpha$ forest, it is
not feasible to fit more general extinction curves that have more free parameters, such
as $R_{V}$ and the prominence of the 2175 \AA\ feature; for example, the extinction curves proposed by
\cite{ccm1989} or \cite{fm1986} (FM)\footnote{Also, \cite{zafar2011b} found only 4 out of
42 GRBs to be better fit with the FM extinction law.}. This is also the reason for including
quasi-simultaneous XRT data to fit the SED, so that the mostly unabsorbed
and well-covered X-ray spectrum for energies $E > 0.8$ keV allows us to
constrain better the spectral index; this means that we reduce the uncertainty for the
optical and NIR spectral slope to the question of
whether a spectral break of 0.5 is required or not.

The absorption of the X-ray spectrum by medium weight metals at energies below $\text{E} <
0.8$ keV is modeled with the XSPEC models {\tt PHABS} \& {\tt ZPHABS}. However, the resulting
host intrinsic hydrogen column densities $N_{\rm H,X}$ is not discussed further here
because we refer to the recent
findings by \cite{buchner2017}, who have used a more more sophisticated model to fit XRT spectra.
These authors find the distribution of $N_{\rm H,X}$ to be consistent with sources being randomly
distributed in an ellipsoidal gas cloud. This is in contrast to previous studies by, for example,
\cite{starling2013} or \cite{campana2015}, who find a positive dependence with redshift (as we do
here), which could be interpreted as increasing absorption from the IGM with distance.

%Added by TeX Support
\input{table/av_lit.tex}

\subsection{Additional data for GRBs at $z > 4$ not observed or
detected with GROND}\label{sec:35}

Our GROND observations for GRB 090429B only led to upper
limits in all seven bands, and we therefore carried out the analysis on \emph{Gemini}/NIRI $JH$
and $K$-band data published in \cite{cucchiara2011}. We did not carry out the fit for
GRB 120923A ourselves, but used the visual extinction of $A_V = 0.06$ mag determined by
\cite{tanvir2017n}, who have used a very similar method for fitting simultaneous XRT and
\emph{Gemini-N}/NIRI data.  Additionally, for two of the bursts that were not observed by
GROND, namely GRB 100513A and 140304A, we used PAIRITEL $JHK$ and RATIR $grizJH$ data for the
analysis, respectively. So in total we analyzed and fitted 22 GRB broadband SEDs, 19 with GROND and
3 with optical and NIR data from different instruments. For the rest of the GRBs at $z > 4$,
we collected $A_V$ measurements from the literature if available; these measurements are listed in Tab. \ref{tab:avlit}.

\section{Results}\label{sec:results}

The results, i.e., the best-fit models and parameters, from fitting the combined
NIR, optical, and X-ray SEDs with the local extinction curves are summarized in the top
portion of Tab. \ref{tab:avs} and a plot containing all the SEDs is shown in Fig. \ref{fig:allsed}.
As mentioned above, the SEDs of the 10 GRBs that show evidence for a small to
medium amount of dust ($A_V>0.1$ mag) were also fitted with the SN extinction curve.
These SEDs are additionally plotted in Fig. \ref{fig:allsed_sn}, in comparison to the
best results from the local extinction curves (zoomed into the optical and NIR regime; the full
plot is moved to the Appendix: Fig. \ref{fig:allsed_sncompl}). The corresponding parameters
are given in the bottom portion of Tab. \ref{tab:avs}.

\begin{figure}
                \centering
                \includegraphics[width=9cm]{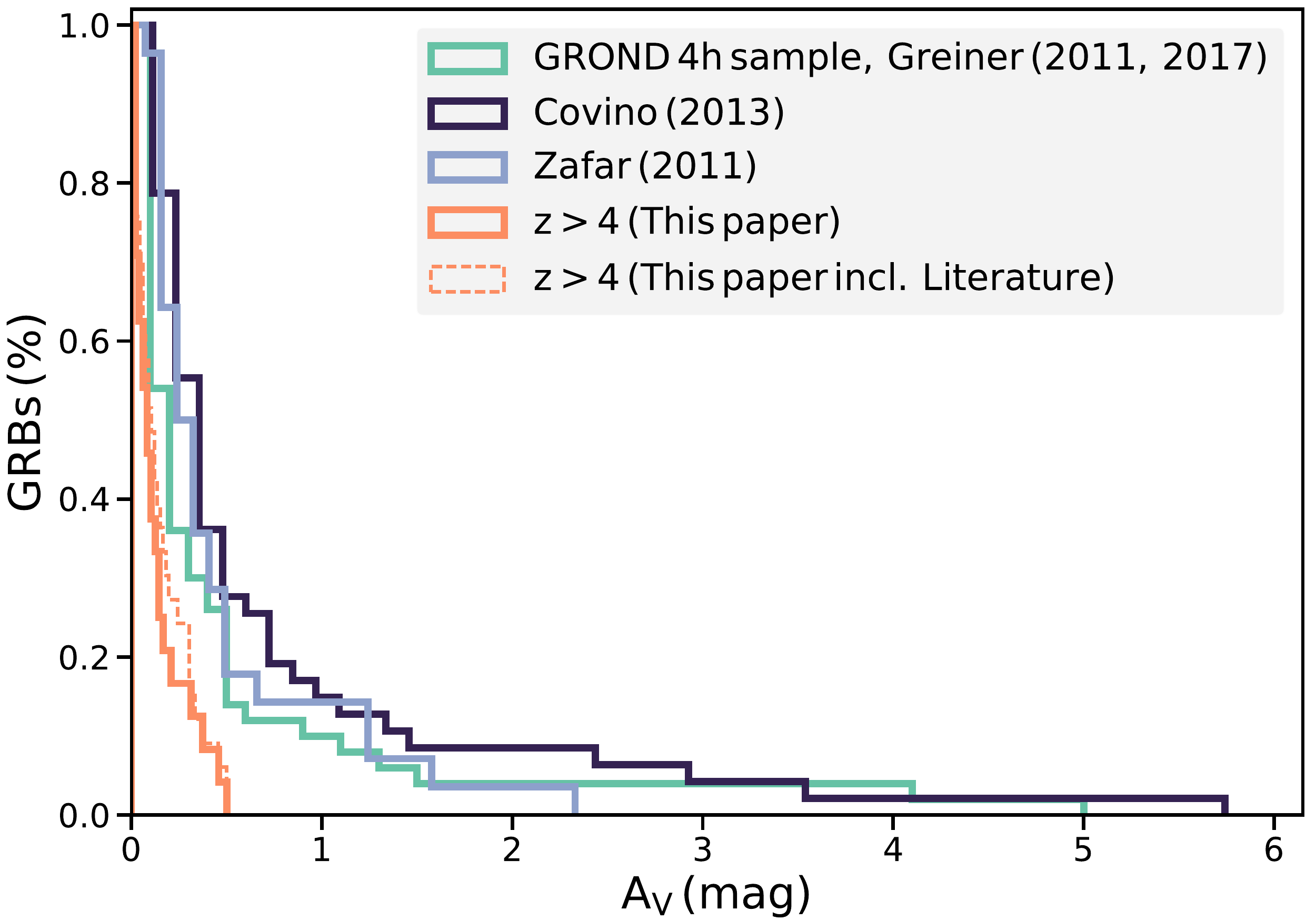}
                \caption{Normalized and reversed cumulative distribution of the host intrinsic dust
        extinction derived for the GRBs with redshift $z > 4$ from this work, compared to the
        BAT6 sample \citep{covino2013} and GROND 4h sample \citep{greiner2011}, which are both
        unbiased samples. We also plot the data from \cite{zafar2011b}; this plot is based on the
        GRB sample from \cite{fynbo2009} and likely biased toward low extinction sightlines.}
                \label{fig:av_dist}
        \end{figure}

With these measurements we increase the number of determined $A_V$ values for GRBs at $z>4$
by a factor of $\sim$\numrange{2}{3}. As previously observed for GRBs at lower redshift, the local
extinction curves provide a good fit to the data, with the featureless SMC extinction curve usually
best describing the observed magnitudes \citep{greiner2011, zafar2011b}. We find that from the 22
modeled GRBs at $z>4$, 16 are best fit with the SMC and only 6 are best fit with the LMC or MW extinction
curve. In contrast to other samples with GRBs at mostly $z<4$, however, we find that all of the
GRBs are only behind small to medium dust column densities ($A_V < 0.5$ mag) within their host
galaxies line of sight, which is also true for the values collected from the literature (see Tab.
\ref{tab:avlit}). The distribution of our best-fit $A_V$ values compared to those from the unbiased
samples from \cite{greiner2011} and \cite{covino2013} and the data from \cite{zafar2011b},
based on the GRB sample from \cite{fynbo2009}, is shown in Fig. \ref{fig:av_dist}. While all
of these samples include a $\sim20$\% of GRBs with $A_V>0.5$ mag (and
$\sim$\numrange{3}{5}\% with $A_V>1.5$ mag), only $\sim 20$\% of our GRBs at redshift
$z>4$ have a visual extinction of $0.2<A_V<0.5$ mag and the remaining $\sim 80$\% have
$A_V<0.2$ mag. A two-sample Kolmogorov-Smirnov test returns a rejection probability of $p\sim68$\%
for the null hypothesis that the $A_V$ values from our GRBs at $z>4$ and those from
\cite{greiner2011} are drawn from the same sample.

Considering the 10 SEDs that were also fitted with the SN extinction curve, we find
that, besides GRB 090205 and 090516, the goodness of fit is comparable to the result
from the local extinction curves, while for two of the bursts, GRB 140428A and 151027B, the SN
extinction curve provides a better fit to the data. In case of GRB 140428A,
the somewhat flat SED between the $H$ and $z'$ band can be identified with the flattening
between (rest-frame) 3,000\,\AA\, and 1,700\,\AA\ of the SN extinction curve ($\chi^2_\mathrm{red} = 1.16$)
or as a small (rest-frame) 2175\,\AA\, feature when fitting the MW extinction curve ($\chi^2_\mathrm{red} = 1.44$).
This result is consistent with that from fitting the GROND SED to determine
a photometric redshift, where we also find that the SN extinction curve is in better
agreement with the data (see App. \ref{app:phot}). For GRB 151027B we find $\chi^2_\mathrm{red}
= 1.05$ for the SN extinction curve, compared to $\chi^2_\mathrm{red} = 1.17$ for the
LMC extinction curve. However, because of the small amount of dust ($A_V \sim 0.10$ mag), 
it is hard to distinguish between the specific features of those extinction curves (see 
Fig. \ref{fig:allsed_sn}).
Also, although the SMC extinction curve provides the best fit
($\chi^2_\mathrm{red} = 1.11$), GRB 071025 can also be well modeled with the SN
extinction curve ($\chi^2_\mathrm{red} = 1.16$), which is in accordance with the results
from \cite{perley2010} and \cite{jang2011} and with our result from fitting the
GROND SED to determine a photometric redshift (see App. \ref{app:phot}). GRB
071025 is, aside from GRB 151112A ($A_V \sim 0.5$ mag), also the burst for which we
find the highest visual extinction (SMC: $A_V\sim 0.45$ mag, SN: $A_V\sim 0.57$ mag).

\section{Discussion and conclusions}\label{sec:discuss}

We have analyzed a sample of 40 GRBs at $z>4$, of which 22 were used to
measure the host intrinsic visual extinction $A_V$ and study the shape of the
dust extinction curves toward the GRB lines of sight. Since all of the bursts,
including the above-mentioned cases of GRB 071025A and 140428A, can be modeled with locally
measured extinction curves, we cannot draw any firm conclusion about whether
an SN-type dust extinction curve is truly required for some of the GRBs.
Other SN-dominated extinction curves have been proposed by, for example,
\cite{bianchi2007, hirashita2008, hirashita2010}, and \cite{nozawa2015}. 
But, since only the extinction curve from \cite{todini2001} and \cite{maiolino2004}
was available for us in an analytical form, we restricted our analysis to their results.
A common feature, however, that these extinction curves share is a strong steepening
in the UV. This steepening is hard to measure at $z > 4$ anyway because of the dominant absorption
from the Lyman-$\alpha$ forest. Also, the low $A_V$ values combined with
a limited wavelength coverage limit the distinction of the various characteristic features.
So from here, we focus the discussion to the somewhat unexpected lack
of highly dust-extinguished GRBs, which can only be explained as a result of one or both
of the following suppositions:

        \begin{enumerate}
        \item At a redshift of $z>4$, when the Universe was less than $\sim1.6$ Gyr
    old, not enough dust was present in young GRB host galaxies to cause visual
    extinctions of $A_V>0.5$ mag.
        \item High dust-extinguished GRBs at $z>4$ are absent in our sample and other samples
    because their optical and NIR afterglow was too faint to be detected.
        \end{enumerate}

        \begin{figure*}
                \centering
                \resizebox{\hsize}{!}{\includegraphics{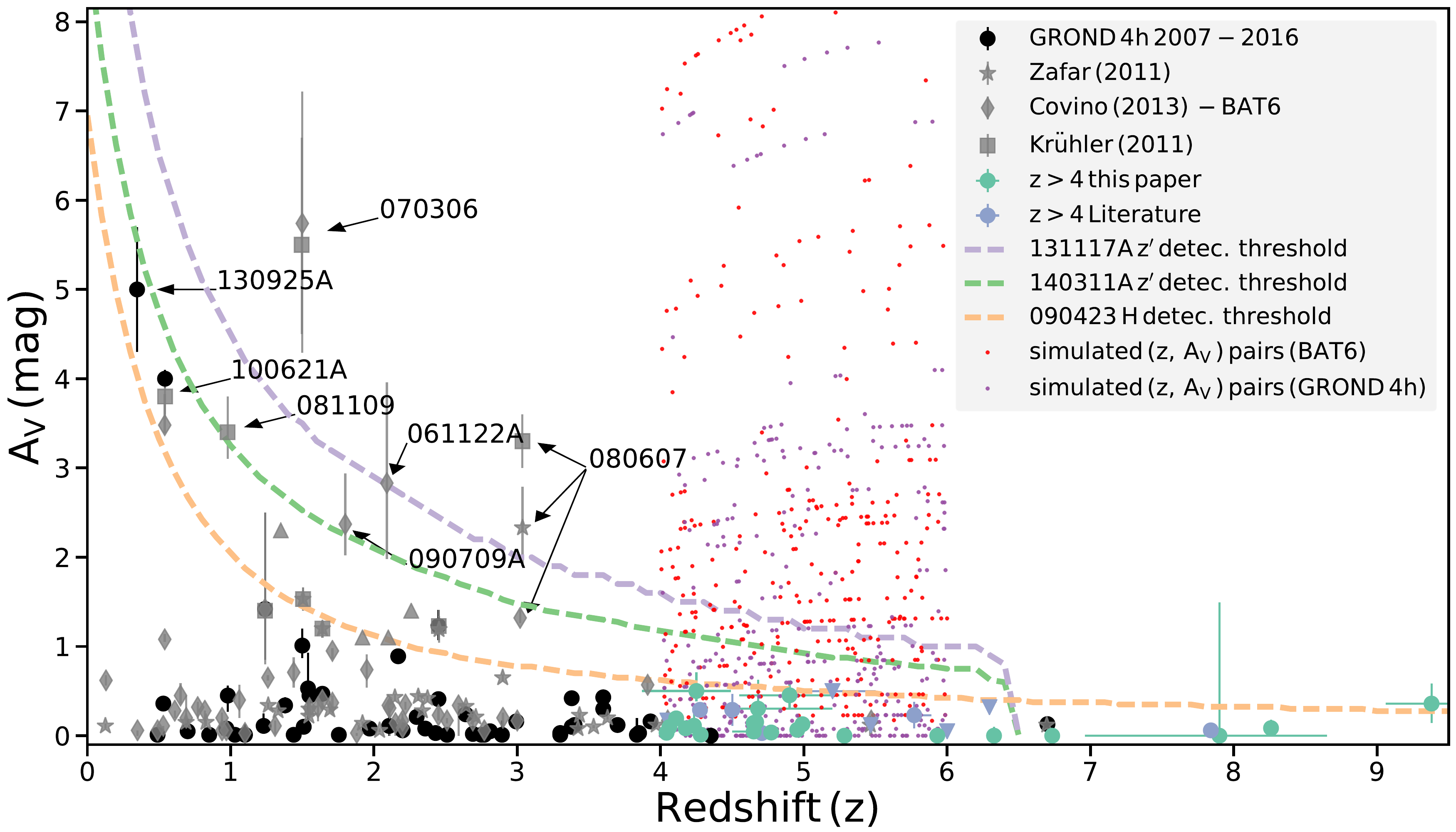}}
                \caption{Host intrinsic visual dust extinction $A_V$ toward the line of sight
        of the GRB plotted against redshift for the GRBs analyzed in this paper and in comparison
        to different samples at mostly lower redshift \citep{covino2013, greiner2011,
        kruehler2011b, zafar2011b}. The dashed lines represent the GROND detection thresholds
        for GRB 131117, 140311A, and 090423 in the given band (see Sec. \ref{sec:discuss}).
        Highly extinguished GRBs are labeled. The red and purple dots represent 500 from
        the $10^{6}$ simulated ($z$, $A_{V}\mathrm{(obs.)}$) pairs.}
                \label{fig:avz}
        \end{figure*}

In order to check if the lack of higher extinguished afterglows can be explained with
an observational bias, we calculated theoretical GROND detection thresholds for three of
the afterglow SEDs from Tab. \ref{tab:avs}, by artificially placing them at
redshifts between $0 < z < 10$. The SED for GRB 131117A corresponds to an epoch observed
at just $\sim 0.1$ hrs after the trigger and
is additionally very flat ($\beta_o\sim0.25$). The SED of GRB 140311A is more representative of the GRBs
studied here; its SED is much steeper ($\beta_o \sim 0.85$) and was created from
the afterglow emission at $T_0 + 9.8$ hrs. For both of these bursts we assume a typical
limiting magnitude in the GROND $z'$ band of $z'_{\mathrm{lim}} = 24.2$ mag. As a more extreme
example, we also calculate the detection threshold for the SED of the very high redshift
GRB 090423, which corresponds to an observation taken at $T_0 + 17.3$ hrs ($\beta_o \sim 0.88$).
Here we assume a typical GROND $H$-band limiting magnitude of $H_{\text{lim}} = 22$ mag.
Our results are represented by the dashed lines in Fig. \ref{fig:avz}, where we plot our
$A_V$ values against redshift in comparison to different GRB samples at mostly lower redshift.
Above these lines, the dust extinction would be so high that these GRBs would not have been
detected by GROND (at the given time the SED was created and if the GRB had been detected from the redshift we artificially placed it at). %\LEt{Please check for intended meaning. This was ambiguously worded. If possible, integrate parenthetical statement into sentence.}
One can see that a few exceptions
fall above these lines. For example, GRB 070306 was observed with a larger telescope and GRB
080607, 061222A, and 090709A were observed within a very short time after the \emph{Swift}/BAT
trigger \citep{cenko2006,jaunsen2008, cenko2010a, perley2011}.

Although the detection thresholds for these burst clearly show that it is much harder to detect
highly extinguished afterglows at $z>4$, the absence of burst with $0.5<A_V<1.5$ mag, i.e., the lack of
data points between these lines given the size of our sample, seems somewhat unexpected. Therefore, to further test to what extent we suffer from an observational bias, we made the following
Monte Carlo simulation. We derived the intrinsic brightness in the GROND $z'$ band for the 17 GRBs at
redshift $4.0<z<6.0$ and distributed random values out of the $A_V$ distribution from the GROND 4h
sample \citep{greiner2011} or the BAT 6 sample \citep{covino2013}, by also putting each GRB at a
random redshift between $4.0<z<6.0$\footnote{Since the extinction from dust increases
from the red to blue wavelengths, this of course increases the extinction in the given
(observer frame) band.}.
We thus assume that the intrinsic brightness of the GRB is unrelated to the absorption by
dust within the host galaxy. In Fig. \ref{fig:avz} we plot 500 from $10^{6}$ simulated ($z$,
$A_V\mathrm{(obs.)}$) pairs for both samples. In case of the $A_V$ distribution from
\citep{greiner2011}, on average we would expect to find $n=0.8\pm 0.9$ bursts with $A_V>0.5$
mag that are brighter than $24.2$ mag in the $z'$ band and $n=2.0\pm 1.3$ in case of the
$A_V$ distribution from the BAT 6 sample. The probabilities of detecting zero GRBs with $A_V>0.5$
mag are $p=43.6$\% and $p=11.4$\%, respectively. To detect at least one burst with $A_V>0.5$
mag at a confidence of $1\sigma$, it would require a sample size of $n=21$ (GROND 4h sample)
or $n=10$ (BAT 6) GRBs. The different outcomes of the two samples can be explained by the
higher percentage of bursts with $0.5<A_V\lesssim2.0$ mag in the sample from \cite{covino2013},
which are the only bursts we are theoretically able to detect at $z>4$, given our brightness
distribution. Although the case is less clear for the $A_V$ distribution from the
GROND 4h sample, our sample size should be big enough to contain at least one GRB with
$A_V>0.5$ mag at least after also including the GRBs that were not observed with GROND.
%\LEt{This was a bit difficult to follow. Please check for intended meaning. }

Hence, these considerations suggest that we partly suffer from an observational bias toward
highly extinguished GRBs, meaning that we cannot expect to easily detect bursts with
$A_{V}>0.5$ mag at $z>4$ with a sensitivity of about $\sim24$ mag (reachable with a 2m telescope). Nevertheless,
our results can be interpreted as evidence that GRB host galaxies at high
redshift are on average less dusty than at $z\sim2$.

To further test the occurrence of $A_{V}>0.5$
mag at $z>4$, very rapid observations at NIR wavelengths are required, which are difficult to achieve
in large numbers from Chile since the South Atlantic Anomaly suppresses the number of
night-time GRBs by more than a factor of two.

\begin{acknowledgements}
The author acknowledges support from a studentship at the
European Southern Observatory in Chile and thanks the many astronomers
who dedicated their time observing the numerous GRBs during the operation
of GROND between 2007 and 2016. We also thank the anonymous referee for
providing valuable and constructive feedback that help improving the presentation
of this work. J.B. and T.K. acknowledge support through
the Sofja Kovalevskaja Award to P.S. from the Alexander von Humboldt
Foundation of Germany. Part of the funding for GROND (both hardware
and personnel) was generously granted from the Leibniz-Prize to Prof.
G. Hasinger (DFG grant HA 1850/28-1).
\end{acknowledgements}

\bibliographystyle{aa}
%\bibliography{dust}
\bibliography{main.bib}

\clearpage 

\begin{appendix}

\section{Photometric redshifts}\label{app:contours}

\begin{figure}
                \resizebox{\hsize}{!}{\includegraphics{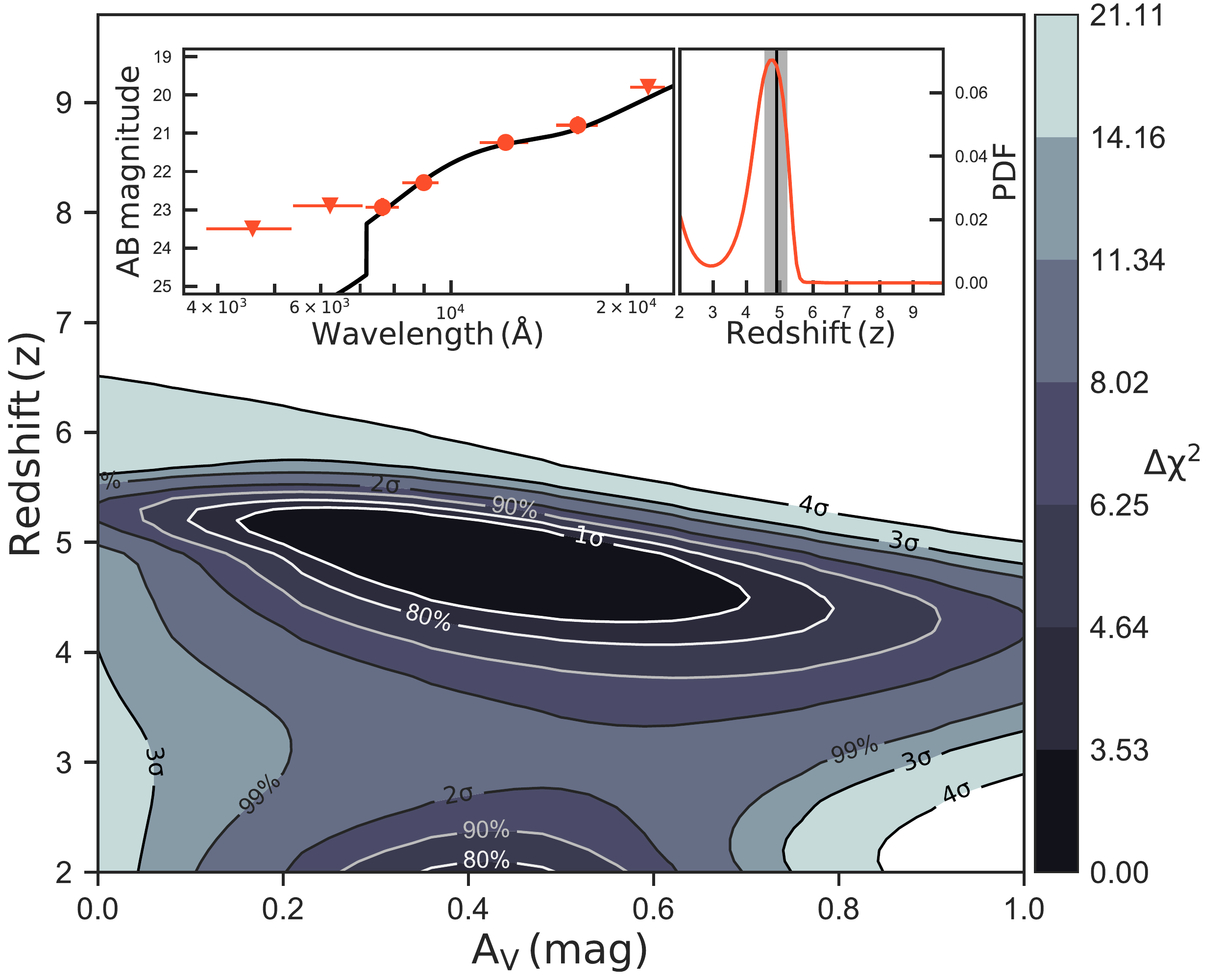}}
                \caption{GRB 071025}
                \label{fig:071025contour}
        \end{figure}

\begin{figure}
                \resizebox{\hsize}{!}{\includegraphics{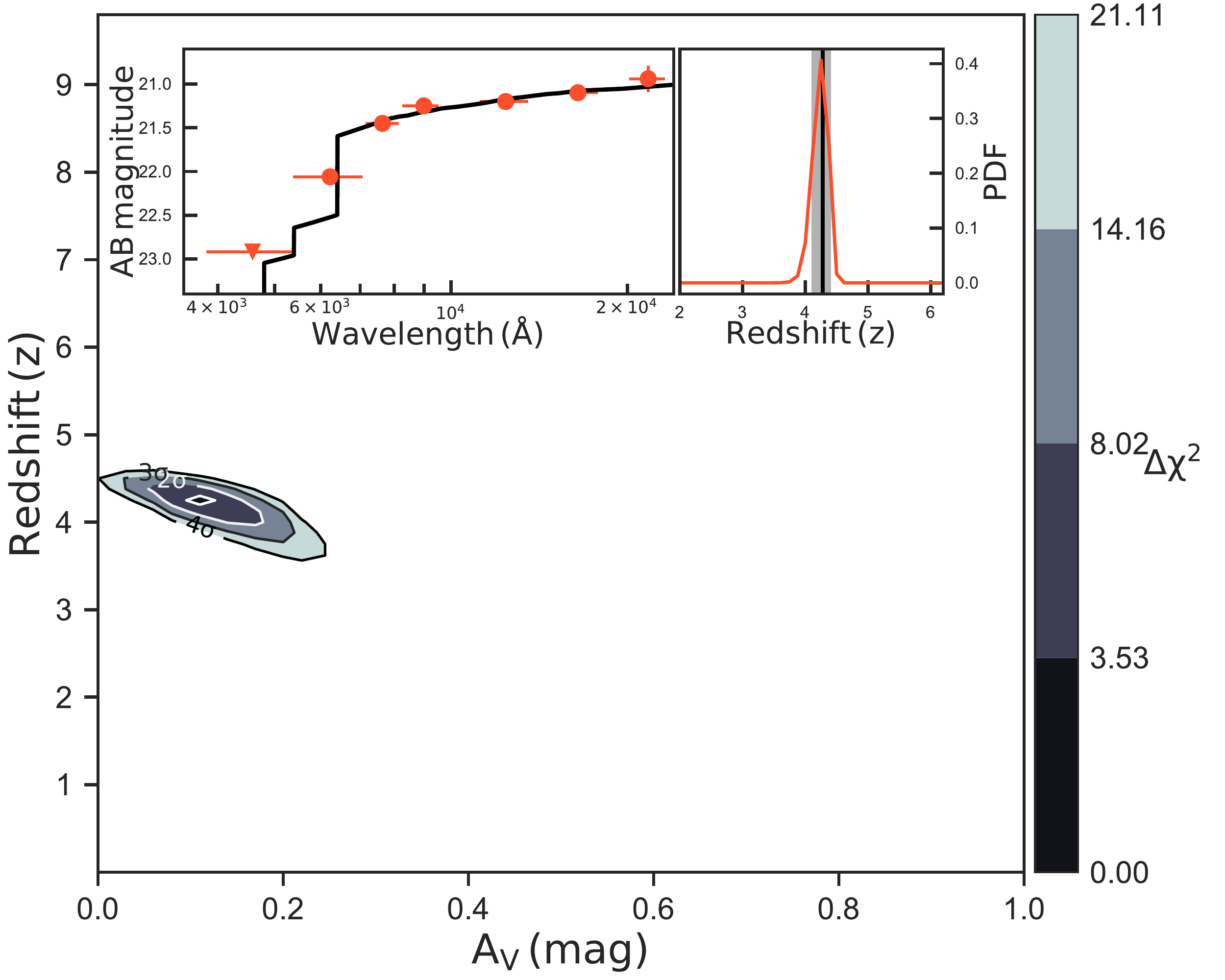}}
                \caption{GRB 080916C}
                \label{fig:080916Ccontour}
        \end{figure}

\begin{figure}
                \resizebox{\hsize}{!}{\includegraphics{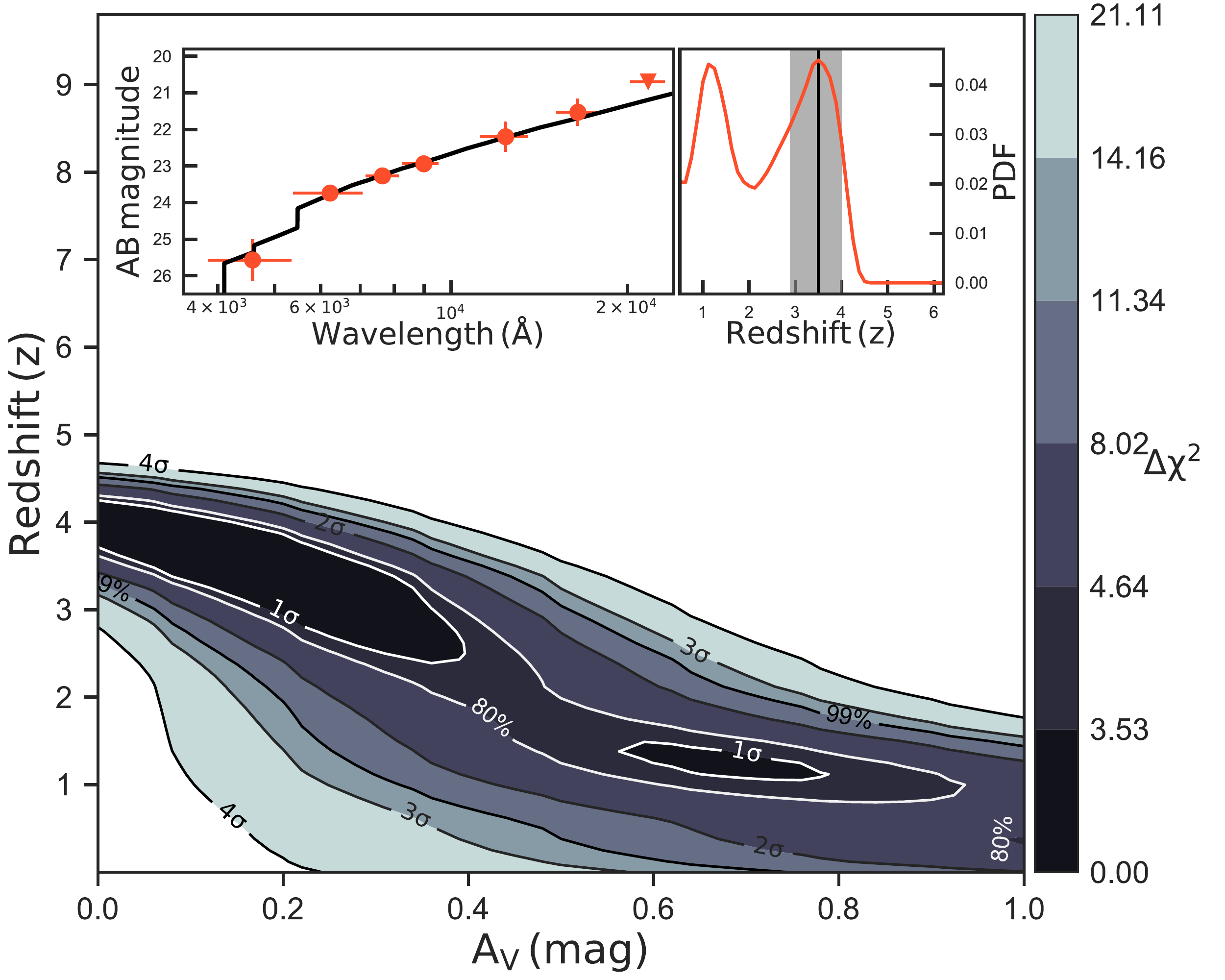}}
                \caption{GRB 100518A}
                \label{fig:100518Acontour}
        \end{figure}

\begin{figure}
                \resizebox{\hsize}{!}{\includegraphics{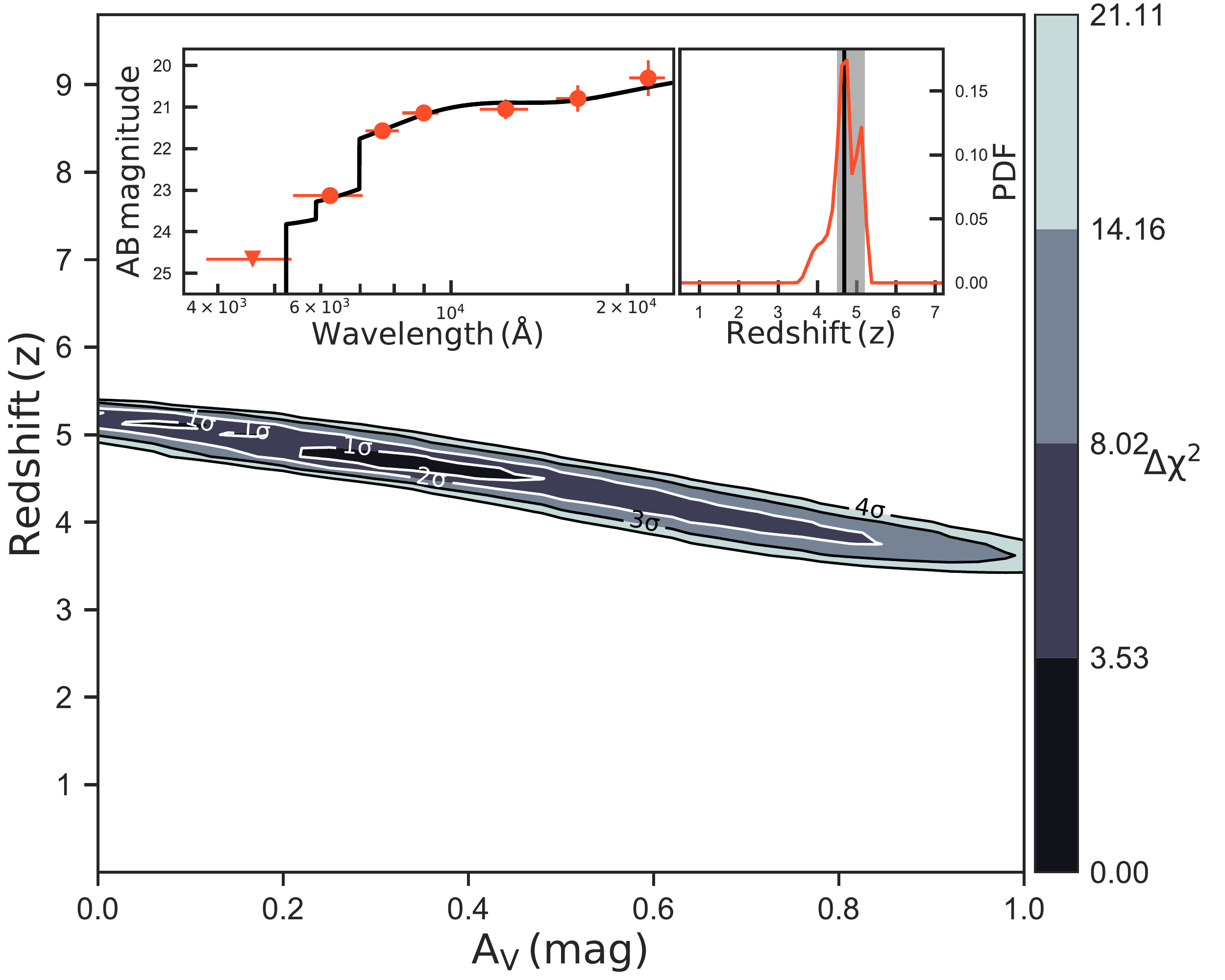}}
                \caption{GRB 140428A}
                \label{fig:140428Acontour}
        \end{figure}

\begin{figure}
                \resizebox{\hsize}{!}{\includegraphics{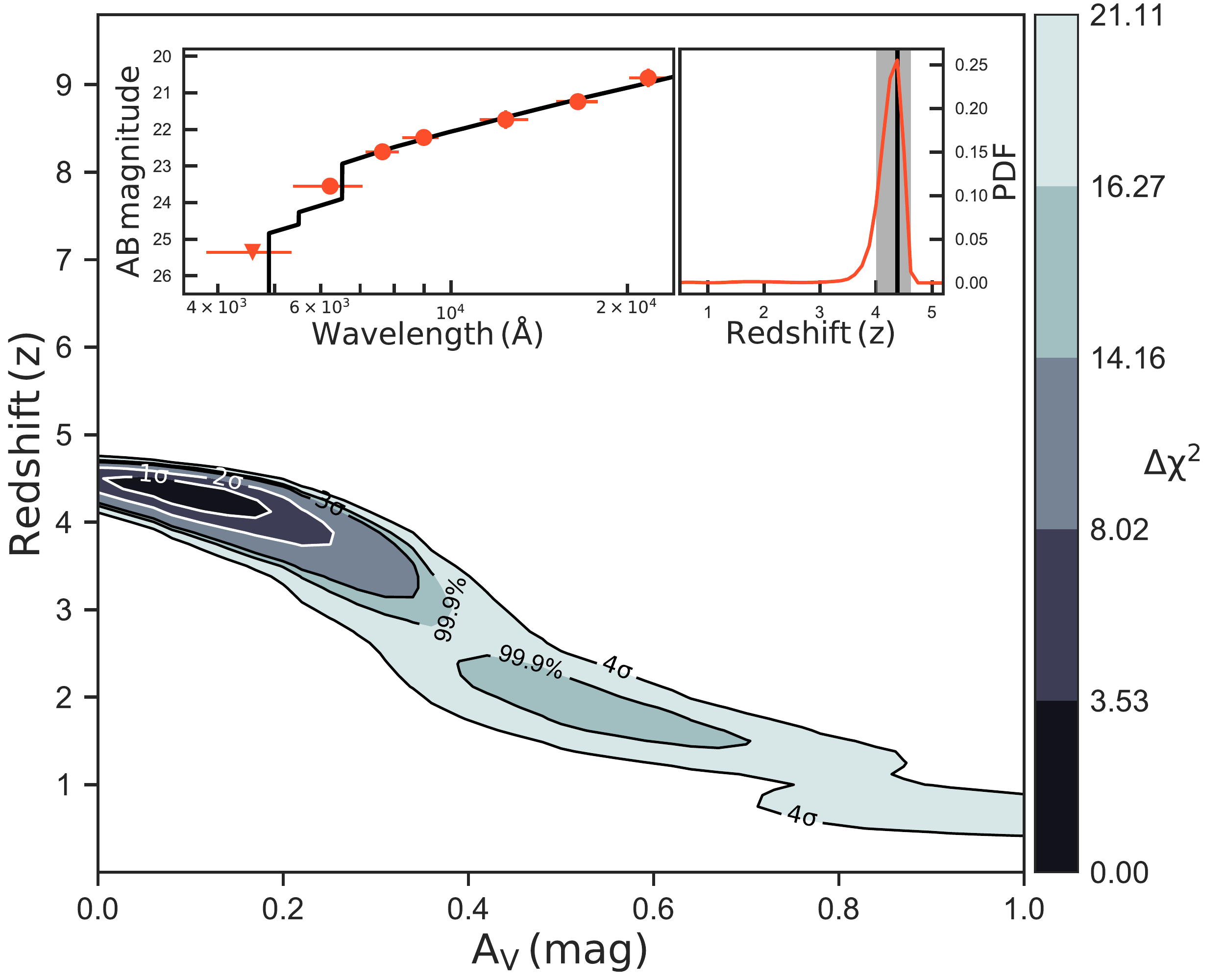}}
                \caption{GRB 151112A}
                \label{fig:151112Acontour}
        \end{figure}

In Fig. \ref{fig:071025contour} to Fig. \ref{fig:151112Acontour} we show  the GROND SEDs
and $A_V$ versus redshift contour plots for GRB 071025, 080916C, 100518A, 140428A, and
151112A.

\subsection{Individual bursts}

\paragraph{GRB 071025} is best fit with the SN-like dust extinction law, $A_V=0.39^{+0.20}_{-0.14}$ mag,
and a photometric redshift of $z_{\mathrm{phot}} = 4.88 \pm 0.35$, which is consistent with the
photometric redshifts of $4.4 < z_{\mathrm{phot}} < 5.2$ and $4.6 < z_{\mathrm{phot}} < 4.85^{+0.05}_{-0.10}$
reported by \cite{perley2010} and \cite{jang2011}, who also find evidence for extinction caused by
supernova-synthesized dust. We  limited the redshift parameter space to $z > 2$ to prevent a low-redshift solution that would be in strong disagreement with the results
from the above-mentioned authors.

\paragraph{GRB 080916C} is best fit with $z_{\mathrm{phot}} = 4.28^{+0.06}_{-0.10}$,
$A_V=0.11^{+0.08}_{-0.09}$ mag and the SMC extinction curve. This is consistent
with the results from \cite{greiner2009080916C}.

\paragraph{GRB 100518A} For GRB 100518A we cannot confirm the previously reported 
redshift of $z_{\mathrm{phot}} =4.0^{+0.3}_{-0.5}$ \citep{greiner2015}. Our GROND
magnitudes, as listed in Tab. \ref{tab:grondmags}, are best fit with a
$z_{\mathrm{phot}}=3.50^{+0.50}_{-0.62}$ and a host intrinsic extinction of $A_V =0.19^
{+1.48}_{-0.19}$ mag, indicating that it is difficult to distinguish between a high redshift
and low visual extinction or vice versa. The high redshift ($z \gtrsim 3$) nature
of the GRB is further supported from late-time observations in the \emph{VLT}/FORS2 R band,
not revealing a host galaxy down to a limit of > 28.7 mag \citep{greiner2015}.

\paragraph{GRB 140428A} is best fit with a photometric redshift of $z_{\mathrm{phot}} = 4.68^{+0.52}_{-0.18}$,
which is in accordance with the estimate of $z \sim 4.7$ from the LRIS spectrum \citep{perley201428b}
and the SN extinction law ($A_V=0.36^{+0.06}_{-0.32}$ mag).

\paragraph{GRB 151112A} is best fit with a photometric redshift of $z_{\mathrm{phot}}
= 4.27^{+0.24}_{-0.38}$ and the SMC extinction law ($A_V=0.08^{+0.34}_{-0.02}$ mag).
This is consistent with the initial estimate from \cite{bolmer2015}.

\section{Light curves of the X-ray, optical, and near-infrared afterglow}\label{app:lcs}

The GROND and XRT observations of the afterglow were supplemented - if available -
by additional observations reported in the literature to create and fit
as well-covered light curves as possible. To fit the afterglow light curves we used
phenomenological models such as a single or smoothly broken power law or a combination
of both (similar to, e.g., \cite{zanioni2013}). Flares and re-brightenings were modeled by
adding a Gaussian component, and a possible contribution from the host galaxy was
modeled by adding a constant term,

\begin{equation}
F_{\nu}(t) \propto t^{-\alpha_1} \left( + \left(\left(\frac{t}{t_b}\right)^{-s\cdot\alpha_2} + \left(\frac{t}{t_b}\right)^{-s\cdot\alpha_3}\right)^{-1/s}\right)
\left(+ e^{-\frac{1}{2}\cdot \left(\frac{t-t_{\mathrm{mid}}}{\sigma_t}\right)^2}\right) \left(+h\right)
.\end{equation}

This method is sufficient to identify regions of temporal (and possibly spectral)
evolution and to rescale XRT and/or GROND data to a common reference time. In most cases
we kept the GROND data fixed and used the usually better covered XRT light curve
to rescale the X-ray spectrum. We only used the model best fit to the GROND
light curves for GRB 080825B, for which the XRT light curve 
consists of only four data points.

\begin{figure}
                \resizebox{\hsize}{!}{\includegraphics{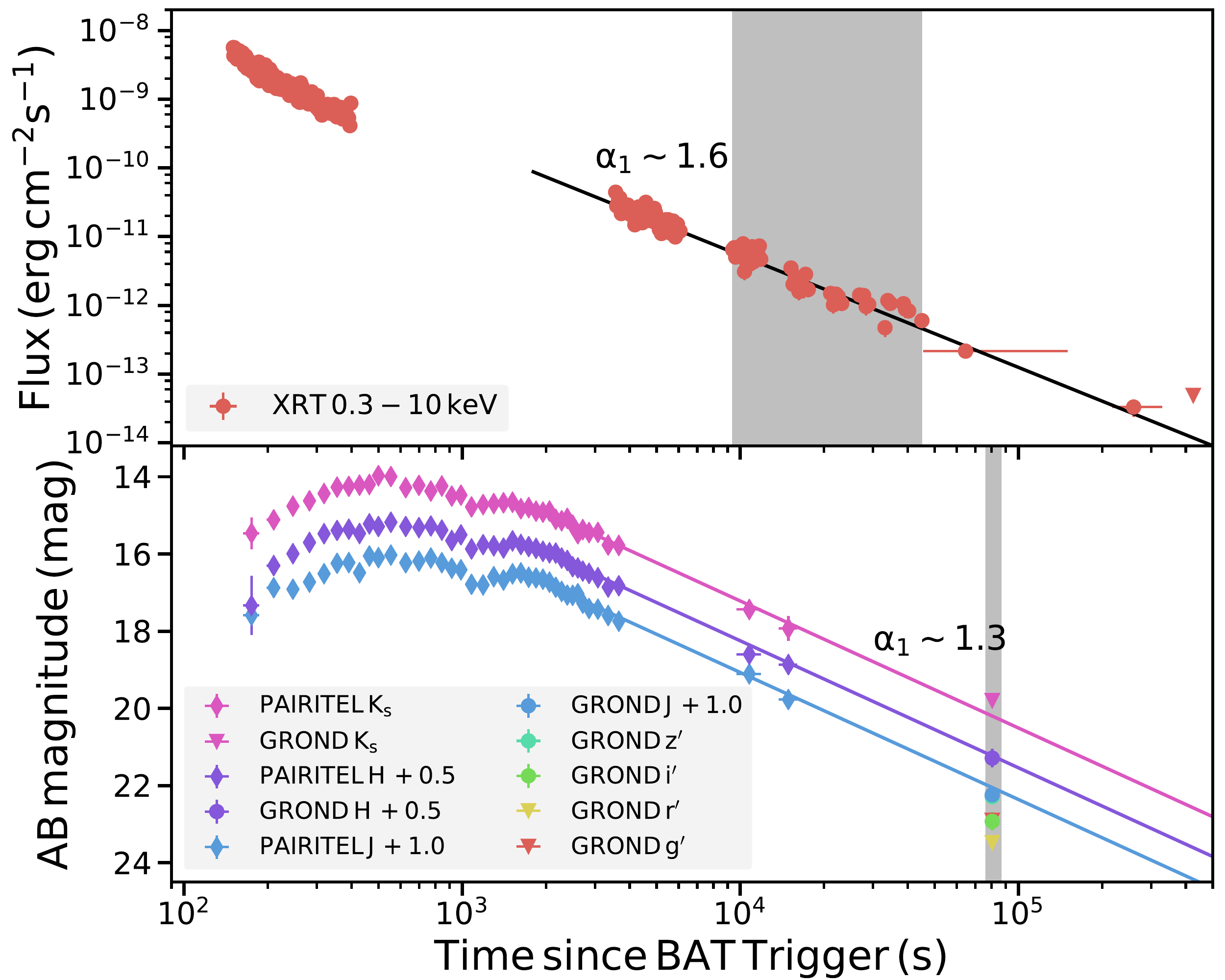}}
                \caption{GRB 071025}
                \label{fig:071025lc}
        \end{figure}

\begin{figure}
                \resizebox{\hsize}{!}{\includegraphics{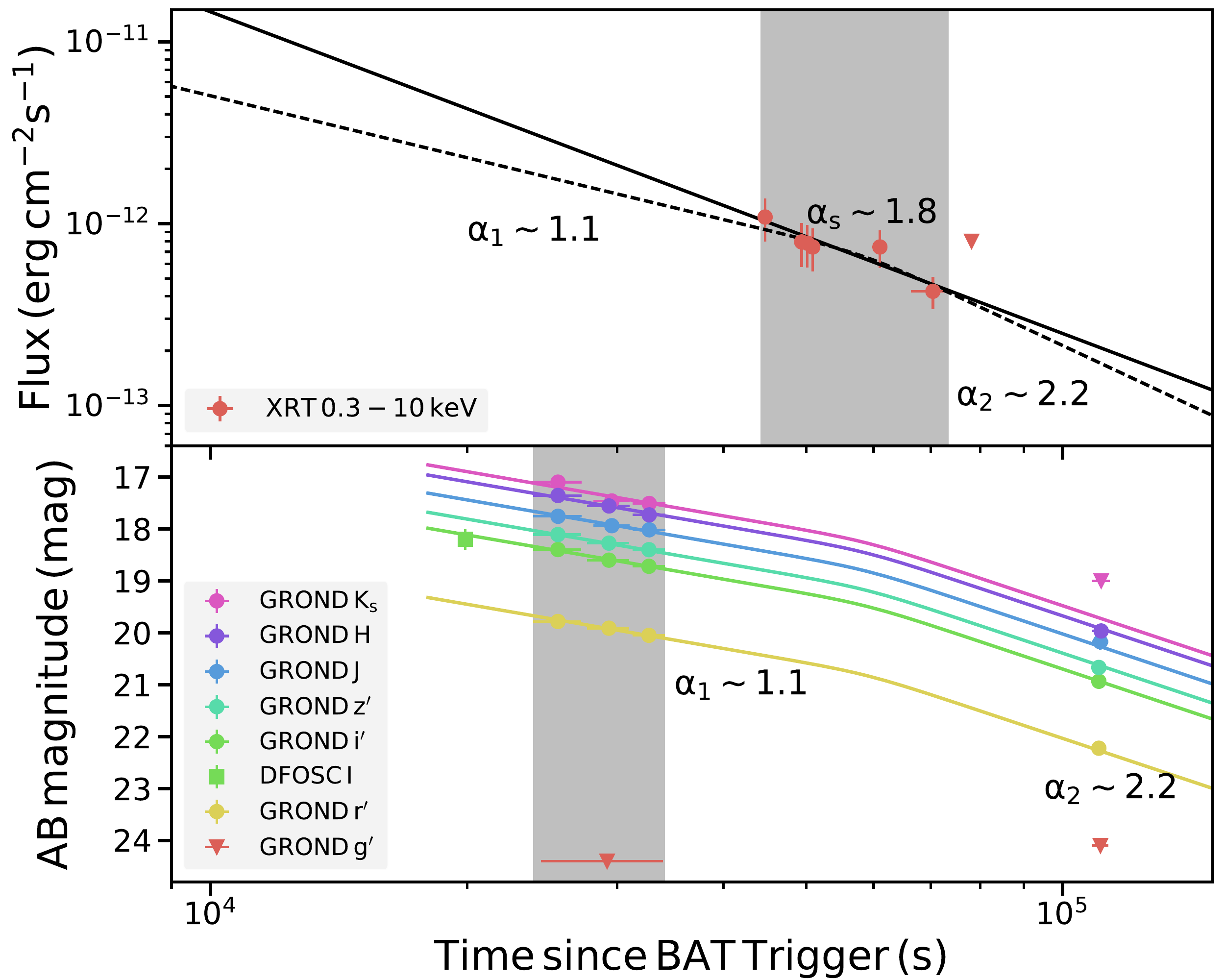}}
                \caption{GRB 080825B}
                \label{fig:080825Blc}
        \end{figure}

\begin{figure}
                \resizebox{\hsize}{!}{\includegraphics{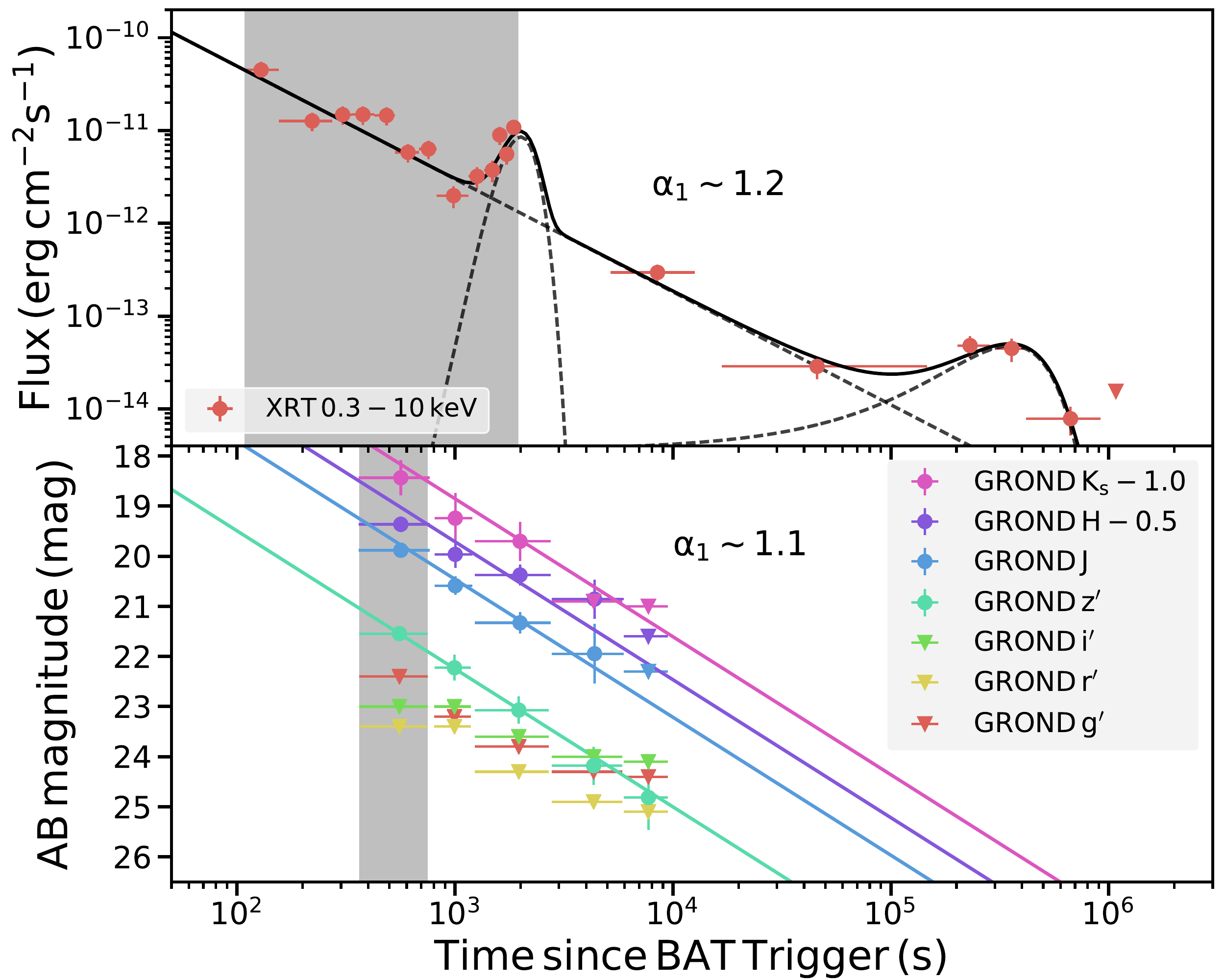}}
                \caption{GRB 080913}
                \label{fig:080913lc}
        \end{figure}

\begin{figure}
                \resizebox{\hsize}{!}{\includegraphics{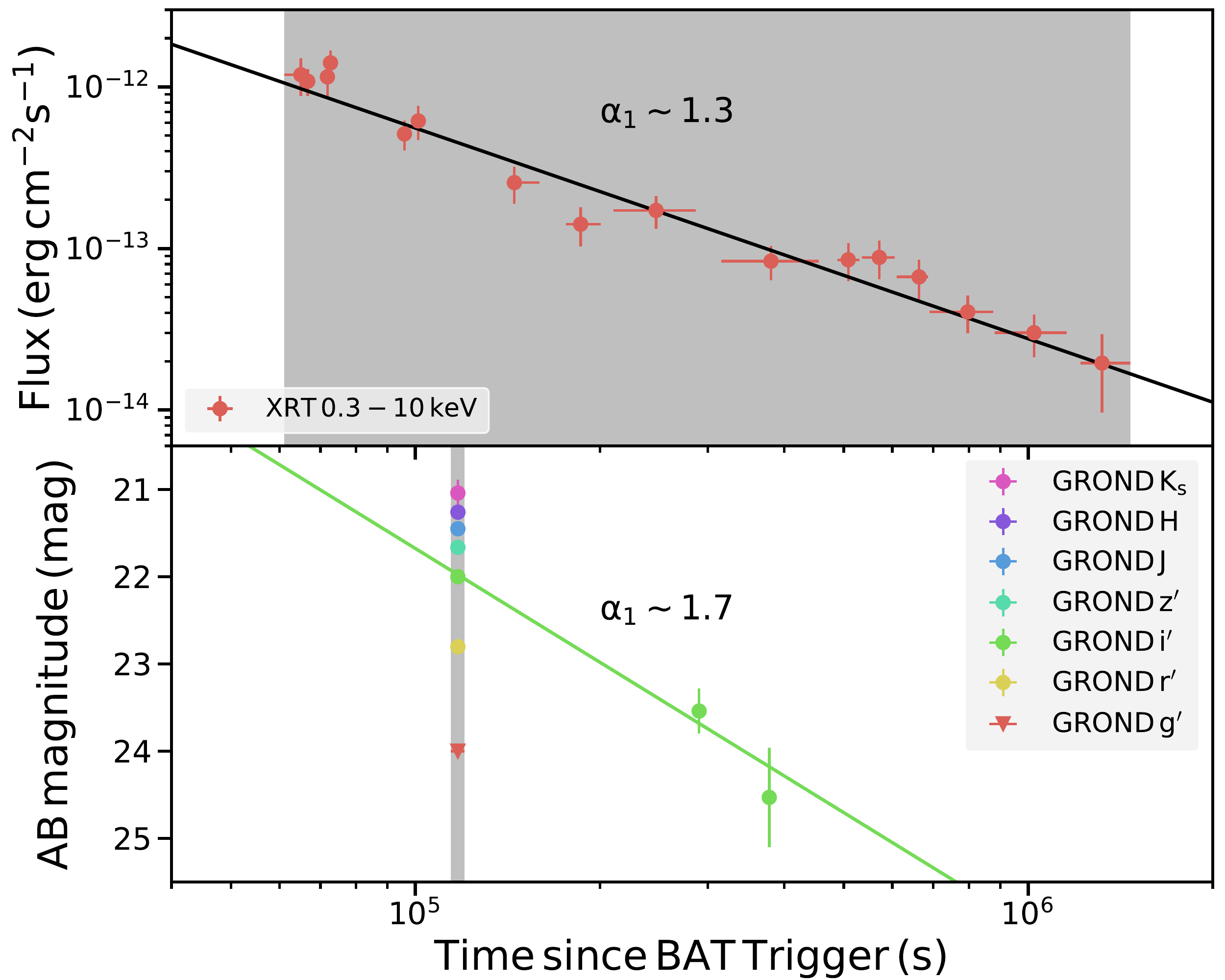}}
                \caption{GRB 080916C}
                \label{fig:080916Clc}
        \end{figure}

\begin{figure}
                \resizebox{\hsize}{!}{\includegraphics{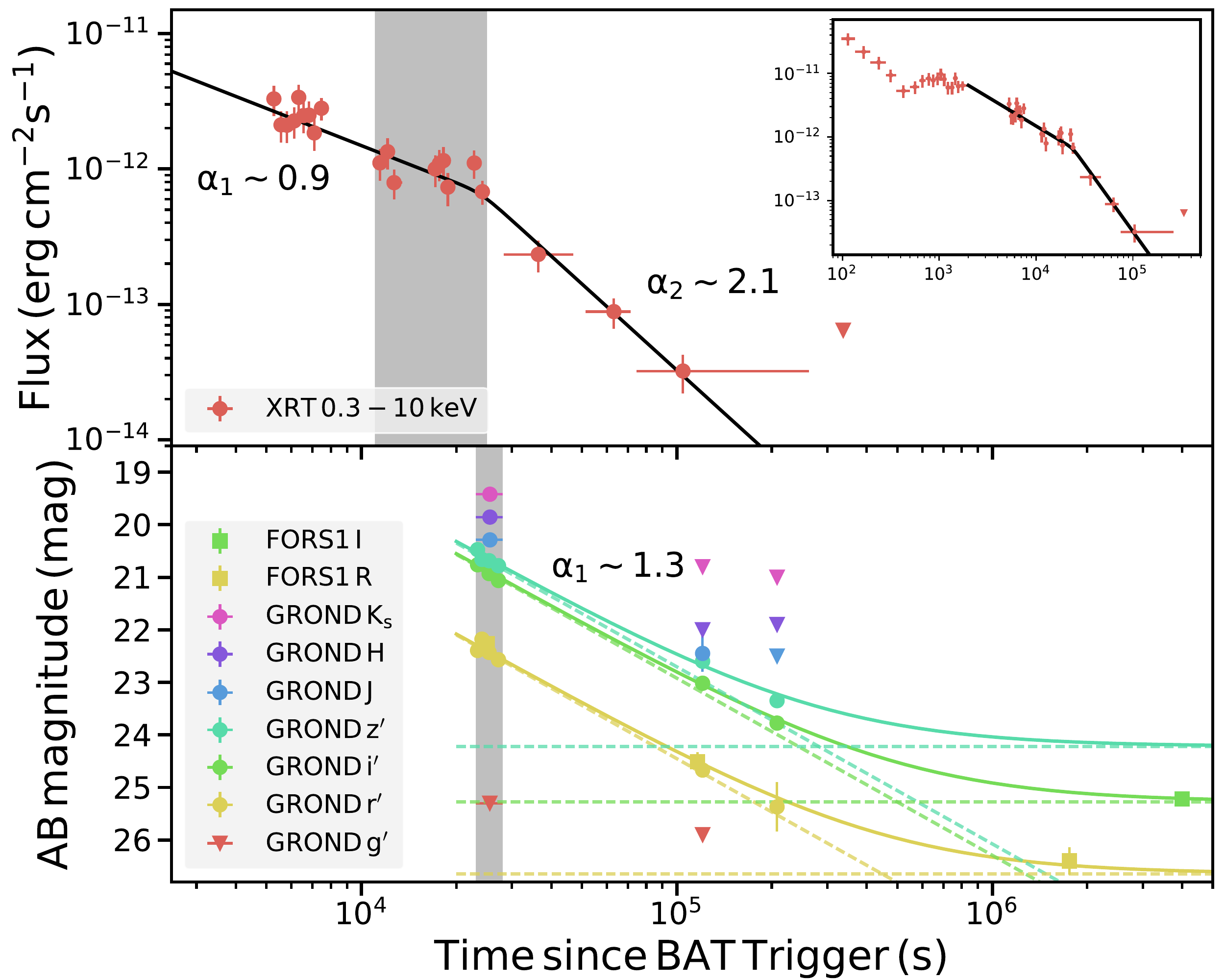}}
                \caption{GRB 090205}
                \label{fig:090205lc}
        \end{figure}

\begin{figure}
                \resizebox{\hsize}{!}{\includegraphics{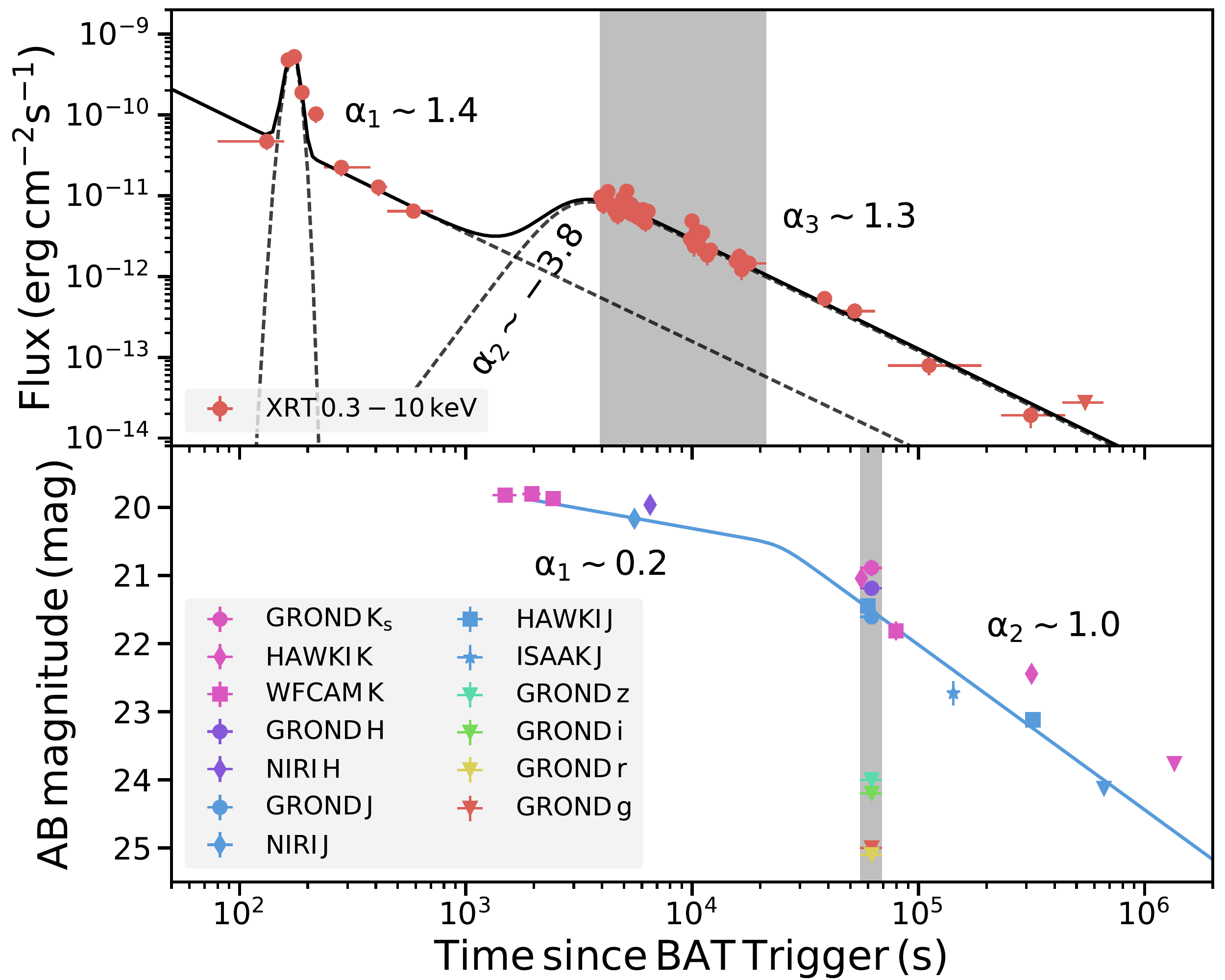}}
                \caption{GRB 090423}
                \label{fig:090423lc}
        \end{figure}

\begin{figure}
                \resizebox{\hsize}{!}{\includegraphics{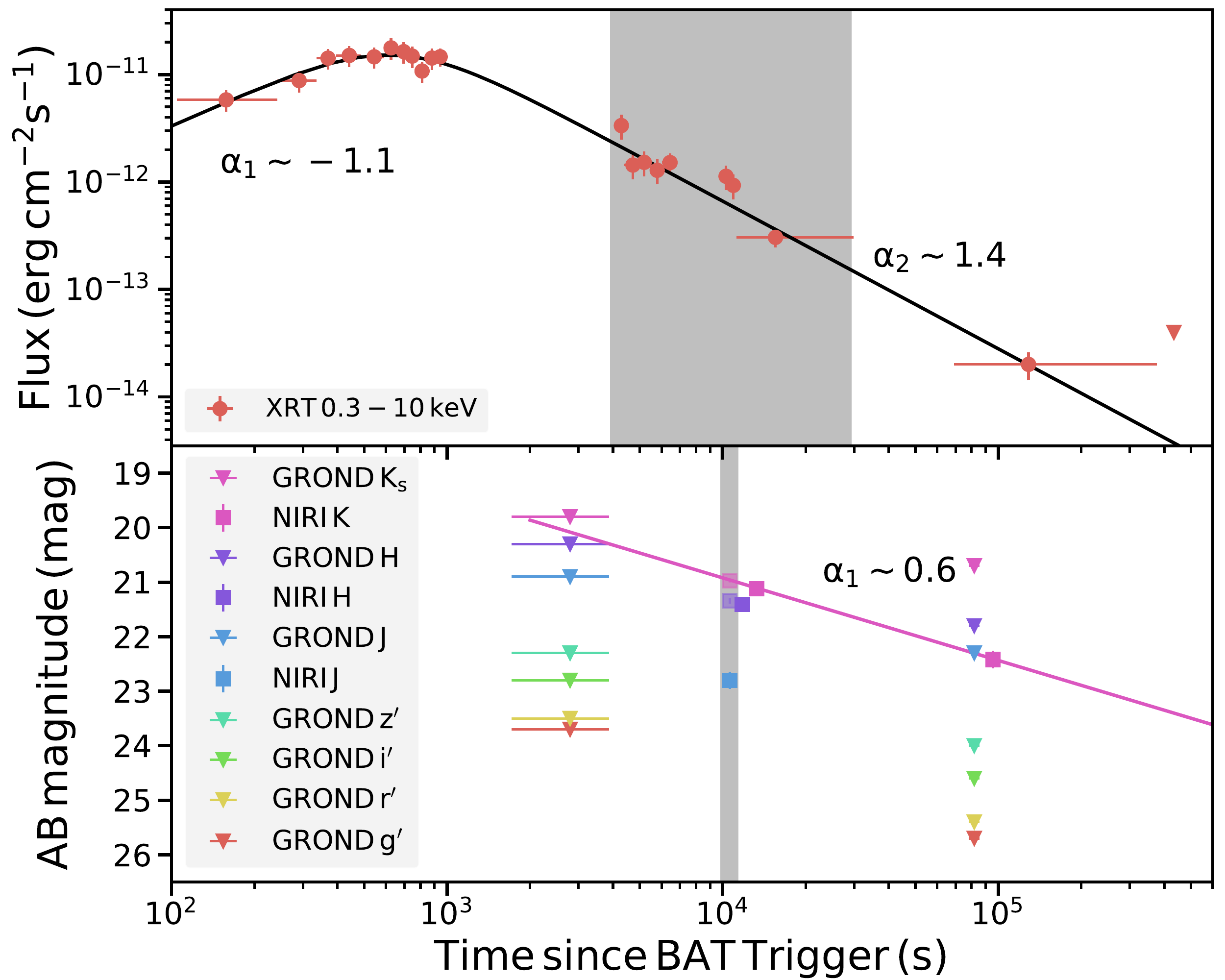}}
                \caption{GRB 090429B}
                \label{fig:090429Blc}
        \end{figure}

\begin{figure}
                \resizebox{\hsize}{!}{\includegraphics{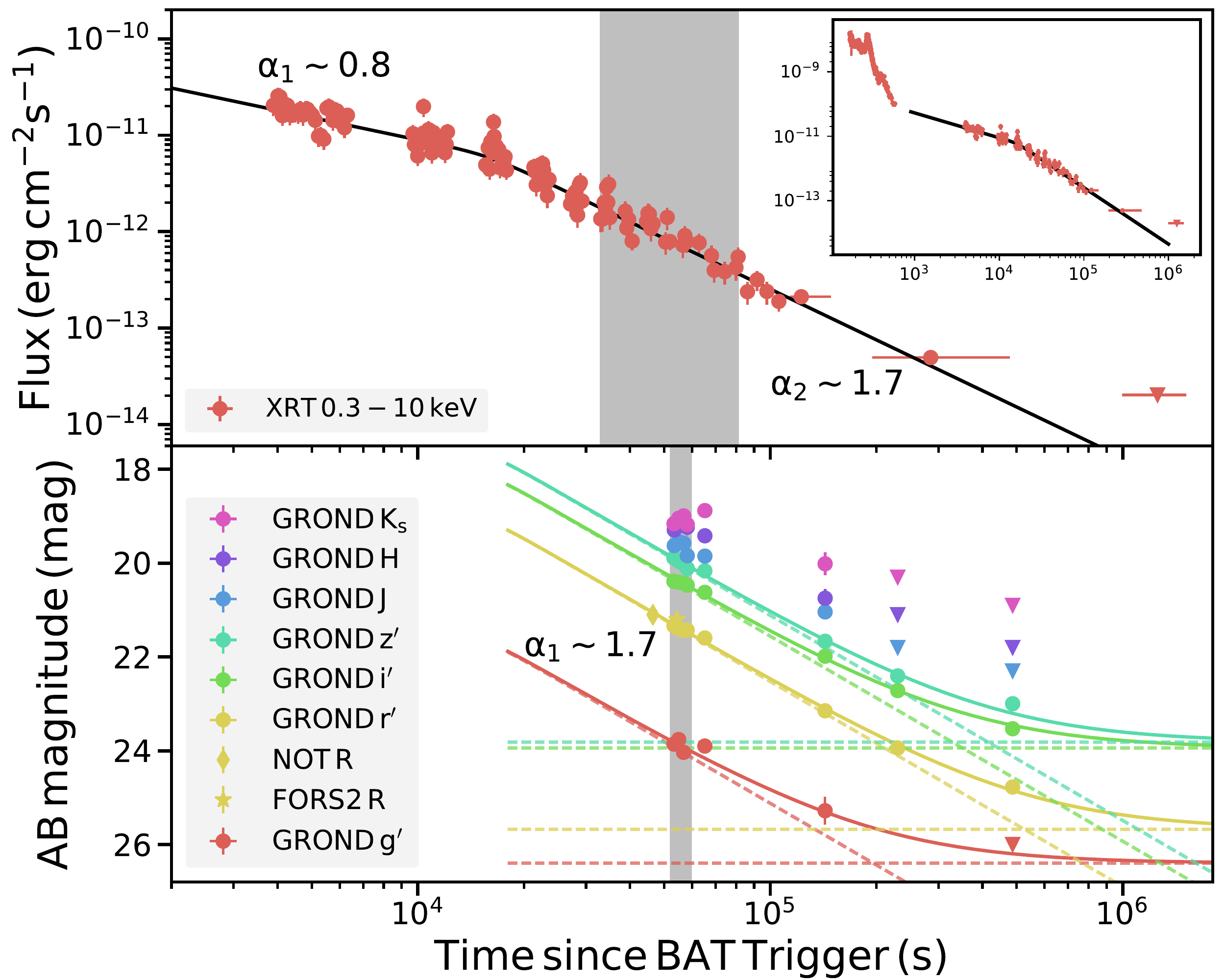}}
                \caption{GRB 090516A}
                \label{fig:090516Alc}
        \end{figure}

\begin{figure}
                \resizebox{\hsize}{!}{\includegraphics{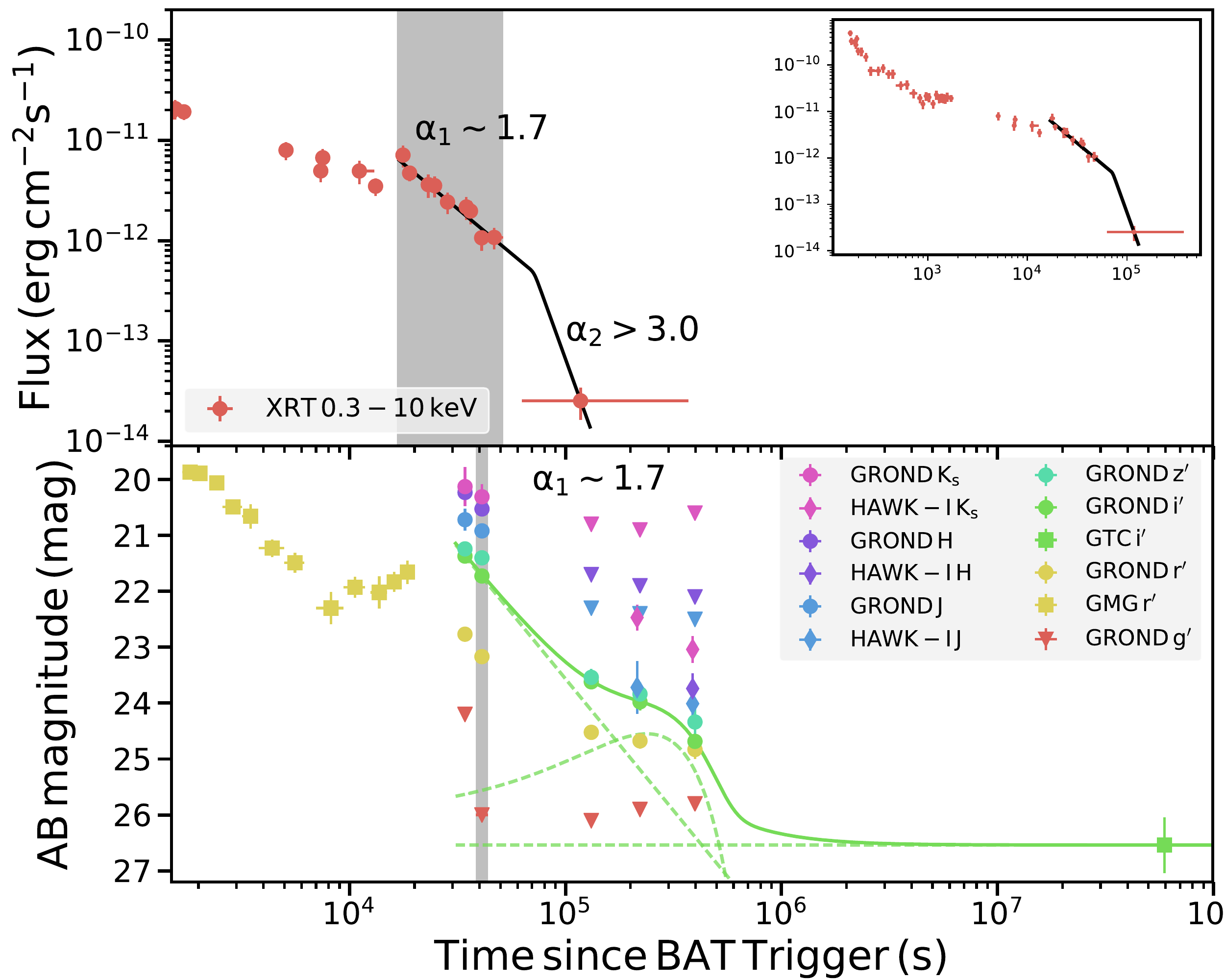}}
                \caption{GRB 100219A}
                \label{fig:100219Alc}
        \end{figure}

\begin{figure}
                \resizebox{\hsize}{!}{\includegraphics{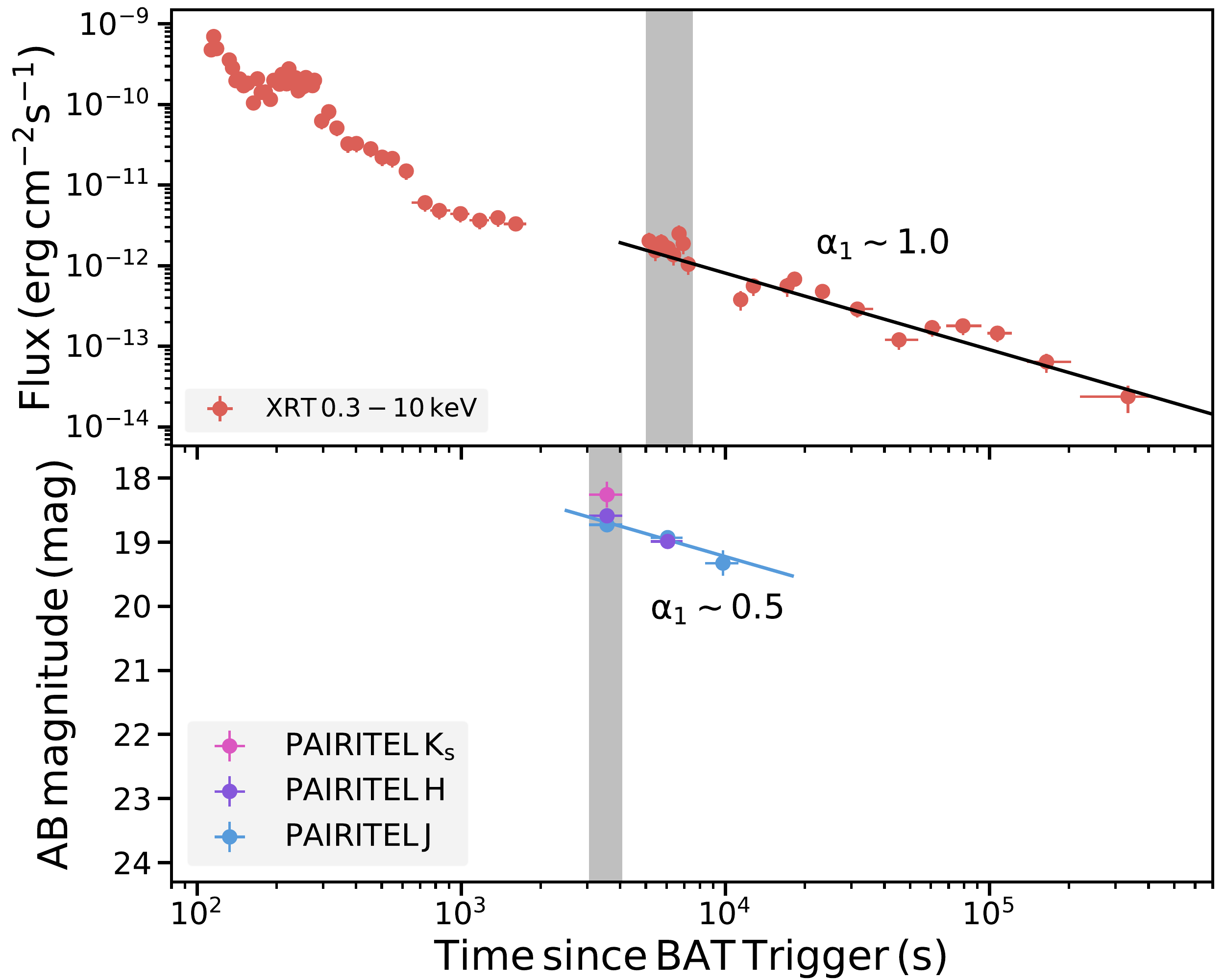}}
                \caption{GRB 100513A}
                \label{fig:100513Alc}
        \end{figure}

\begin{figure}
                \resizebox{\hsize}{!}{\includegraphics{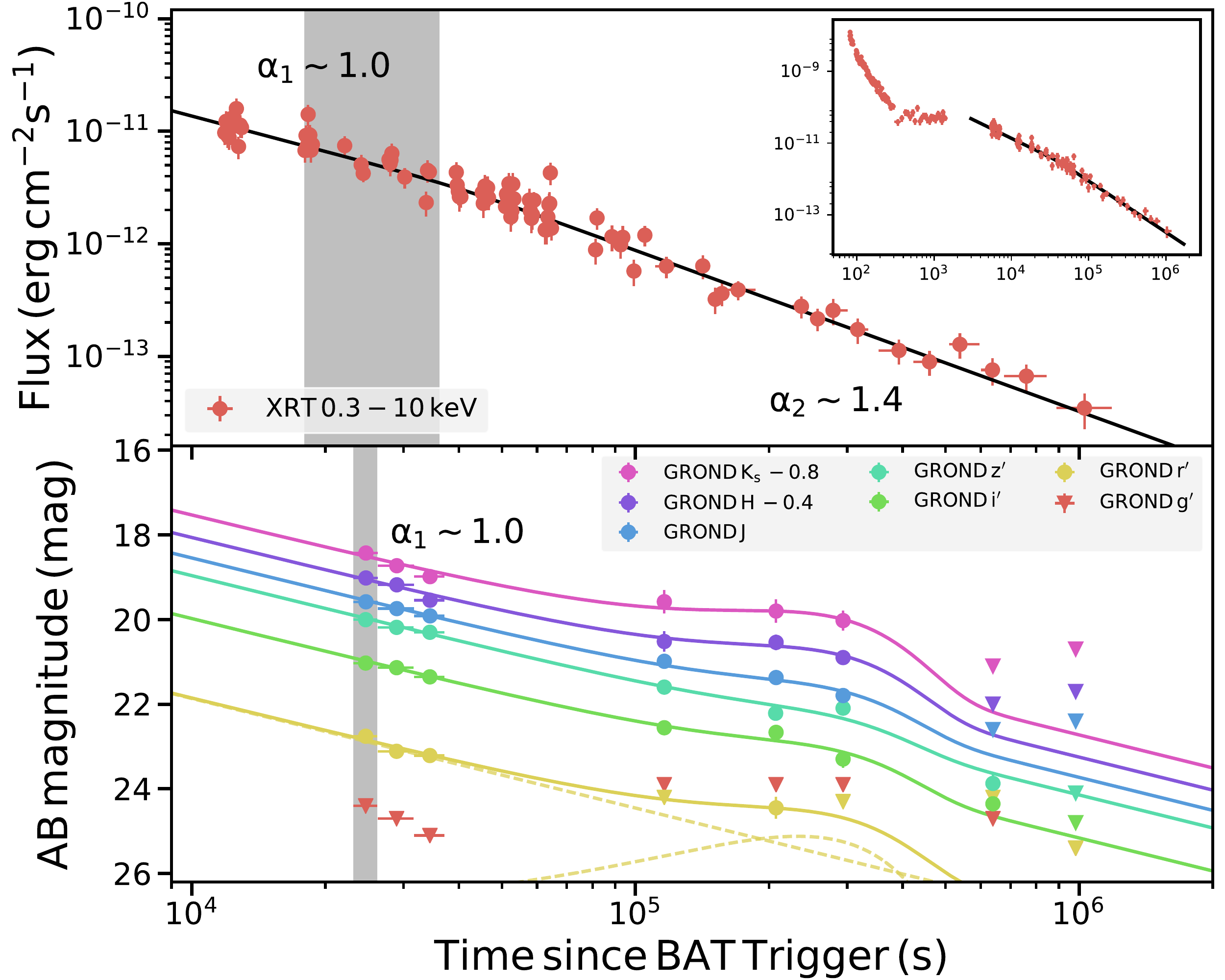}}
                \caption{GRB 111008A}
                \label{fig:111008Alc}
        \end{figure}

\begin{figure}
                \resizebox{\hsize}{!}{\includegraphics{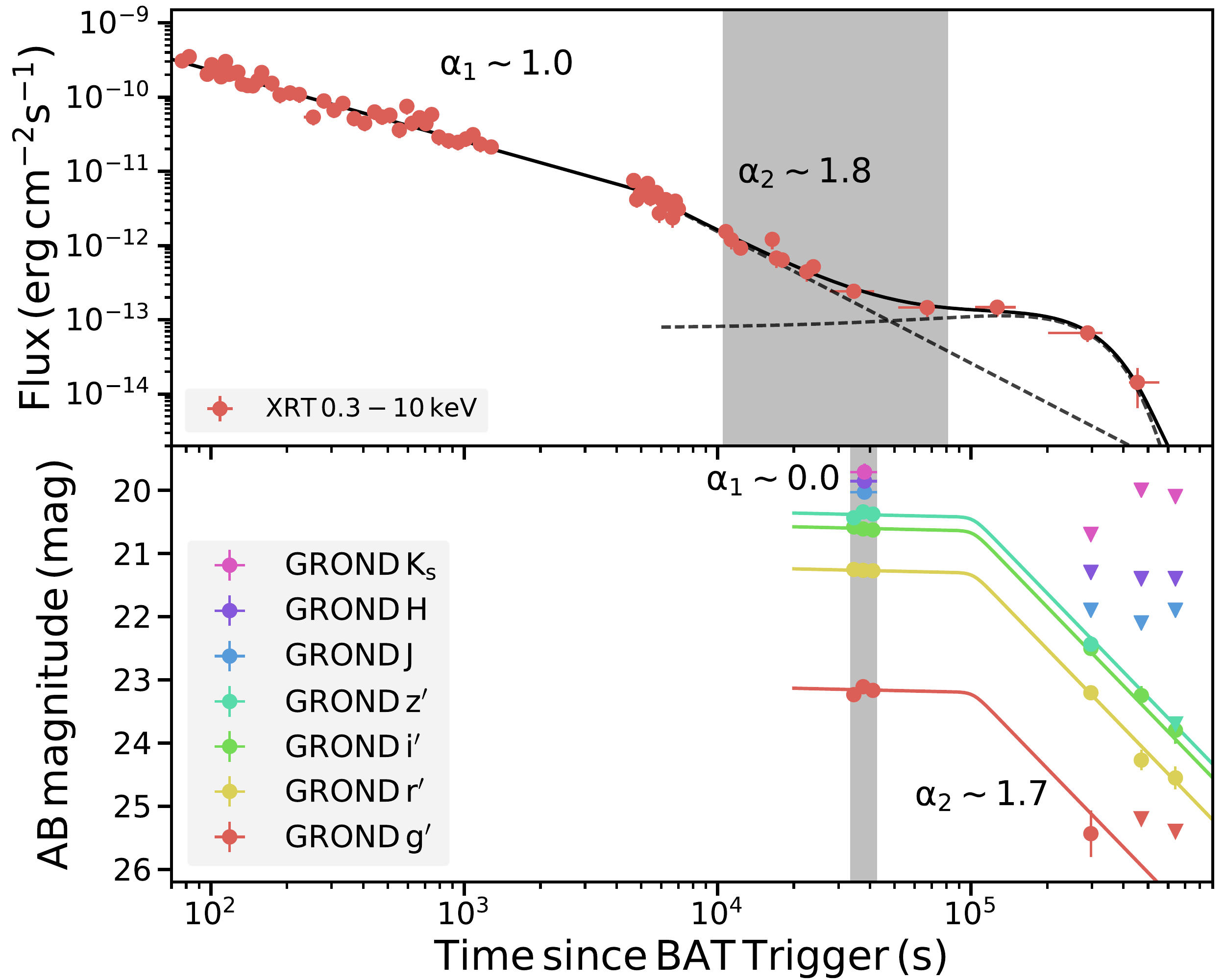}}
                \caption{GRB 120712A}
                \label{fig:120712Alc}
        \end{figure}

\begin{figure}
                \resizebox{\hsize}{!}{\includegraphics{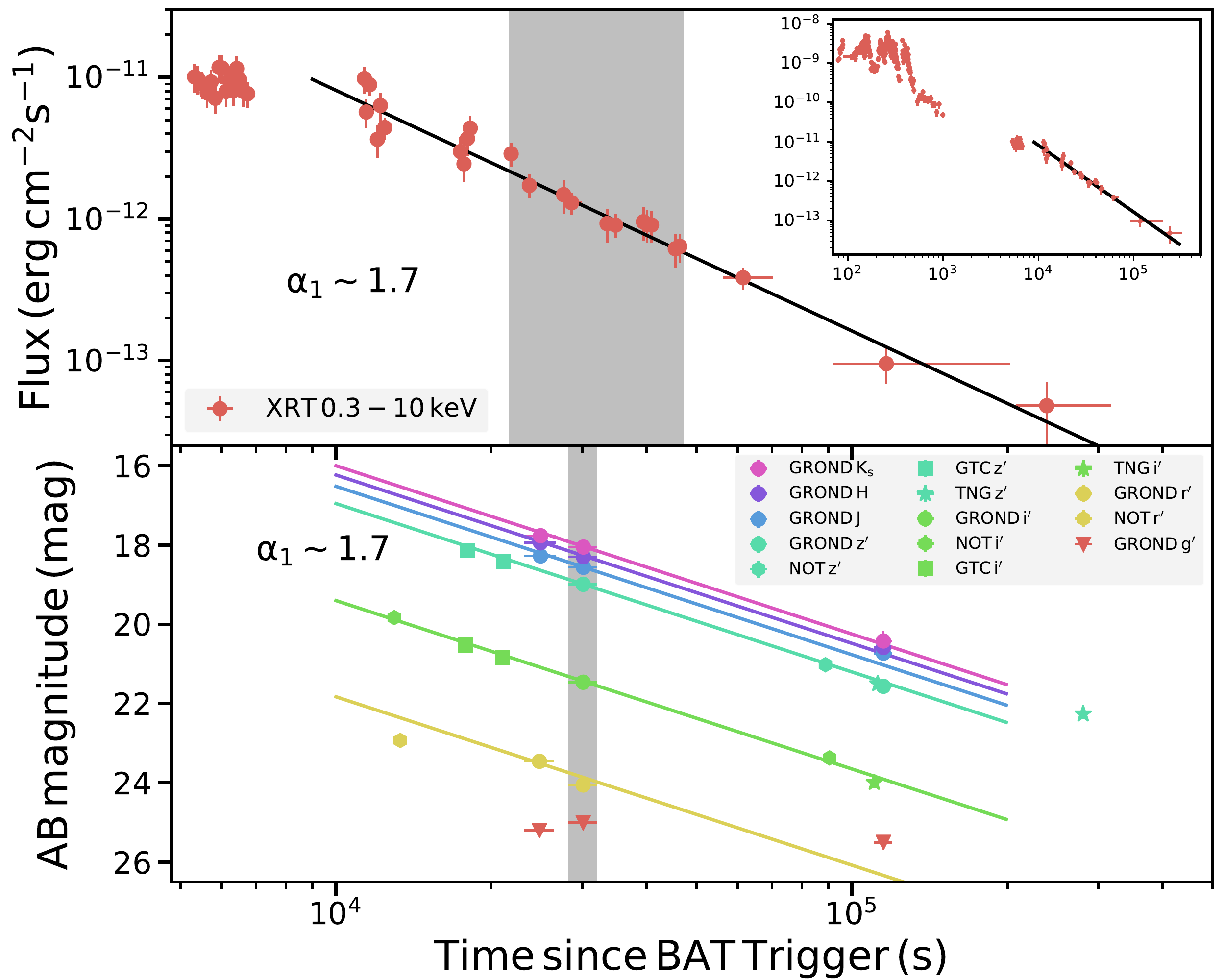}}
                \caption{GRB 130606A}
                \label{fig:130606Alc}
        \end{figure}

\begin{figure}
                \resizebox{\hsize}{!}{\includegraphics{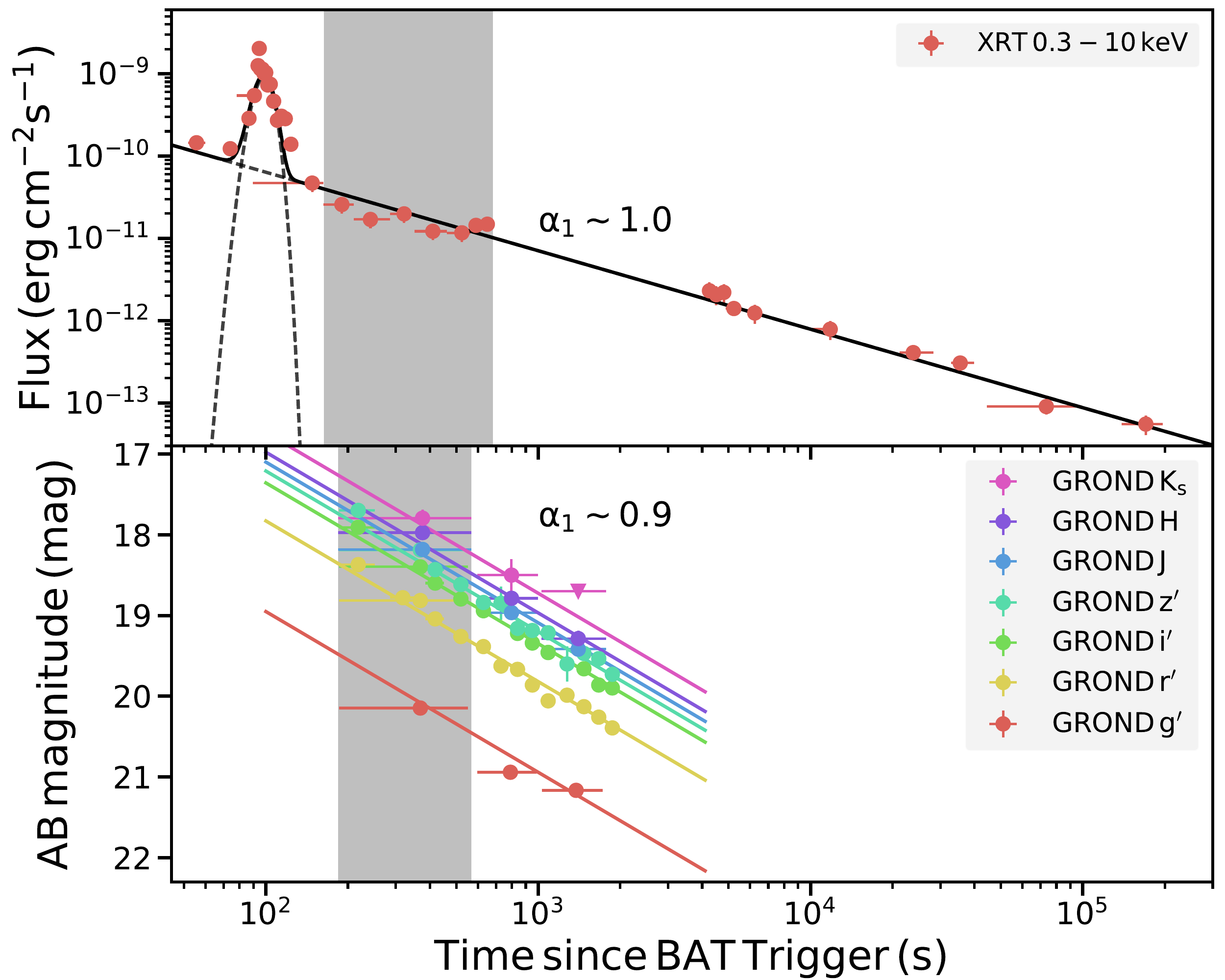}}
                \caption{GRB 131117A}
                \label{fig:131117Alc}
        \end{figure}

\begin{figure}
                \resizebox{\hsize}{!}{\includegraphics{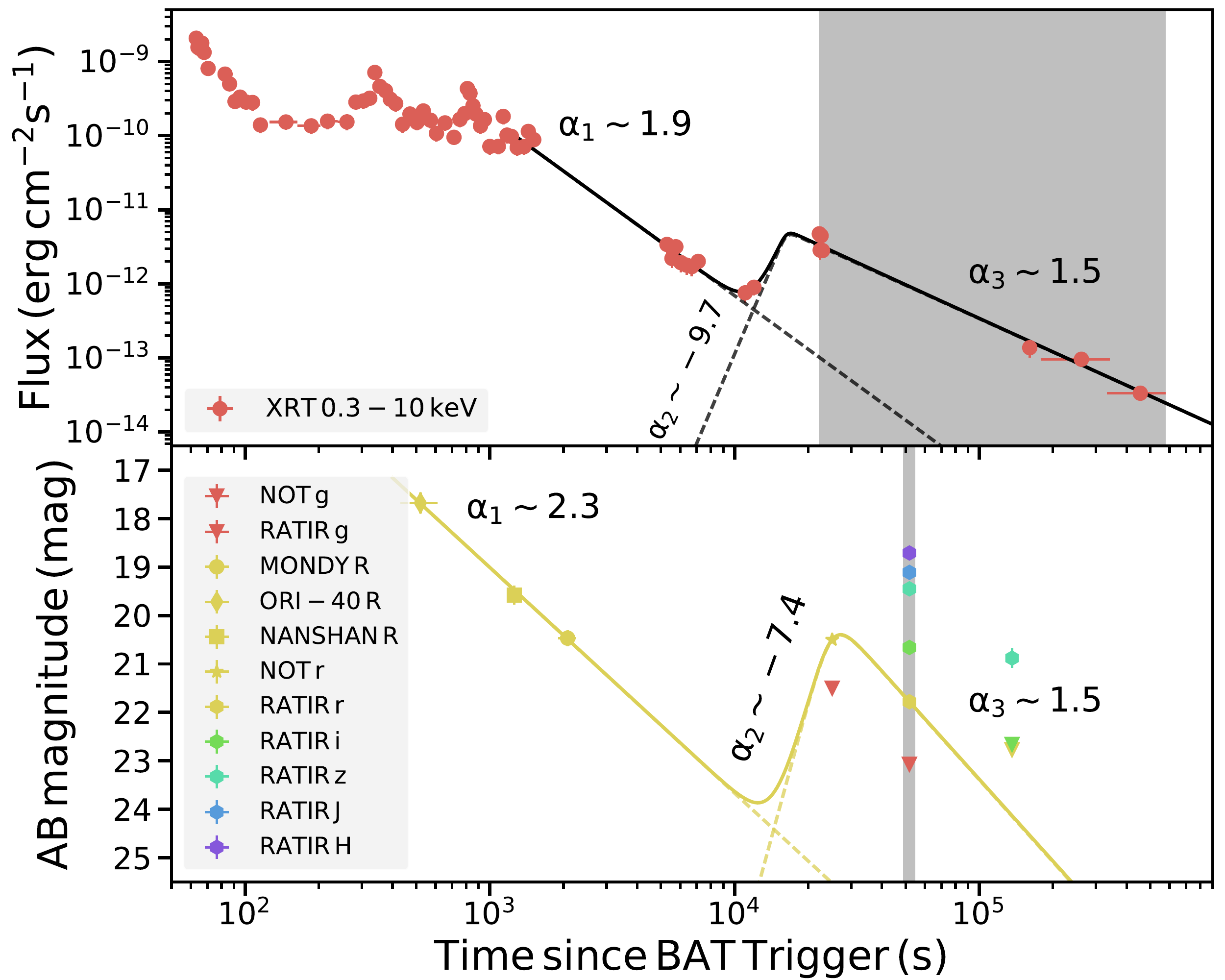}}
                \caption{GRB 140304A}
                \label{fig:140304Alc}
        \end{figure}

\begin{figure}
                \resizebox{\hsize}{!}{\includegraphics{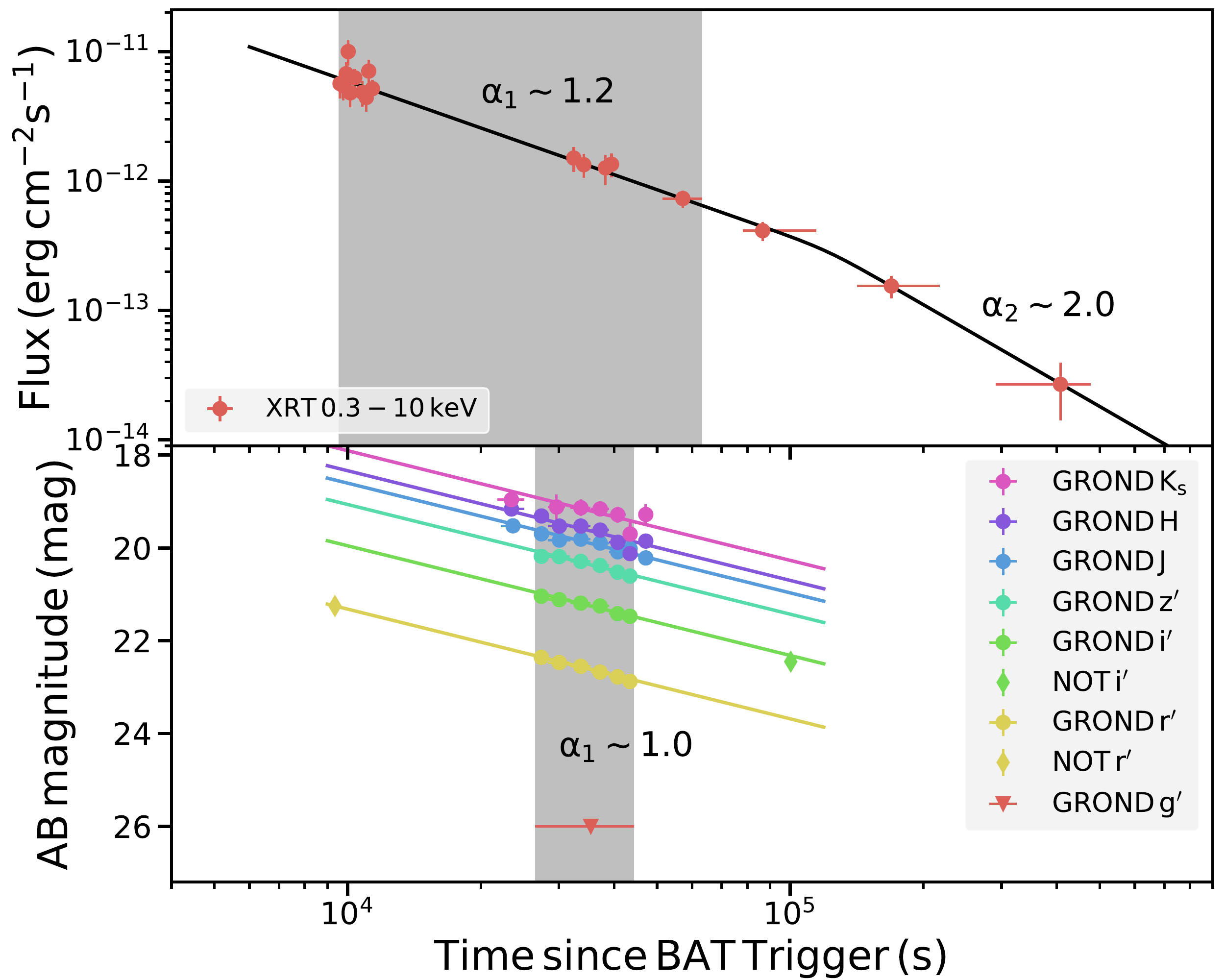}}
                \caption{GRB 140311A}
                \label{fig:140311Alc}
        \end{figure}

\begin{figure}
                \resizebox{\hsize}{!}{\includegraphics{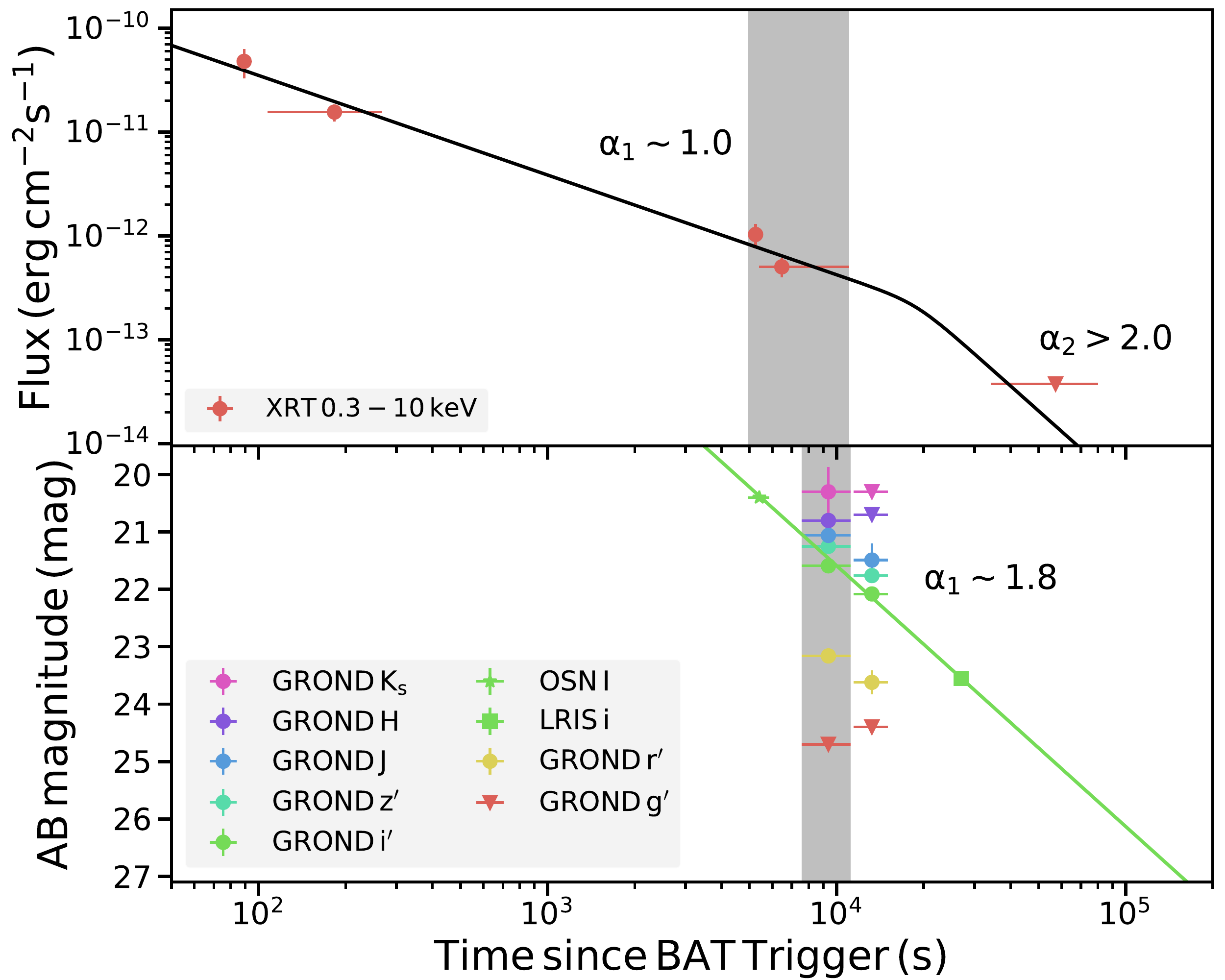}}
                \caption{GRB 140428A}
                \label{fig:140428Alc}
        \end{figure}

\begin{figure}
                \resizebox{\hsize}{!}{\includegraphics{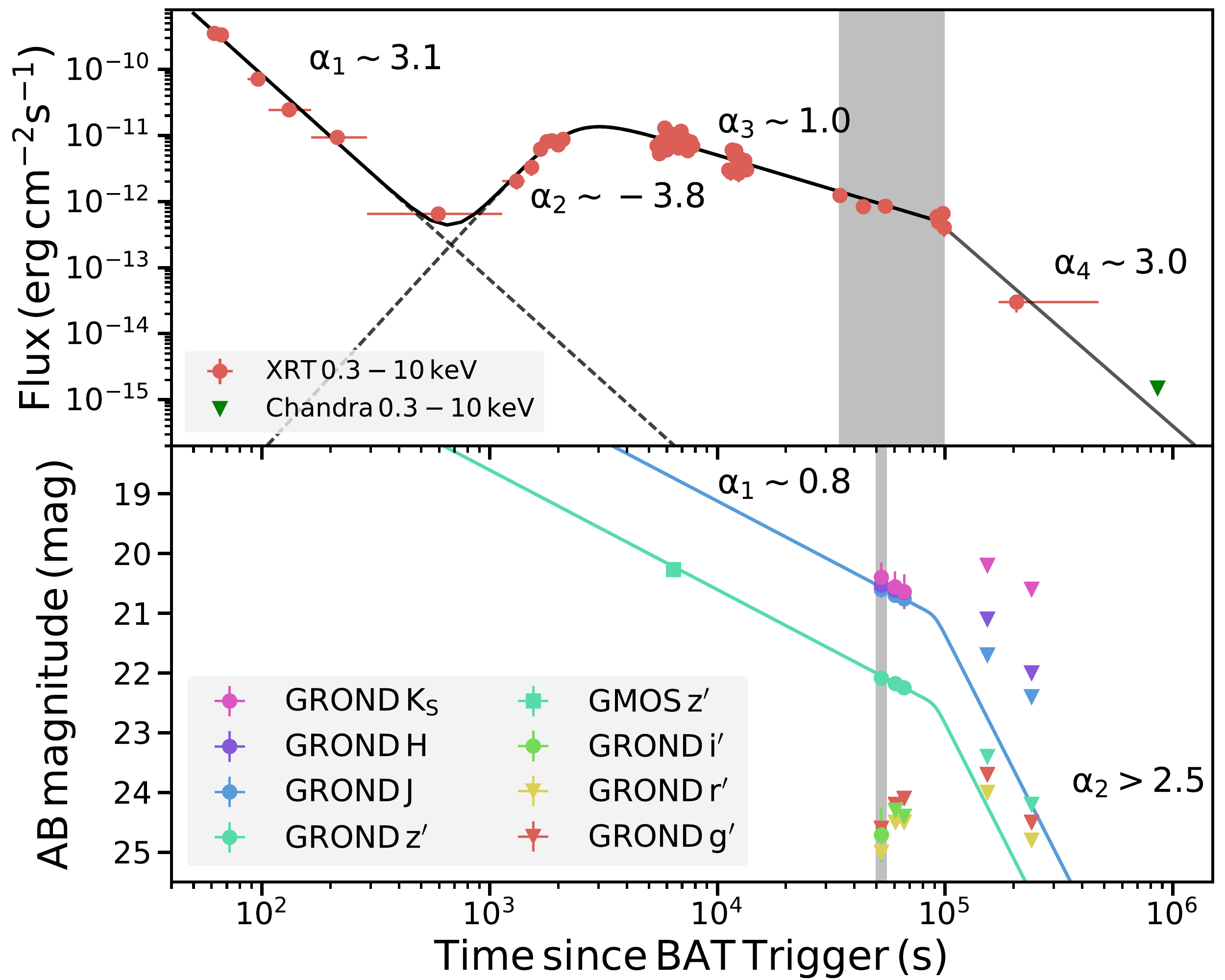}}
                \caption{GRB 140515A}
                \label{fig:140515Alc}
        \end{figure}

\begin{figure}
                \resizebox{\hsize}{!}{\includegraphics{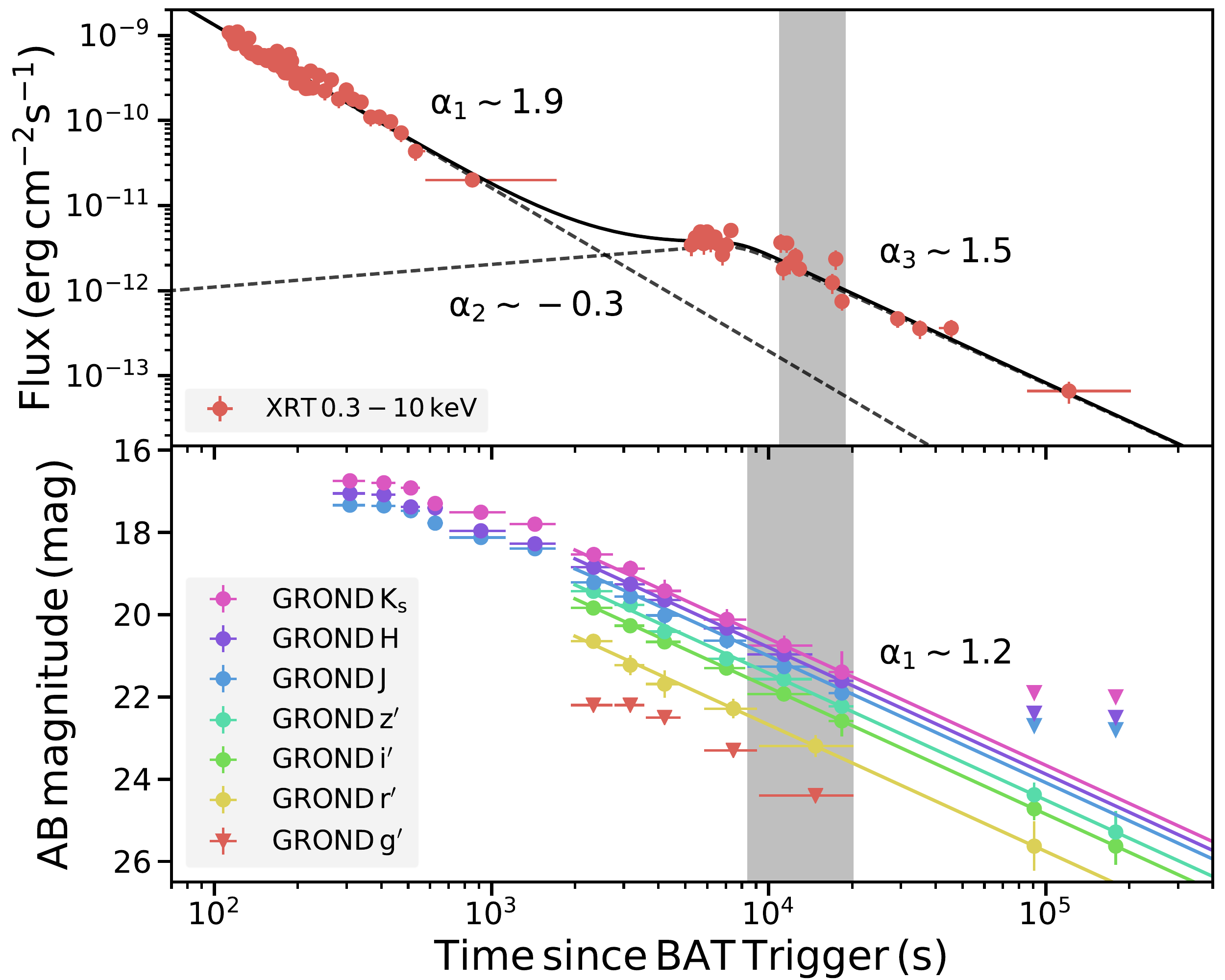}}
                \caption{GRB 140614A}
                \label{fig:140614Alc}
        \end{figure}

\begin{figure}
                \resizebox{\hsize}{!}{\includegraphics{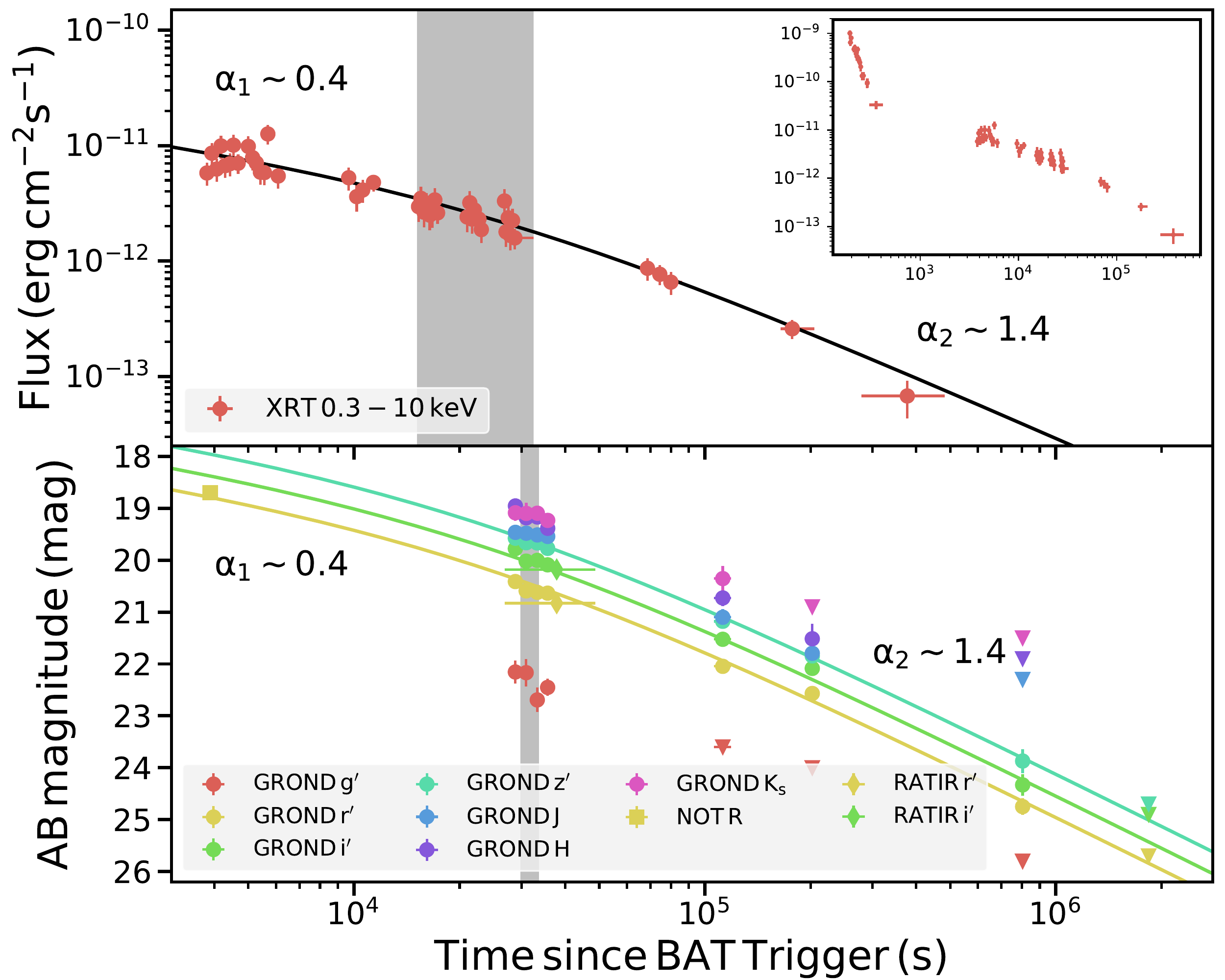}}
                \caption{GRB 151027B}
                \label{fig:151027Blc}
        \end{figure}

\begin{figure}
                \resizebox{\hsize}{!}{\includegraphics{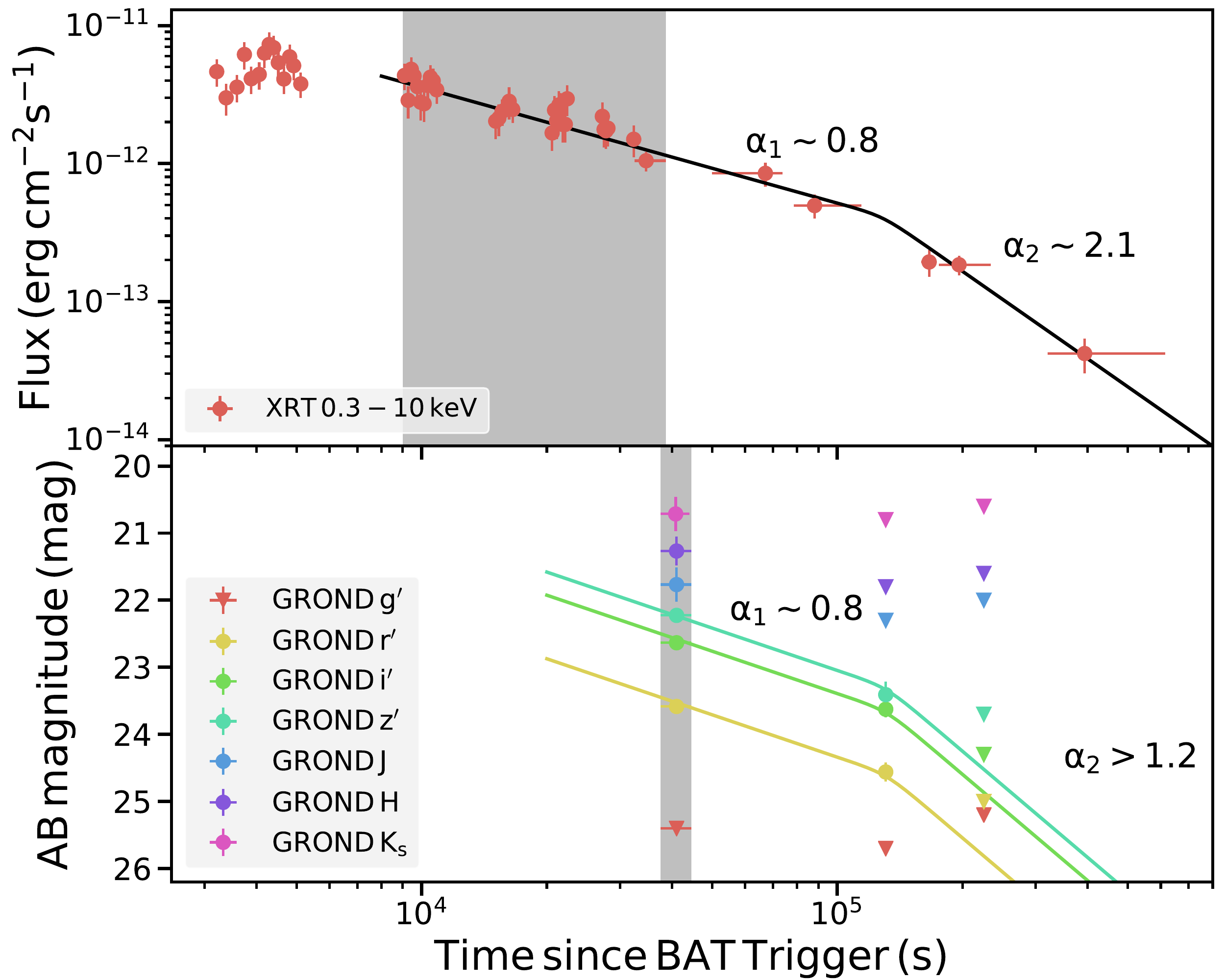}}
                \caption{GRB 151112A}
                \label{fig:151112Alc}
       \end{figure}

All the light curves and best-fit models are shown in Fig. \ref{fig:071025lc} to
Fig. \ref{fig:151112Alc}. In each case, the gray shaded areas indicate the time
intervals, which were chosen to create the quasi-simultaneous X-ray to NIR
broadband SEDs. Additionally, the best-fit models and parameters as well as the references for
the data collected from the literature are given in Tab. \ref{tab:lcfit}.

Finally, for some of the GRBs, the GROND data are already published in
various papers (GRB 080913 in \cite{greiner2009080913}, GRB 080916C in
\cite{greiner2009080916C}, 090423 in \cite{tanvir2009}, GRB 090429B in \cite{cucchiara2011},
and GRB 100219A in \cite{thoene2013}). However, because of to improvements of the GROND data
reduction pipeline and for reasons of consistency, we decided to re-analyze the data. The
complete set of GROND magnitudes is given in Tab. \ref{tab:grondmags}.

\clearpage

\clearpage

\section{Additional tables and figures}\label{app:figtab}

        \begin{figure*}
    \sidecaption
        \includegraphics[width=12.0cm]{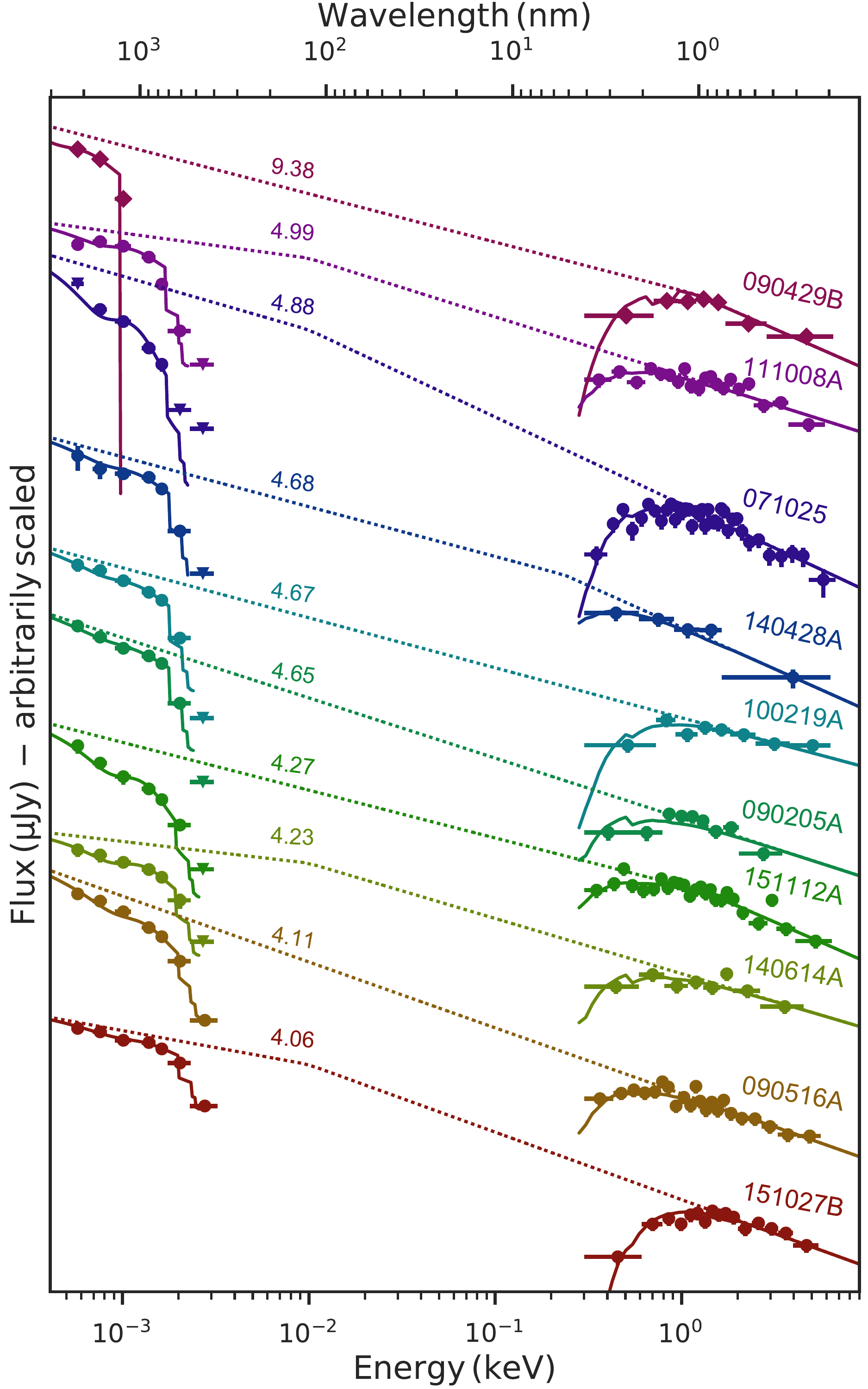}
        \caption{Same as Fig. \ref{fig:allsed}, but for the 10 GRBs we additionally
    fitted with the SN-like dust extinction curve. These GRBs showed evidence
    of a medium amount of dust ($A_V > 0.1$ mag), when fitted with the local
    dust extinction curves.}
        \label{fig:allsed_sncompl}
        \end{figure*}

\onecolumn

%Added by TeX Support
\input{table/lc_fit.tex}

\clearpage

\section{GROND photometry}\label{app:phot}

%Added by TeX Support
\input{table/phot.tex}

\end{appendix}

\end{document}

%% file: table/sample.tex
\begin{table*}
   \caption{All GRBs with a spectroscopic or photometric redshift 
   estimate of $z \geq 4$ (until the 1.\ of January 2017).}
	\vspace{-0.3cm}
	\centering
    \renewcommand{\tabcolsep}{0.21cm}
	\renewcommand*{\arraystretch}{1.13}
    \begin{tabular}{c c c c c c c c c c}
	\hline \hline
   \multicolumn{1}{c}{GRB}& Right Ascension & Declination & Error & \multicolumn{1}{c}{Redshift} & Ref. &
   \multicolumn{1}{c}{$T_{90}$} & $E_{B-V}$ & $N_{\element[ ][]{H}}$ &
   \multicolumn{1}{c}{$T_{\text{GROND}}$} \\
   \multicolumn{1}{c}{(yymmdd\#)}&  \multicolumn{1}{c}{(J2000)} & \multicolumn{1}{c}{(J2000)} &  &
   \multicolumn{1}{c}{(z)} & & \multicolumn{1}{c}{(s)} & (\text{mag}) & $(10^{21}\text{cm}^{-2})$ &
   \multicolumn{1}{c}{(hrs)} \\
      \hline
\multicolumn{9}{c}{GRBs with $z > 4$ observed with GROND} \\
071025 		& +23:40:17.11 & +31:46:42.74	& 0\farcs{20} & $4.88^{+0.35}_{-0.35}$ & (1) & $109\pm 2$		& 0.06	& 0.55	& 21.69 \\
080825B 	& +13:56:48.29 & $-$68:57:18.57 & 0\farcs{25} & $4.31^{+0.14}_{-0.15}$ & (2) & 50	   			& 0.21	& 2.09	& 6.64  \\
080913  	& +04:22:54.73 & $-$25:07:45.98	& 0\farcs{31} & 6.733		 		   & (3) & $8\pm1$			& 0.04	& 0.33	& 0.10  \\
080916C 	& +07:59:23.32 & $-$56:38:17.92	& 0\farcs{37} & $4.28^{+0.6}_{-0.10}$ & (1) & $\sim 60$		& 0.28	& 1.50	& 31.74 \\
090205 		& +14:43:38.68 & $-$27:51:10.10	& 0\farcs{30} & 4.650 				   & (4) & $8.8\pm 1.8$		& 0.10	& 0.80	& 6.39 	\\
090423  	& +09:55:33.27 & +18:08:58.06	& 0\farcs{29} & 8.26 				   & (5) & $10.3\pm 1.1$	& 0.03	& 0.29	& 15.29 \\
090429B 	& +14:02:40.10 & +32:10:14.20 	& 0\farcs{30} & $9.38^{+0.14}_{-0.32}$ & (6) & $5.5\pm 1.0$ 	& 0.01	& 0.12	& 21.72 \\
090516  	& +09:13:02.60 & $-$11:51:14.90	& 0\farcs{30} & 4.109 				   & (7) & $210\pm 65$ 		& 0.04 	& 0.45	& 14.60 \\
100219A 	& +10:16:48.51 & $-$12:34:00.50	& 0\farcs{29} & 4.667 			       & (8) & $18.8\pm 5.0$	& 0.07	& 0.65	& 9.20  \\
100905A		& +02:06:12.04 & +14:55:45.80	& 0\farcs{31} & $7.88^{+0.75}_{-0.94}$ & (1) & $3.4 \pm 0.5$	& 0.05	& 0.54	& 16.02 \\
111008A 	& +04:01:48.24 & $-$32:42:32.87	& 0\farcs{29} & 4.990 				   & (9) & $63.46\pm 2.19$	& 0.01	& 0.10	& 6.43 	\\
120712A 	& +11:18:21.24 & $-$20:02:01.41	& 0\farcs{30} & 4.175 			       & (10)& $14.7\pm 3.3$	& 0.04	& 0.36	& 9.25 	\\
120923A		& +20:15:10.78 & +06:13:16.30 	& 0\farcs{30} & 7.84 				   & (11)& $27.2\pm 3.0$	& 0.13	& 0.98	& 18.95	\\
130606A 	& +16:37:35.13 & +29:47:46.61	& 0\farcs{20} & 5.913 				   & (12)& $276.58\pm 19.31$& 0.02	& 0.20	& 6.43	\\
131117A 	& +22:09:19.37 & $-$31:45:44.22	& 0\farcs{40} & 4.042 				   & (13)& $11.00\pm 3.16$	& 0.02	& 0.15	& 0.05	\\
140311A 	& +13:57:13.27 & +00:38:32.11	& 0\farcs{14} &	4.954 				   & (14)& $71.4\pm 9.5$ 	& 0.03 	& 0.25	& 11.72	\\
140428A 	& +12:57:28.38 & +28:23:06.88	& 0\farcs{18} &	$4.68^{+0.52}_{-0.18}$ & (1) & $17.42\pm 5.90$	& 0.01	& 0.09	& 2.05	\\
140515A 	& +12:24:15.52 & +15:06:16.62	& 0\farcs{24} & 6.327 				   & (15)& $23.4\pm 2.1$	& 0.02	& 0.23	& 13.76 \\
140614A 	& +15:24:40.66 & $-$79:07:43.20	& 0\farcs{30} & 4.233 			       & (16)& $720\pm 120$		& 0.11	& 0.82	& 0.54 	\\
151027B 	& +05:04:52.69 & $-$06:27:01.07 & 0\farcs{25} & 4.062 				   & (17)& $80.00\pm 35.78$ & 0.18 	& 0.58 	& 7.76	\\
151112A 	& +00:08:12.75 & $-$61:39:48.47 & 0\farcs{36} &$4.27^{+0.24}_{-0.38}$  & (1) & $19.32\pm 31.24$ & 0.01 	& 0.18	& 10.43	\\ \hline
\multicolumn{9}{c}{GRBs with $z > 4$ observed with GROND that have been excluded from the sample} 									\\
080129  	& +07:01:08.20 & $-$07:50:46.51 & 0\farcs{28} & 4.349 				   & (18)& $48\pm 10$		& 0.87	& 6.42 	& 0.10  \\
100518A 	& +20:19:09.33 & $-$24:33:16.57 & 0\farcs{28} & $3.50^{+0.50}_{-0.62}$ & (1) & $25$ 			& 0.07 	& 0.63 	& 17.13 \\
131227A 	& +04:29:30.84 & +28:52:58.92	& 0\farcs{30} & 5.3	 				   & (19)& $18.0\pm1.6$		& 0.90	& 1.51	& 21.39 \\ \hline
\multicolumn{9}{c}{All other GRBs with a redshift estimate of $z > 4$:} \\
000131 		& +06:13:31.1 & $-$51:56:41.7 	& 1\farcs{1}  & $4.500$ 			   & (20)& $96.3$ 			& 0.05 	& 0.41  & -     \\
050502B 	& +09:30:10.1 &	+16:59:47.9   	& 1\farcs{4}  & $5.2^{+0.3}_{-0.3}$    & (21)& $17.5\pm 0.2$ 	& 0.03  & 0.36  & - 	\\
050505 		& +09:27:03.3 &	+30:16:24.2   	& 1\farcs{4}  & $4.275$ 			   & (22)& $60\pm 2$   		& 0.02  & 0.17  & - 	\\
050814		& +17:36:45.4 &	+46:20:21.8   	& 1\farcs{4}  & $5.77^{+0.12}_{-0.12}$ & (23)& $65^{+40}_{-20}$ & 0.02 	& 0.23	& - 	\\
050904		& +00:54:50.9 &	+14:05:09.3   	& 3\farcs{5}  & $6.295$ 			   & (24)& $225\pm 10$  	& 0.05  & 0.45  & - 	\\
050922B 	& +00:23:13.4 &	$-$05:36:17.3 	& 1\farcs{7}  & $4.5\pm 0.5$ 		   & (25)& $250\pm 20$		& 0.03  & 0.31  & - 	\\
060206 		& +13:31:43.4 &	+35:03:02.8   	& 1\farcs{5}  & $4.048$ 			   & (26)& $7\pm 2$      	& 0.01  & 0.09  & - 	\\
060223A 	& +03:40:49.6 & $-$17:07:49.8 	& 1\farcs{5}  & $4.406$ 			   & (27)& $11\pm 2$ 		& 0.10 	& 0.69 	& - 	\\
060510B 	& +15:56:29.2 &	+78:34:11.8   	& 1\farcs{5}  & $4.941$ 			   & (28)& $276\pm 10$   	& 0.04  & 0.41  & - 	\\ 
060522   	& +21:31:44.9 & +02:53:09.9   	& 1\farcs{4}  & $5.11$ 				   & (29)& $69\pm 5$     	& 0.05  & 0.42  & - 	\\
060927 		& +21:58:12.0 &	+05:21:49.0   	& 1\farcs{6}  & $5.467$ 			   & (30)& $22.6\pm 0.3$  	& 0.05  & 0.46  & - 	\\ 
100302A 	& +13:02:03.8 &	+74:35:23.7   	& 1\farcs{5}  & $4.813$ 			   & (31)& $17.9\pm 1.7$ 	& 0.02  & 0.19  & - 	\\
100513A 	& +11:18:26.8 & +03:37:40.8   	& 1\farcs{4}  & $4.772$ 			   & (32)& $84\pm 21$ 		& 0.05  & 0.42  & - 	\\
120521C  	& +14:17:08.8 &	+42:08:41.5   	& 1\farcs{6}  & $6.0$ 				   & (33)& $26.7\pm 4.4$ 	& 0.01  & 0.11  & - 	\\
140304A 	& +02:02:34.3 &	+33:28:25.7   	& 1\farcs{5}  & $5.283$ 			   & (34)& $15.6\pm 1.9$ 	& 0.07  & 0.60  & - 	\\
140518A 	& +15:09:00.6 & +42:25:05.7   	& 2\farcs{7}  & $4.707$ 			   & (35)& $60.5\pm 2.4$ 	& 0.01  & 0.15  & - 	\\ \hline
   \end{tabular}
   \tablefoot{Columns 5--10 are the redshift (photometric when given with errors, spectroscopic otherwise), corresponding
   reference, time interval $T_{90}$ over which 90\% of the total background-subtracted counts are observed (GRBs
   with $T_{90} > 2$ s are classified as long gamma-ray bursts (LGRBs)), galactic foreground reddening from \cite{sf2011},
   as retrieved from the NASA Extragalactic database (NED, \url{http://irsa.ipac.caltech.edu/applications/DUST/}), total
   galactic neutral hydrogen column density provided by \cite{k2005}, and start of the GROND observations in hours after
   the GRB Trigger (rounded to two decimal places).}
   \tablebib{
	(1)~this work; (2)~\citet{kruehler2011a}; (3)~\citet{patel2010}; (4)~\citet{fugazza2009}; (5)~\citet{tanvir2009};
    (6)~\citet{cucchiara2011}; (7)~\citet{ugarte2009a}; (8)~\citet{mao2012}; (9)~\citet{wiersema2011}; (10)~\citet{xu2012};
    (11)~\citet{tanvir2017n}; (12)~\citet{xu130606A}; (13)~\citet{hartoog2013}; (14)~\citet{tanvir2014}; (15)~\citet{chornock2014};
    (16)~\citet{kruehler2014}; (17)~\citet{xu2015}; (18)~\citet{greiner2009080129}; (19)~\citet{cucchiara20132}; (20)~\citet{andersen2000};
    (21)~\citet{afonso2011}; (22)~\citet{berger2006}; (23)~\citet{curran2008}; (24)~\citet{kawai2006}; (25)~\citet{schulze2015};
    (26)~\citet{fynbo2006}; (27)~\citet{chary2007}; (28)~\citet{price2007}; (29)~\citet{cenko2006a}; (30)~\citet{ruiz2007};
    (31)~\citet{chornock2010}; (32)~\citet{cenko2010b}; (33)~\citet{laskar2014}; (34)~\citet{jeong2014}; (35)~\citet{chornock2014}.
    }
   \label{tab:sample}
\end{table*}

%% file: table/av.tex
\begin{table*}[!ht]
	\renewcommand*{\arraystretch}{1.36}
    \renewcommand{\tabcolsep}{0.17cm}
	\centering
	\caption{Summary of the best-fit models and parameters of the 22 GRBs analyzed in this
    paper. \label{tab:avs}}
	\begin{tabular}{lccccccclcl}
		\hline \hline
		GRB & Redshift& Model& Ext. Curve&$\beta_{\mathrm{o}}$\tablefootmark{a}& $A_V$ & $E_{\mathrm{break}}$& Norm.& \multicolumn{1}{c}{$\chi^2$} & 
        $\chi^2_{\mathrm{red}}$ &$T_{\mathrm{SED}}$\tablefootmark{b} \\ [1.5pt]
		(yymmdd\#) & fixed to &     &  &   & (mag) & (keV) & (keV/$\text{cm}^{2}$/s/keV) & \multicolumn{1}{c}{(d.o.f.)} & & (hrs) \\ \hline
		071025		& 4.900	& bknpow	& smc	& $0.77_{-0.08}^{+0.17}$ & $0.45_{-0.14}^{+0.13}$ & $1.45\cdot10^{-2}$ & $2.78\cdot10^{-4}$ & 40.1 (36) & 1.11 & 22.4 \\
		080825B		& 4.310	& bknpow	& smc	& $0.67_{-0.03}^{+0.19}$ & $0.05_{-0.02}^{+0.02}$ & $9.93\cdot10^{-3}$ & $1.82\cdot10^{-3}$ & 2.9 (5) 	& 0.58 & 12.4 \\
		080913		& 6.733	& pow		& smc\tablefootmark{d}	& $0.59_{-0.02}^{+0.02}$ & $0.00_{-0.00}^{+0.00}$ & $.............$    & $1.02\cdot10^{-3}$ & 9.9 (12) 	& 0.83 & 0.2  \\
		080916C		& 4.280	& bknpow	& smc\tablefootmark{d}	& $0.40_{-0.02}^{+0.18}$ & $0.01_{-0.01}^{+0.03}$ & $9.99\cdot10^{-3}$ & $1.15\cdot10^{-3}$ & 12.3 (13) & 0.95 & 32.6 \\
		090205A		& 4.650	& bknpow	& lmc	& $0.88_{-0.24}^{+0.04}$ & $0.14_{-0.05}^{+0.04}$ & $1.24\cdot10^{+0}$ & $1.61\cdot10^{-4}$ & 9.2 (9) 	& 1.02 & 7.1  \\
		090423A		& 8.260	& pow		& smc	& $0.88_{-0.04}^{+0.04}$ & $0.08_{-0.09}^{+0.09}$ & $.............$    & $4.37\cdot10^{-5}$ & 28.0 (31) & 0.90 & 17.3 \\
		090429B\tablefootmark{c}&9.380&bknpow&lmc&$0.77_{-0.29}^{+0.07}$ & $0.36_{-0.21}^{+0.22}$ & $1.21\cdot10^{+0}$ & $1.81\cdot10^{-4}$ & 2.2 (4) 	& 0.55 & 2.9  \\
		090516A		& 4.109	& pow		& smc	& $0.97_{-0.02}^{+0.02}$ & $0.19_{-0.03}^{+0.03}$ & $.............$    & $1.62\cdot10^{-4}$ & 17.1 (23) & 0.74 & 15.3 \\
		100219A		& 4.667	& pow		& smc	& $0.74_{-0.03}^{+0.03}$ & $0.15_{-0.05}^{+0.04}$ & $.............$    & $2.22\cdot10^{-4}$ & 4.2 (9)	& 0.47 & 11.4 \\
		100513A\tablefootmark{c}&4.772&bknpow&smc&$0.50_{-0.35}^{+0.09}$ & $0.04^{+0.24}_{-0.04}$ & $9.94\cdot10^{-3}$ & $6.30\cdot10^{-3}$ & 0.8 (3) 	& 0.27 & 1.0  \\
		100905A		& 7.880	& bknpow	& mw\tablefootmark{d}	& $0.79_{-0.33}^{+0.35}$ & $0.00_{-0.00}^{+1.49}$ & $9.30\cdot10^{-1}$ & $4.10\cdot10^{-5}$ & 0.9 (3)	& 0.30 & 16.0 \\
		111008A		& 4.990	& bknpow	& smc	& $0.35_{-0.01}^{+0.20}$ & $0.13_{-0.07}^{+0.03}$ & $9.98\cdot10^{-3}$ & $9.28\cdot10^{-3}$ & 27.0 (19) & 1.42 & 6.8  \\
		120712A		& 4.175 & bknpow	& smc	& $0.61_{-0.02}^{+0.24}$ & $0.08_{-0.08}^{+0.03}$ & $9.99\cdot10^{-3}$ & $9.49\cdot10^{-4}$ & 9.5 (9) 	& 1.06 & 10.6 \\
		130606A		& 5.931	& pow		& smc\tablefootmark{d}	& $1.01_{-0.02}^{+0.02}$ & $0.00_{-0.00}^{+0.02}$ & $.............$    & $1.83\cdot10^{-4}$ & 7.1 (11) 	& 0.65 & 8.4  \\
		131117A		& 4.042	& bknpow	& smc	& $0.35_{-0.02}^{+0.14}$ & $0.03_{-0.03}^{+0.03}$ & $1.00\cdot10^{-2}$ & $2.51\cdot10^{-2}$ & 10.4 (8) 	& 1.30 & 0.1  \\
		140304A\tablefootmark{c}&5.283&pow& smc\tablefootmark{d}	& $0.95_{-0.03}^{+0.04}$ & $0.01_{-0.01}^{+0.05}$ & $.............$    & $2.04\cdot10^{-4}$ & 10.4 (6) 	& 1.33 & 14.4 \\
		140311A		& 4.954	& pow		& smc	& $0.85_{-0.02}^{+0.02}$ & $0.07_{-0.03}^{+0.03}$ & $.............$    & $2.08\cdot10^{-4}$ & 20.6 (17) & 1.21 & 9.8  \\
		140428A		& 4.680	& bknpow	& mw	& $0.86_{-0.30}^{+0.09}$ & $0.30_{-0.23}^{+0.32}$ & $9.33\cdot10^{-1}$ & $1.25\cdot10^{-4}$ & 7.2 (5) 	& 1.44 & 2.6  \\
		140515A		& 6.327	& pow		& smc\tablefootmark{d}	& $0.77_{-0.03}^{+0.04}$ & $0.00_{-0.00}^{+0.10}$ & $.............$	   & $1.36\cdot10^{-4}$ & 3.7 (6) 	& 0.62 & 14.6 \\
		140614A		& 4.233	& bknpow	& smc	& $0.60_{-0.05}^{+0.05}$ & $0.11_{-0.05}^{+0.17}$ & $1.89\cdot10^{+0}$ & $2.83\cdot10^{-4}$ & 5.2 (8) 	& 0.65 & 4.1  \\
		151027B		& 4.062	& bknpow	& lmc	& $0.50_{-0.02}^{+0.08}$ & $0.10_{-0.06}^{+0.05}$ & $1.00\cdot10^{-2}$ & $4.21\cdot10^{-3}$ & 18.7 (16) & 1.17 & 8.8  \\
		151112A		& 4.270	& bknpow	& lmc	& $0.71_{-0.06}^{+0.05}$ & $0.50_{-0.11}^{+0.21}$ & $1.36\cdot10^{+0}$ & $2.59\cdot10^{-4}$ & 21.4 (25) & 0.86 & 11.4 \\ \hline
		071025		& 4.900	& bknpow	& sn	& $0.80_{-0.07}^{+0.16}$ & $0.57_{-0.21}^{+0.18}$ & $9.99\cdot10^{-3}$ & $3.40\cdot10^{-4}$ & 41.9 (36) & 1.16 & 22.4 \\  
  		090205		& 4.900	& pow		& sn	& $0.90_{-0.03}^{+0.03}$ & $0.10_{-0.03}^{+0.03}$ & $...$ 			   & $1.31\cdot10^{-4}$ & 13.7 (10) & 1.37 & 7.1  \\ 
        090429B\tablefootmark{c}&9.380&bknpow&sn& $0.72_{-0.27}^{+0.06}$ & $0.19_{-0.14}^{+0.22}$ & $1.19\cdot10^{+0}$ & $1.75\cdot10^{-4}$ & 2.2 (4) 	& 0.55 & 2.9  \\
        090516A		& 4.109	& pow		& sn	& $0.98_{-0.02}^{+0.02}$ & $0.24_{-0.05}^{+0.04}$ & $.............$    & $1.63\cdot10^{-4}$ & 26.4 (23) & 1.15 & 15.3 \\
        100219A		& 4.667	& pow		& sn	& $0.75_{-0.04}^{+0.04}$ & $0.16_{-0.05}^{+0.05}$ & $.............$    & $2.21\cdot10^{-4}$ & 5.9 (9) 	& 0.66 & 11.4 \\
        111008A		& 4.990	& bknpow	& sn	& $0.38_{-0.02}^{+0.21}$ & $0.18_{-0.15}^{+0.05}$ & $9.96\cdot10^{-3}$ & $9.47\cdot10^{-3}$ & 28.1 (19) & 1.48 & 6.8  \\
		140428A		& 4.680	& bknpow	& sn	& $0.74_{-0.24}^{+0.15}$ & $0.15_{-0.11}^{+0.13}$ & $2.36\cdot10^{-1}$ & $2.36\cdot10^{-3}$ & 5.8 (5) 	& 1.16 & 2.6  \\   
		140614A		& 4.233	& bknpow	& sn	& $0.32_{-0.32}^{+0.31}$ & $0.27_{-0.19}^{+0.09}$ & $9.94\cdot10^{-3}$ & $2.89\cdot10^{-3}$ & 5.6 (8) 	& 0.70 & 4.1  \\ 
  		151027B		& 4.062	& bknpow	& sn	& $0.51_{-0.02}^{+0.07}$ & $0.09_{-0.05}^{+0.03}$ & $9.93\cdot10^{-3}$ & $3.31\cdot10^{-3}$ & 16.8 (16) & 1.05 & 8.8  \\  
        151112A		& 4.270	& bknpow	& sn	& $0.71_{-0.07}^{+0.06}$ & $0.41_{-0.12}^{+0.11}$ & $1.35\cdot10^{+0}$ & $2.59\cdot10^{-4}$ & 23.0 (25) & 0.92 & 11.4 \\
\hline
\end{tabular}
		\tablefoot{In the top section of the table we list the best-fit parameters from fitting the local extinction curves. In case we found evidence for a medium amount
        of dust ($A_V > 0.1$ mag), we also fitted the SN extinction curve, for which the corresponding best-fit parameters are listed at the bottom part of the table.
        \tablefoottext{a}{Spectral slope in the NIR/optical wavelength regime. In case the SED is best-fit with a
        broken power law (bknpow) the slope in the X-ray is fixed to $\beta_{\mathrm{x}} = \beta_{\mathrm{o}} + 0.5$, otherwise
        (pow) $\beta_{\mathrm{x}} = \beta_{\mathrm{o}}$ (see Section \ref{sec:34})}
        \tablefoottext{b}{The common reference time where the broadband SED was created -- in hrs
        after the prompt trigger.}
		\tablefoottext{c}{Fit not performed with GROND data (see Section \ref{sec:35}).}
        \tablefoottext{d}{The extinction curve can basically not be identified since there is
        no evidence for extinction or the $A_V$ is consistent with zero. However, we list the model
        resulting in the lowest $\chi^2_{\mathrm{red}}$.}
		}
\end{table*}

%		100205A\tablefootmark{b} & 12.00 & bknpow	& smc	& $0.47$ 	&	$-$ 				   & $8.36\cdot10^{-2}$ & $2.76\cdot10^{-4}$ &2.6 \\

%% file: table/av_lit.tex
\begin{table}[h]
	\caption{Host intrinsic visual extinction for the GRBs at $z > 4$ that
    were not observed by GROND. All values are collected from the literature.}
	\label{tab:avlit}     
	\centering
	\renewcommand{\tabcolsep}{0.45cm}
	\renewcommand*{\arraystretch}{1.05}
	\begin{tabular}{c c c c}  
	\hline\hline
		GRB 	& Redshift\tablefootmark{a} & 	$A_V$  & Reference	\\
		yymmdd\#&  (z) 					& 	(mag)		&	\\ \hline
		000131 	& $4.500$ 				& $0.29^{+0.18}_{-0.18}$ & (1)		\\
		050502B & $5.2^{+0.3}_{-0.3}$	& $<0.5$ 				 & (2)	 	\\
		050505 	& $4.275$ 				& $0.29^{+0.06}_{-0.06}$ & (3)		\\
		050814	& $5.77^{+0.12}_{-0.12}$& $0.23^{+0.15}_{-0.15}$ & (1)		\\
		050904	& $6.295$ 				& $<0.32$  				 & (4)(5)	\\
		050922B & $4.5\pm 0.5$ 			& ...  					 & ...		\\
		060206 	& $4.048$				& $<0.17$ 				 & (6)		\\
		060223A & $4.406$ 				& ... 	 				 & ...		\\
		060510B & $4.941$ 				& ... 	 				 & ...		\\
		060522  & $5.11$   				& ... 	 				 & ...		\\
		060927 	& $5.467$   			& $<0.12$ 				 & (7)		\\
		100302A & $4.813$				& ...	 			     & ...		\\
		120521C & $6.0$					& $<0.05$ 				 & (8)		\\
		120923A\tablefootmark{b} & $7.84$& $0.06$ 				 & (9) 		\\
		140518A & $4.707$  				& $0.03^{+0.02}_{-0.02}$ & (9)		\\ \hline 
	\end{tabular}
	\tablefoot{
	\tablefoottext{a}{Photometric if given with errors, spectroscopic otherwise.}
    \tablefoottext{b}{Observed by GROND but not detected.}
    }
    \tablebib{
	(1)~\citet{curran2008}; (2)~\citet{afonso2011}; (3)~\citet{hurkett2006}; (4)~\citet{zafar2010};
    (5)~\citet{stratta2011}; (6)~\citet{covino2013}; (7)~\citet{zafar2011a}; (8)~\citet{laskar2014};
    (9)~\citet{tanvir2017n}; (10)~\citet{little2015}.
    }    
\end{table}

%% file: table/lc_fit.tex
\begingroup
\renewcommand*{\arraystretch}{1.545}
\begin{longtable}{c c c c c c c c}
\caption{\label{tab:lcfit}Light curve models and parameters used to flux normalize XRT X-ray and GROND NIR/optical data
to a common reference time.}\\
\hline\hline
	 			&	Model 	& $\alpha_1$  	  & $\alpha_2$	  & $\alpha_3$	&  $t_{\mathrm{break}}$ & host 	& additional data\tablefootmark{a} 	\\ 
                &			&				  &				  &				&	(ks)	& (y/n)	& (reference)			\\ \hline
\hline
\endfirsthead
\caption{continued.}\\
\hline\hline
	 			&	Model 	& $\alpha_1$  	  & $\alpha_2$	  & $\alpha_3$	&  $t_{b}$ & host 	& add. data\tablefootmark{a}		\\ 
                &			&				  &				  &				&	(ks)	& (y/n)	& (reference)			\\ \hline
\hline
\endhead
\hline
\endfoot
   	\multicolumn{8}{c}{GRB 071025} \\	
    NIR/opt. &  	PL 		& $1.32\pm0.11$   & $...$		  & $...$ 		&  $...$	& no	& PAIRTEL $JHK_s$ (1) \\
    X-ray		& 	PL		& $1.60\pm0.01$   & $...$		  & $...$ 		&  $...$	& 		&					\\ \hline
    \multicolumn{8}{c}{GRB 080825B\tablefootmark{b}} \\
    NIR/opt. &  	BRPL 	& $1.1\pm0.1$ 	  & $2.2\pm0.6$	  & $...$ 		&  $59\pm 24$& no	& DFOSC $I$	(2) 	\\
    X-ray		& 	PL		& $1.8\pm0.6$ 	  & $...$		  & $...$ 		&  $...$	& 		&					\\ \hline 
	\multicolumn{8}{c}{GRB 080913} \\	
    NIR/opt. &  	PL 		& $1.13\pm0.08$ 	& $...$		  & $...$ 		&  $...$	& no	&  					\\
    X-ray		& 	PL		& $1.22\pm0.06$   	& $...$		  & $...$ 		&  $...$	& 		&  					\\ \hline  
    \multicolumn{8}{c}{GRB 080916C} \\
    NIR/opt. &  	PL 		& $1.73\pm 0.39$ & $...$	  	  & $...$ 		&  $...$	& no	& 	 				\\
    X-ray		& 	PL		& $1.30\pm 0.09$ & $...$		  & $...$ 		&  $...$	& 		&					\\ \hline     
    \multicolumn{8}{c}{GRB 090205} \\
    NIR/opt. &  	PL 		& $1.35 \pm 0.04$ & $...$		  & $...$ 		&  $...$ 	& yes	& FORS1 $RI$ (3)	\\
    X-ray		& 	BRPL	& $0.91 \pm 0.03$ & $2.12\pm0.07$ & $...$ 		&  $23\pm 1$& 		&					\\ \hline   
	\multicolumn{8}{c}{GRB 090423} \\
    NIR/opt. & BRPL 		& $0.05\pm0.01$ & $1.00\pm 0.03$  & $...$ 		&  $24\pm8$ 	& no	& HAWKI $KJ$ / WFCAM $K$	\\
    X-ray		& PL + BRPL	& $1.38\pm0.04$ & $-3.75\pm4.20$  & $1.34\pm0.15$&  $3.0\pm4.6$& 		& NIRI $HJ$ / ISAAK $J$ (4)(5)\\ \hline   
	\multicolumn{8}{c}{GRB 090429B\tablefootmark{c}} \\
    NIR/opt. &  	PL 		& $0.61\pm 0.08$ & $...$		  & $...$ 		&  $...$ 	& no	& NIRI $JHK$ (6) 			\\
    X-ray		& 	BRPL	& $-1.14\pm0.59$ & $1.37\pm0.10$  & $...$ 		&  $0.65\pm0.18$	& 		&				\\ \hline     
    \multicolumn{8}{c}{GRB 090516A} \\
    NIR/opt. &  	PL 		& $1.72 \pm 0.05$ & $...$		  & $...$ 		&  $...$ 	& yes	& FORS2 $R$ (7) / NOT $R$ (8) \\
    X-ray		& 	BRPL	& $0.78 \pm 0.06$ & $1.75\pm0.05$ & $...$ 		&  $16\pm 1$& 		&				\\ \hline  
    \multicolumn{8}{c}{GRB 100219A} \\
    NIR/opt. 	&  	PL 		& $1.68\pm0.46$ & $...$		  	& $...$ 		&  $...$ 	& yes	&  HAWKI $KHJ$ / GMG $r'$			\\
    X-ray		& 	BRPL	& $1.72\pm0.25$ & $>3.0$ 	  	& $...$ 		&  $...$	& 		&   GTC $i'$ (9)				\\ \hline    
    \multicolumn{8}{c}{GRB 100513A} \\
    NIR/opt. &  	PL 		& $0.48\pm0.17$ & $...$		  	  & $...$ 		&  $...$ 	& no	& PAIRITEL $JHK_s$ (10) \\
    X-ray		& 	PL		& $0.96\pm0.06$ & $...$ 		  & $...$ 		&  $...$	& 		& 				\\ \hline 
    \multicolumn{8}{c}{GRB 100905A} \\
    NIR/opt. &  	PL		& $0.60\pm0.06$ & $...$ & $...$ & $...$ & no	&  		\\
    X-ray		& 	PL 		& $0.88\pm0.03$ & $...$ & $...$ & $...$ & 		&  		\\ \hline
    \multicolumn{8}{c}{GRB 111008A} \\
    NIR/opt. &  	PL 		& $1.02 \pm 0.06$ & $...$		  & $...$ 		&  $...$ 	& no	& 			\\
    X-ray		& 	BRPL	& $0.97 \pm 0.09$ & $1.42\pm0.05$ & $...$ 		&  $41\pm 11$& 		& 			\\ \hline
    \multicolumn{8}{c}{GRB 120712A} \\
    NIR/opt. &  	BRPL 	& $0.04\pm0.08$ & $1.66\pm0.16$ & $...$ 		&  $102\pm13$ & no	&  			\\
    X-ray		& 	BRPL	& $0.95\pm0.03$ & $1.80\pm0.11$ & $...$ 		&  $6\pm 1$& 		& 			\\ \hline 
    \multicolumn{8}{c}{GRB 130606A} \\
    NIR/opt. &  	PL		& $1.71\pm0.03$ & $...$ & $...$ & $...$ & no	&  	NOT $r'i'z'$ / TNG $i'z'$ (11)	\\
    X-ray		& 	PL 		& $1.71\pm0.03$ & $...$ & $...$ & $...$ & 		&  	GTC $i'z'$ (12)					\\ \hline
    \multicolumn{8}{c}{GRB 131117A} \\
    NIR/opt. &  	PL		& $0.86\pm0.01$ & $...$ & $...$ & $...$ & no	&  			\\
    X-ray		& 	PL + G 	& $0.95\pm0.02$ & $...$ & $...$ & $...$ & 		&  		\\ \hline
    \multicolumn{8}{c}{GRB 140304A} \\
    NIR/opt. &  	PL + BRPL& $2.25\pm0.24$ & $-7.42\pm5.94$ & $1.86\pm0.20$ &  $25\pm2$ 		& no	&  RATIR $grizJH$ (13) \\
    X-ray		& 	PL + BRPL& $1.93\pm0.14$ & $-9.71\pm 6.89$ & $1.50\pm0.09$ &  $16\pm 2$		& 		&  Nanshan $R$ (14) / NOT $gr$ (15) \\
    \multicolumn{7}{c}{ }																				&  MONDY $R$(16) / ORI-40 $R$ (17) \\ \hline
    \multicolumn{8}{c}{GRB 140311A} \\
    NIR/opt. &  	PL 		& $0.95\pm0.04$ & $...$ & $...$ 		&  $...$ 		& no	&  NOT $r'i'$ (18)				\\
    X-ray		& 	BRPL	& $1.20\pm0.07$ & $1.97\pm0.58$ & $...$ &  $121\pm 65$	& 		& 							\\ \hline
    \multicolumn{8}{c}{GRB 140428A} \\
    NIR/opt. &  	PL 		& $1.81\pm0.08$ & $...$  & $...$ 		&  $...$ 	& no	& OSN $I$ (19) 			\\
    X-ray		& 	BRPL	& $0.96\pm0.07$ & $>2.0$ & $...$ 		&  $19\pm 9$ 	& 		& 	LRIS $i$ (20)	\\ \hline
    \multicolumn{8}{c}{GRB 140515A} \\
    NIR/opt. &   BRPL		& $0.77\pm0.25$ & $>2.5$ 		 & $...$ 			&  $...$ 		& no	& GMOS $z'$ (21)		\\
    X-ray	& 	PL + BRPL& $3.10\pm0.19$ & $-3.77\pm0.69$ & $1.03\pm0.05$	&  $2.3\pm 0.2$ &	& Chandra (22)		\\ \hline  
    \multicolumn{8}{c}{GRB 140614A} \\
    NIR/opt. &  	PL 		& $1.20\pm 0.04$ & $...$ 	     & $...$ 		&  $...$ 		& no	&  							\\
    X-ray		& 	PL + BRPL& $1.92\pm0.07$ & $-0.27\pm0.32$ & $1.49\pm0.08$&  $7\pm1$ 	& 		& 				\\ \hline
    \multicolumn{8}{c}{GRB 151027B} \\
    NIR/opt. &  	BRPL 	& $0.44\pm0.19$ & $1.44\pm0.14$ & $...$ 		&  $34\pm28$ 	& no	& NOT $R$ (23) 	\\
    X-ray		& 	BRPL	& $0.44\pm0.19$ & $1.44\pm0.14$ & $...$ 		&  $34\pm28$ 	& 		& RATIR $r'i'$ (24) \\ \hline
    \multicolumn{8}{c}{GRB 151112A} \\
    NIR/opt. &  	BRPL 	& $0.84\pm0.06$ & $2.10\pm0.29$ & $...$ 		&  $128\pm 14$ 	& no	&  				\\
    X-ray		& 	BRPL	& $0.84\pm0.06$ & $>1.2$ 		& $...$ 		&  $128\pm 14$ 	& 		& 				\\
\end{longtable}
\tablefoot{\tablefoottext{a}{\textbf{GROND optical and near-infrared data were extended} with data collected from GCNs and refereed publications.
The magnitudes were, if necessary, converted to the AB system.}
\tablefoottext{b}{The XRT light curve is also consistent with the model for the NIR/optical afterglow (see Fig. \ref{fig:080825Blc}).}
\tablefoottext{b}{We used the temporal decay slope as derived from the NIRI $K$ band data to re-scale the NIRI $J$, $H$, and $K$-band
observation to a common reference time (see Fig. \ref{fig:090429Blc}).}
}
   \tablebib{
	(1)~\citet{perley2010}; (2)~\citet{thoene2008}; (3)~\citet{davanzo2010}; (4)~\citet{tanvir2009}; (5)~\citet{salvaterra2009};
    (6)~\citet{cucchiara2011}; (7)~\citet{gorosabel2009}; (8)~\citet{ugarte2009a}; (9)~\citet{thoene2013};
    (10)~\citet{morgan2010}; (11)\citet{hartoog2015};
    (12)~\citet{tirado2013}; (13)~\citet{butler2014}; (14)~\citet{xu2014a}; (15)~\citet{ugarte2014}; (16)~\citet{volnova2014};
    (17)~\citet{volnova2014b}; (18)~\citet{xu2014}; (19)~\citet{aceituno2014}; (20)~\citet{perley201428a};
    (21)~\citet{melandri2015}; (22)~\citet{margutti2014}; (23)~\citet{malesani2015}; (24)~\citet{watson2015}
    .}
\endgroup

%	\hline 
%	\end{tabular}
%	\tablefoot{
%	\tablefoottext{a}{bla}
%	}
%	\tablebib{
%	.}    
%\end{table*}

%% file: table/phot.tex
\begingroup
\renewcommand*{\arraystretch}{1.23}
\begin{longtable}{l c c c c c c c}
\caption{\label{tab:grondmags}GROND photometry of the GRB afterglows.}\\
\hline\hline
\multicolumn{1}{c}{$T_{\text{GROND}}$\tablefootmark{a}}   & \multicolumn{7}{c}{AB magnitude\tablefootmark{b}} \\
\multicolumn{1}{c}{(s)}       & $g'$  & $r'$ & $i'$ & $z'$ & $J$ & $H$ & $K_s$ \\ 
\hline
\endfirsthead
\caption{continued.}\\
\hline\hline
\multicolumn{1}{c}{$\text{T}_{\text{GROND}}$\tablefootmark{a}}   & \multicolumn{7}{c}{AB magnitude\tablefootmark{b}} \\
\multicolumn{1}{c}{(s)}       & $g'$  & $r'$ & $i'$ & $z'$ & $J$ & $H$ & $K_s$ \\ 
\hline
\endhead
\hline
\endfoot
\multicolumn{8}{c}{GRB 071025B / $T_0 =$ 04:08:54 UT (MJD $= 54398.17285$) (1)} \\
$80533\pm 2452$     & $ >25.2$ 			& $ >24.5$ 			& $22.93\pm 0.24$ 	& $22.29\pm 0.13$ 	& $21.24\pm 0.18$ 	& $20.79\pm 0.24$ 	& $>19.8$ 			\\ \hline
%\multicolumn{8}{c}{GRB 080129 / $T_0 =$ 06:06:45 UT (MJD $ = 54494.25469$) (2)} \\
%$167910\pm 3395$	& $>24.9$			& $23.45\pm 0.10$	& $22.12\pm	0.05$	& $21.44\pm	0.05$ 	& $20.75\pm	0.10$ 	& $20.19\pm 0.11$ 	& $19.53\pm 0.10$	\\ 
%$246744\pm 727$		& $>25.3$			& $24.40\pm 0.12$	& $23.06\pm	0.19$	& $22.12\pm	0.13$ 	& $21.36\pm 0.03$ 	& $>20.7$ 			& $>20.5$			\\ 
%$328339\pm 1761$	& $>25.4$			& $24.84\pm 0.13$	& $24.04\pm 0.39$	& $23.12\pm	0.23$ 	& $>21.7$		  	& $>21.0$			& $>20.6$			\\ \hline
\multicolumn{8}{c}{GRB 080825B / $T_0 =$ 17:46:40 UT (MJD $ = 54703.74074$) (2)} \\
$25566\pm 1670$		& $>24.6$ 			& $19.78\pm 0.03$	& $18.40\pm 0.03$	& $18.11\pm 0.03$	& $17.76\pm 0.06$	& $17.36\pm 0.05$	& $17.10\pm 0.08$	\\		
$29344\pm 1670$ 	& $>24.0$			& $19.91\pm 0.03$	& $18.60\pm 0.03$	& $18.27\pm 0.03$	& $17.94\pm 0.06$	& $17.56\pm 0.06$	& $17.46\pm 0.08$	\\
$32695\pm 1431$ 	& $>24.3$			& $20.05\pm 0.03$	& $18.72\pm 0.03$	& $18.40\pm 0.03$	& $18.02\pm 0.06$	& $17.73\pm 0.06$	& $17.51\pm 0.08$	\\
$110261\pm 1981$ 	& $>24.1$			& $22.22\pm 0.04$	& $20.93\pm 0.04$	& $20.67\pm 0.05$	& $20.17\pm 0.15$	& $19.97\pm 0.14$	& $>18.8$			\\ \hdashline
$29201\pm 4779$		& $>24.4$			& $19.94\pm 0.03$ 	& $18.62\pm 0.03$ 	& $18.30\pm 0.03$ 	& $17.92\pm 0.06$ 	& $17.57\pm 0.05$ 	& $17.41\pm 0.08$	\\ \hline
\multicolumn{8}{c}{GRB 080913 / $T_0 =$ 06:46:54 UT (MJD $ = 54722.28257$) (3)} \\
$556\pm	193$		& $>22.4$ 			& $>23.4$ 			& $>23.0$			& $21.55\pm 0.14$	& $19.88\pm 0.10$ 	& $19.86\pm	0.16$ 	& $19.44\pm 0.35$	\\
$994\pm	189$		& $>23.2$ 			& $>23.4$ 			& $>23.0$			& $22.23\pm	0.26$	& $20.59\pm 0.18$	& $20.46\pm	0.27$ 	& $20.24\pm 0.50$	\\
$1962\pm 730$		& $>23.8$ 			& $>24.3$ 			& $>23.6$			& $23.07\pm	0.27$ 	& $21.33\pm 0.21$	& $20.88\pm	0.21$ 	& $20.70\pm 0.39$	\\
$4326\pm 1542$		& $>24.3$ 			& $>24.9$ 			& $>24.0$			& $24.18\pm	0.38$ 	& $21.95\pm 0.60$	& $21.36\pm	0.39$ 	& $>20.9$			\\
$7715\pm 1769$		& $>24.4$ 			& $>25.1$ 			& $>24.1$			& $24.81\pm	0.65$	& $>22.3$			& $>21.6$			& $>21.0$			\\ \hline
\multicolumn{8}{c}{GRB 080916C / $T_0 =$ 00:12:45 UT (MJD $ = 54725.00885$) (4)} \\
$117308\pm 3045$  	& $>24.0$     		& $22.80\pm 0.07$ 	& $22.00\pm 0.05$ 	& $21.66\pm 0.05$ 	& $21.45\pm 0.06$ 	& $21.26\pm 0.08$ 	& $21.04\pm 0.15$   \\ \hline
\multicolumn{8}{c}{GRB 090205 / $T_0 =$ 23:03:14 UT (MJD $= 54867.96058$) (5)} \\
$23348\pm 340$ 		& $>22.9$			& $22.39\pm 0.14$	& $20.76\pm 0.06$ 	& $20.48\pm 0.05$ 	& $20.16\pm 0.13$ 	& $19.84\pm 0.15$ 	& $19.25\pm 0.26$ 	\\
$24111\pm 342$ 		& $>23.1$			& $22.18\pm 0.10$	& $20.75\pm 0.05$ 	& $20.66\pm 0.05$	& $20.17\pm 0.12$ 	& $19.66\pm 0.12$ 	& $19.60\pm 0.29$ 	\\
$25396\pm 860$ 		& $>24.2$			& $22.43\pm 0.06$	& $20.93\pm 0.04$ 	& $20.69\pm 0.04$	& $20.49\pm 0.10$ 	& $19.89\pm 0.09$ 	& $19.87\pm 0.28$ 	\\
$27201\pm 858$ 		& $>25.1$			& $22.57\pm 0.05$	& $21.06\pm 0.03$	& $20.78\pm 0.03$	& $20.34\pm 0.09$ 	& $20.05\pm 0.13$ 	& $19.70\pm 0.21$ 	\\
$120692\pm 2656$ 	& $>25.9$			& $24.67\pm 0.13$	& $23.02\pm 0.07$ 	& $22.60\pm 0.08$	& $>22.6$ 			& $>22.0$ 			& $>20.8$ 			\\
$207844\pm 2677$ 	& $>25.2$			& $25.37\pm 0.47$	& $23.78\pm 0.15$ 	& $23.35\pm 0.13$	& $>22.5$ 			& $>21.9$ 			& $>21.0$ 			\\ \hdashline
$25534\pm 2525$ 	& $>25.3$ 			& $22.50\pm 0.04$ 	& $20.97\pm 0.03$ 	& $20.64\pm 0.03$ 	& $20.29\pm 0.07$ 	& $19.85\pm 0.08$ 	& $19.42\pm 0.12$ 	\\ \hline
\multicolumn{8}{c}{GRB 090423 / $T_0 =$ 07:55:19 UT (MJD$ = 54944.33008$) (6)} \\
$62136\pm 7076$ 	& $>25.0$ 			& $>25.1$ 			& $>24.2$ 			& $>24.0$ 			& $21.61\pm 0.11$	& $21.19\pm 0.11$ 	& $20.89\pm 0.12$	\\
$150527\pm 3557$	& $>24.6$ 			& $>24.8$ 			& $>24.7$ 			& $>23.2$ 			& $>22.5$			& $>21.9$ 			& $>21.1$			\\ \hline
\multicolumn{8}{c}{GRB 090429B / $T_0 =$ 05:30:03 UT (MJD$ = 54950.22920$) (7)} \\
$1902\pm 1081$ 		& $>23.7$ 			& $>23.5$ 			& $>22.8$ 			& $>22.3$ 			& $>20.9$			& $>20.3$ 			& $>19.8$			\\
$81805\pm 3629$		& $>25.7$ 			& $>25.4$ 			& $>24.6$ 			& $>24.0$ 			& $>22.3$			& $>21.8$	 		& $>20.7$			\\ \hline
\multicolumn{8}{c}{GRB 090516 / $T_0 = $ 08:27:50 UT (MJD$ = 54967.35266$) (8)} \\
$53290\pm 724$		& $23.86\pm 0.14$	& $21.34\pm 0.03$ 	& $20.39\pm 0.04$ 	& $19.89\pm 0.04$ 	& $19.63\pm 0.10$ 	& $19.29\pm 0.10$ 	& $19.16\pm 0.16$ 	\\	
$54959\pm 859$ 		& $23.76\pm 0.10$	& $21.40\pm 0.03$ 	& $20.41\pm 0.03$ 	& $19.96\pm 0.03$ 	& $19.54\pm 0.08$ 	& $19.17\pm 0.09$ 	& $19.04\pm 0.13$ 	\\	
$56764\pm 861$ 		& $24.04\pm 0.13$	& $21.43\pm 0.03$ 	& $20.43\pm 0.03$ 	& $20.01\pm 0.03$ 	& $19.57\pm 0.08$ 	& $19.20\pm 0.10$ 	& $18.99\pm 0.14$ 	\\
$58112\pm 175$ 		& $>23.6$			& $21.43\pm 0.06$ 	& $20.47\pm 0.05$ 	& $20.11\pm 0.05$ 	& $19.84\pm 0.14$ 	& $19.23\pm 0.16$ 	& $19.18\pm 0.23$ 	\\
$65203\pm 862$ 		& $23.90\pm 0.29$	& $21.60\pm 0.04$ 	& $20.62\pm 0.03$ 	& $20.17\pm 0.04$ 	& $19.85\pm 0.09$ 	& $19.42\pm 0.10$ 	& $18.88\pm 0.14$ 	\\
$143076\pm 859$ 	& $25.28\pm 0.29$	& $23.15\pm 0.06$ 	& $21.99\pm 0.05$ 	& $21.67\pm 0.06$ 	& $21.00\pm 0.15$ 	& $20.75\pm 0.20$ 	& $20.01\pm 0.25$	\\
$229907\pm 859$ 	& $>25.3$			& $23.94\pm 0.15$ 	& $22.72\pm 0.09$ 	& $22.41\pm 0.13$ 	& $>21.8$			& $>21.1$ 			& $>20.3$			\\
$487152\pm 3319$ 	& $>26.0$ 			& $24.78\pm 0.15$ 	& $23.53\pm 0.13$ 	& $23.00\pm 0.15$ 	& $>22.3$ 			& $>21.8$ 			& $>20.9$			\\	
%$24008050\pm 3576$	& $>26.3$  			& $>26.1$ 			& $>25.2$ 			& $>24.6$ 			& $>22.6$			& $>22.0$ 			& $>21.0$			\\
\hdashline
$55095\pm 2530$ 	& $23.87\pm 0.07$	& $21.40\pm 0.03$ 	& $20.41\pm 0.03$	& $19.92\pm 0.03$	& $19.61\pm 0.07$	& $19.18\pm 0.07$	& $19.00\pm 0.10$ 	\\ \hline
%\multicolumn{8}{c}{GRB 100205A} \\
%$9479\pm 597$     	& $>23.3$   		& $>23.5$ 			& $>23.0$ 			& $>22.6$ 			& $>21.9$ 			& $> 21.8$ 			& $>21.7$ 			\\ \hline
\multicolumn{8}{c}{GRB 100219A / $T_0 =$ 15:15:46 (MJD $= 55246.63606$) (9)} \\
$34248\pm 1130$ 	& $>24.2$ 			& $22.77\pm 0.09$ 	& $21.37\pm 0.06$ 	& $21.24\pm 0.08$ 	& $20.72\pm 0.20$ 	& $20.23\pm 0.19$ 	& $20.12\pm 0.34$ 	\\
$41043\pm 2656$		& $>26.0$			& $23.17\pm 0.05$ 	& $21.73\pm 0.04$ 	& $21.40\pm 0.04$ 	& $20.92\pm 0.12$ 	& $20.53\pm 0.13$ 	& $20.31\pm 0.23$ 	\\
$131666\pm 2672$	& $>26.1$			& $24.52\pm 0.12$	& $23.62\pm 0.10$ 	& $23.54\pm 0.15$ 	& $>22.3$ 			& $>21.7$ 			& $>20.8$ 			\\
$221279\pm 2528$	& $>25.9$ 			& $24.68\pm 0.14$ 	& $23.98\pm 0.15$ 	& $22.84\pm 0.20$	& $>22.4$			& $>21.9$ 			& $>20.9$ 			\\
$397579\pm 3575$	& $>25.8$			& $24.83\pm 0.17$ 	& $24.69\pm 0.27$ 	& $24.34\pm 0.32$	& $>22.5$			& $>22.1$ 			& $>20.6$ 			\\ \hline
\multicolumn{8}{c}{GRB 100518A / $T_0 =$ 11:33:35 UT (MJD $=55334.48166$) (10)} \\
$71558\pm 9906$ 	& $25.57\pm 0.57$	& $23.74\pm 0.13$ 	& $23.27\pm 0.18$ 	& $22.94\pm 0.17$ 	& $22.20\pm 0.42$  	& $21.53\pm 0.37$ 	& $>20.3$ 			\\ \hline
\multicolumn{8}{c}{GRB 100905A / $T_0 =$ 15:08:14 UT (MJD $= 55444.63072$) (11)} \\
$57687\pm 9809$		& $>25.9$ 			& $>26.1$ 			& $>25.2$ 			& $>24.7$			& $21.85\pm 0.20$ 	& $21.72\pm 0.26$ 	& $21.50\pm 0.43$ 	\\ \hline
\multicolumn{8}{c}{GRB 111008A / $T_0 =$ 22:12:58 UT ($\text{MJD} = 55842.92567$) (12)} \\
$24653\pm 1512$ 	& $>24.4$ 			& $22.76\pm 0.09$ 	& $21.03\pm 0.04$ 	& $20.00\pm 0.03$	& $19.58\pm 0.06$ 	& $19.42\pm 0.07$ 	& $19.23\pm 0.10$	\\ 		
$28944\pm 2675$ 	& $>24.7$			& $23.11\pm 0.09$	& $21.14\pm 0.04$	& $20.18\pm 0.03$ 	& $19.74\pm 0.06$	& $19.58\pm 0.08$ 	& $19.53\pm 0.14$	\\
$34402\pm 2682$ 	& $>25.1$			& $23.21\pm 0.09$	& $21.36\pm 0.04$	& $20.30\pm 0.03$	& $19.92\pm 0.06$	& $19.95\pm 0.10$ 	& $19.79\pm 0.15$	\\
$116116\pm 1762$ 	& $>23.9$			& $>24.2$			& $22.56\pm 0.11$	& $21.60\pm 0.06$	& $20.99\pm 0.15$	& $20.92\pm 0.25$ 	& $20.38\pm 0.28$	\\
$207405\pm 3588$ 	& $>23.9$			& $24.45\pm 0.26$	& $22.67\pm 0.16$ 	& $22.21\pm 0.11$	& $21.37\pm 0.16$	& $20.94\pm 0.18$	& $20.60\pm 0.28$	\\
$293685\pm 3582$ 	& $>23.9$			& $>24.3$			& $23.30\pm 0.20$	& $22.09\pm 0.11$ 	& $21.80\pm 0.17$	& $21.30\pm 0.18$	& $20.83\pm 0.24$	\\
$639133\pm 3576$ 	& $>24.7$			& $>24.2$			& $24.35\pm 0.34$	& $23.87\pm 0.27$	& $>22.6$			& $>22.0$			& $>21.1$			\\
$982066\pm 1771$ 	& $>25.4$			& $>25.4$			& $>24.8$			& $>24.1$			& $>22.4$			& $>21.7$			& $>20.7$			\\ \hline
\multicolumn{8}{c}{GRB 120712A / $T_0 =$ 13:42:27 UT ($\text{MJD} = 56120.57115$) (13)} \\
$34495\pm 1191$ 	& $23.24\pm 0.09$	& $21.26\pm 0.03$ 	& $20.58\pm 0.04$ 	& $20.44\pm 0.04$ 	& $20.02\pm 0.09$ 	& $19.90\pm 0.11$	& $19.67\pm 0.18$	\\ 
$37541\pm 1771$ 	& $23.11\pm 0.06$ 	& $21.27\pm 0.03$	& $20.61\pm 0.03$ 	& $20.34\pm 0.03$	& $20.12\pm 0.08$	& $19.78\pm 0.10$	& $19.89\pm 0.22$	\\
$41032\pm 1633$ 	& $23.17\pm 0.07$	& $21.28\pm 0.03$	& $20.63\pm 0.03$ 	& $20.38\pm 0.04$	& $19.97\pm 0.08$	& $19.95\pm 0.11$	& $19.34\pm 0.17$	\\
$297391\pm 2680$	& $25.44\pm 0.37$	& $23.21\pm 0.07$	& $22.50\pm 0.08$	& $22.44\pm 0.14$	& $>21.9$ 			& $>21.3$			& $>20.7$ 			\\
$470194\pm 2447$ 	& $>25.2$			& $24.27\pm 0.16$	& $23.25\pm 0.15$	& $...$				& $>22.1$ 			& $>21.4$			& $>20.0$ 			\\
$641263\pm 1767$ 	& $>25.4$			& $24.55\pm 0.19$	& $23.80\pm 0.22$	& $>23.7$			& $>21.9$ 			& $>21.4$			& $>20.1$ 			\\ \hdashline
$37984\pm 4681$ 	& $23.17\pm 0.05$	& $21.28\pm 0.03$ 	& $20.62\pm 0.03$ 	& $20.42\pm 0.03$	& $20.03\pm 0.07$	& $19.86\pm 0.08$	& $19.71\pm 0.14$	\\ \hline
\multicolumn{8}{c}{GRB 120923A / $T_0 =$ 05:16:06 UT ($\text{MJD} = 56193.21951$) (14)} \\
$72311\pm 4059$		& $>23.8$ 			& $>24.3$			& $>23.7$			& $>23.7$			& $>22.1$ 			& $>21.7$			& $>20.7$ 			\\ \hline
\multicolumn{8}{c}{GRB 130606A / $T_0 =$ 21:04:39 UT ($\text{MJD} = 56449.87823$) (15)} \\
$24785\pm 1650$		& $>25.2$			& $23.46\pm 0.09$	& $...$				& $...$				& $18.28\pm 0.05$ 	& $17.95\pm 0.05$	& $17.77\pm 0.13$	\\ 
$30132\pm 1937$		& $>25.0$			& $24.05\pm 0.17$	& $21.46\pm 0.05$	& $18.99\pm 0.03$ 	& $18.56\pm 0.05$	& $18.30\pm 0.06$	& $18.05\pm 0.08$	\\
$114988\pm 4498$	& $>25.5$			& $>25.0$			& $...$				& $21.56\pm 0.06$	& $20.73\pm 0.13$	& $20.58\pm 0.26$	& $20.42\pm 0.25$	\\ \hline
\multicolumn{8}{c}{GRB 131117A / $T_0 =$ 00:34:04 UT ($\text{MJD} = 56613.02366$) (16)} \\
$218\pm 33$ 		& $...$ 			& $18.37\pm 0.03$ 	& $17.91\pm 0.03$ 	& $17.70\pm 0.04$ 	& $...$ 			& $...$ 			& $...$ 			\\
$418\pm 33$ 		& $...$ 			& $19.04\pm 0.04$ 	& $18.60\pm 0.04$ 	& $18.43\pm 0.04$ 	& $...$ 			& $...$ 			& $...$ 			\\
$520\pm 33$ 		& $...$ 			& $19.26\pm 0.04$ 	& $18.80\pm 0.04$ 	& $18.62\pm 0.05$ 	& $...$ 			& $...$ 			& $...$ 			\\
$629\pm 33$ 		& $...$ 			& $19.39\pm 0.04$ 	& $18.94\pm 0.04$ 	& $18.84\pm 0.04$ 	& $...$ 			& $...$ 			& $...$ 			\\
$730\pm 33$ 		& $...$				& $19.63\pm 0.04$ 	& $...$ 			& $18.85\pm 0.21$ 	& $...$ 			& $...$ 			& $...$ 			\\
$840\pm 33$ 		& $...$ 			& $19.67\pm 0.05$ 	& $19.22\pm 0.04$ 	& $19.16\pm 0.04$ 	& $...$ 			& $...$ 			& $...$ 			\\
$951\pm 33$ 		& $...$ 			& $19.86\pm 0.05$ 	& $19.34\pm 0.04$ 	& $19.19\pm 0.04$ 	& $...$ 			& $...$ 			& $...$ 			\\
$1087\pm 57$ 		& $...$ 			& $20.06\pm 0.04$ 	& $19.46\pm 0.04$ 	& $19.21\pm 0.04$ 	& $...$ 			& $...$ 			& $...$ 			\\
$1276\pm 57$ 		& $...$ 			& $19.99\pm 0.04$ 	& $...$ 			& $19.60\pm 0.22$ 	& $...$ 			& $...$ 			& $...$ 			\\
$1472\pm 57$ 		& $...$ 			& $20.13\pm 0.04$ 	& $19.66\pm 0.05$ 	& $19.47\pm 0.05$ 	& $...$ 			& $...$ 			& $...$ 			\\
$1668\pm 57$ 		& $...$ 			& $20.26\pm 0.05$ 	& $19.86\pm 0.05$ 	& $19.53\pm 0.05$ 	& $...$ 			& $...$ 			& $...$ 			\\
$1871\pm 57$ 		& $...$ 			& $20.39\pm 0.05$ 	& $19.90\pm 0.05$ 	& $19.73\pm 0.04$ 	& $...$ 			& $...$ 			& $...$ 			\\
%$2873\pm 346$ 		& $...$ 			& $20.60\pm 0.04$ 	& $20.24\pm 0.04$ 	& $20.06\pm 0.04$ 	& $...$ 			& $...$				& $...$ 			\\
%$3660\pm 350$ 		& $...$ 			& $20.92\pm 0.06$ 	& $20.50\pm 0.04$ 	& $20.27\pm 0.04$ 	& $...$ 			& $...$ 			& $...$				\\
%$4458\pm 356$ 		& $...$ 			& $21.10\pm 0.06$ 	& $20.60\pm 0.05$ 	& $20.45\pm 0.05$ 	& $...$ 			& $...$ 			& $...$ 			\\
%$5244\pm 350$ 		& $...$ 			& $21.28\pm 0.07$ 	& $20.81\pm 0.06$ 	& $20.49\pm 0.05$ 	& $...$ 			& $...$ 			& $...$ 			\\
%$6024\pm 344$ 		& $...$ 			& $21.41\pm 0.08$ 	& $20.92\pm 0.07$ 	& $20.61\pm 0.06$ 	& $...$ 			& $...$ 			& $...$ 			\\
%$6787\pm 340$ 		& $...$ 			& $21.50\pm 0.09$ 	& $20.99\pm 0.07$ 	& $20.88\pm 0.07$ 	& $...$ 			& $...$ 			& $...$ 			\\
%$7546\pm 340$ 		& $...$ 			& $21.59\pm 0.09$ 	& $21.10\pm 0.07$ 	& $20.92\pm 0.08$ 	& $...$ 			& $...$ 			& $...$	 			\\
%$8305\pm 340$ 		& $...$ 			& $21.61\pm 0.09$ 	& $21.20\pm 0.08$ 	& $20.98\pm 0.08$ 	& $...$ 			& $...$ 			& $...$ 			\\
%$9071\pm 340$ 		& $...$ 			& $21.72\pm 0.10$ 	& $21.33\pm 0.09$ 	& $21.13\pm 0.07$ 	& $...$ 			& $...$ 			& $...$ 			\\
%$11529\pm 2039$ 	& $22.99\pm 0.36$ 	& $21.90\pm 0.07$ 	& $21.50\pm 0.06$ 	& $21.32\pm 0.05$ 	& $21.20\pm 0.13$ 	& $21.08\pm 0.18$	& $>20.4$ 			\\
%$89138\pm 1780$ 	& $...$ 			& $23.73\pm 0.12$ 	& $23.21\pm 0.11$ 	& $23.11\pm 0.12$ 	& $>22.0$ 			& $>21.5$ 			& $>20.3$ 			\\
\hdashline
$369\pm 184$ 		& $20.15\pm 0.05$ 	& $18.82\pm 0.03$ 	& $18.40\pm 0.03$ 	& $18.19\pm 0.04$ 	& $18.18\pm 0.06$ 	& $17.97\pm 0.06$ 	& $17.80\pm 0.11$ 	\\
$790\pm 194$ 		& $20.94\pm 0.09$ 	& $...$ 			& $...$ 			& $...$ 			& $18.96\pm 0.08$ 	& $18.79\pm 0.08$ 	& $18.50\pm 0.20$ 	\\ 
$1378\pm 348$		& $21.17\pm 0.08$ 	& $...$ 			& $...$ 			& $...$ 			& $19.42\pm 0.08$ 	& $19.29\pm 0.10$ 	& $>18.7$ 			\\
%$4085\pm 1560$ 		& $...$				& $...$ 			& $...$ 			& $...$ 			& $20.31\pm 0.08$ 	& $20.03\pm 0.11$ 	& $>19.5$ 			\\
%$7571\pm 1892$  	& $...$				& $...$ 			& $...$ 			& $...$ 			& $20.80\pm 0.09$ 	& $20.70\pm 0.14$ 	& $>19.9$ 			\\
%$5777\pm 3250$ 		& $22.38\pm 0.09$	& $...$ 			& $...$ 			& $...$ 			& $...$				& $...$				& $...$				\\
\hline
\multicolumn{8}{c}{GRB 140311A / $T_0 = $ 21:05:16 UT ($\text{MJD} = 56727.87866$) (17)} \\
$27387\pm 865$ 		& $> 23.9$			& $22.36\pm 0.07$ 	& $21.04\pm 0.04$	& $20.18\pm 0.03$	& $19.70\pm 0.07$	& $19.31\pm 0.09$	& $18.96\pm 0.21$	\\
$30065\pm 1727$ 	& $> 23.9$			& $22.47\pm 0.05$	& $21.11\pm 0.03$	& $20.19\pm 0.03$	& $19.83\pm 0.06$	& $19.53\pm 0.09$	& $...$				\\
$33620\pm 1733$		& $> 24.2$			& $22.55\pm 0.05$	& $21.19\pm 0.03$	& $20.29\pm 0.03$	& $19.81\pm 0.06$ 	& $19.53\pm 0.08$ 	& $19.13\pm 0.18$ 	\\
$37169\pm 1734$ 	& $> 25.7$			& $22.68\pm 0.04$	& $21.25\pm 0.03$	& $20.38\pm 0.03$  	& $19.90\pm 0.06$	& $19.62\pm 0.08$ 	& $19.16\pm 0.16$	\\
$40733\pm 1739$		& $> 25.5$			& $22.78\pm 0.04$	& $21.42\pm 0.03$ 	& $20.52\pm 0.03$	& $20.08\pm 0.07$ 	& $19.88\pm 0.09$ 	& $19.28\pm 0.18$ 	\\ 
$43437\pm 881$ 		& $> 25.0$			& $22.88\pm 0.07$	& $21.47\pm 0.04$	& $20.61\pm 0.04$	& $20.02\pm 0.09$	& $20.12\pm 0.15$ 	& $19.70\pm 0.32$	\\
$47144\pm 1219$ 	& \multicolumn{4}{c}{NIR only}													& $20.22\pm 0.11$ 	& $19.85\pm 0.14$ 	& $19.28\pm 0.22$ 	\\ 
%$112330\pm 1111$	& $> 24.0$			& $23.55\pm 0.23$	& $21.97\pm 0.10$	& $21.05\pm 0.06$	& $20.83\pm 0.18$	& $20.45\pm 0.19$ 	& $...$				\\
%$120114\pm 1521$	& $> 24.5$ 			& $23.37\pm 0.13$ 	& $21.96\pm 0.07$	& $21.02\pm 0.05$ 	& $20.52\pm 0.12$ 	& $20.20\pm 0.12$ 	& $19.73\pm 0.25$	\\
%$123624\pm 1906$ 	& $> 25.4$			& $23.25\pm 0.10$ 	& $21.95\pm 0.06$	& $20.91\pm 0.04$	& $20.35\pm 0.10$	& $20.13\pm 0.12$	& $19.89\pm 0.25$	\\
%$128233\pm 2608$ 	& $> 25.4$ 			& $23.46\pm 0.07$	& $21.96\pm 0.05$	& $21.04\pm 0.04$ 	& $20.59\pm 0.10$ 	& $20.31\pm 0.10$ 	& $19.77\pm 0.19$	\\
%$209062\pm 1770$	& $> 23.8$ 			& $>24.1$			& $23.52\pm 0.30$	& $22.91\pm 0.21$	& $>22.0$			& $>21.4$			& $>20.5$			\\
\hdashline
$35420\pm 8898$		& $> 26.0$			& $22.64\pm 0.03$	& $21.24\pm 0.03$	& $20.32\pm 0.03$	& $19.87\pm 0.05$	& $19.60\pm 0.06$	& $19.27\pm 0.11$	\\ \hline
\multicolumn{8}{c}{GRB 140428A / $T_0 =$ 22:40:50 UT ($\text{MJD} = 56775.94502$) (18)} \\
$9363\pm 1800$		& $> 24.7$			& $23.16\pm 0.12$ 	& $21.59\pm 0.07$ 	& $21.15\pm 0.08$	& $21.06\pm 0.24$	& $20.80\pm 0.32$	& $20.30\pm 0.43$ 	\\
$13242\pm 1782$		& $> 24.4$			& $23.62\pm 0.21$	& $22.08\pm 0.11$	& $21.76\pm 0.13$	& $21.49\pm 0.29$ 	& $>20.7$			& $>20.4$			\\
$618771\pm 1907$	& $> 23.9$			& $> 23.7$			& $>23.2$			& $> 23.2$			& $>21.6$			& $>21.1$			& $>20.1$			\\ \hline 
\multicolumn{8}{c}{GRB 140515A / $T_0 =$ 09:12:36 UT ($\text{MJD} = 56792.38375$) (19)} \\
$52538\pm 3000$ 	& $>24.6$ 			& $>25.0$			& $24.71\pm 0.45$ 	& $22.09\pm 0.06$ 	& $20.60\pm 0.14$ 	& $20.53\pm 0.18$ 	& $20.40\pm 0.25$	\\	
$60572\pm 2435$ 	& $>24.2$ 			& $>24.5$ 			& $>24.3$ 			& $22.18\pm 0.06$ 	& $20.70\pm 0.11$ 	& $20.62\pm 0.15$ 	& $20.56\pm 0.26$	\\
$66119\pm 2693$		& $>24.1$ 			& $>24.5$			& $>24.4$ 			& $22.25\pm 0.07$ 	& $20.76\pm 0.11$ 	& $20.65\pm 0.14$ 	& $20.64\pm 0.29$	\\
$153522\pm 1787$	& $>23.7$ 			& $>24.0$ 			& $>23.5$ 			& $>23.4$ 			& $>21.7$ 			& $>21.1$ 			& $>20.2$			\\	
$240133\pm 3814$	& $>24.5$ 			& $>24.8$ 			& $>24.5$ 			& $>24.2$ 			& $>22.4$ 			& $>22.0$			& $>20.6$			\\ \hline
\multicolumn{8}{c}{GRB 140614A / $T_0 =$ 01:04:59 UT ($\text{MJD} = 56822.04513$) (20)} \\
%$308\pm 41$ 		& \multicolumn{4}{c}{NIR only}													& $17.34\pm 0.07$	& $17.06\pm 0.09$	& $16.75\pm 0.10$	\\
%$409\pm 41$ 		& \multicolumn{4}{c}{NIR only}													& $17.36\pm 0.08$	& $17.09\pm 0.08$	& $16.80\pm 0.10$	\\
%$511\pm 41$ 		& \multicolumn{4}{c}{NIR only}													& $17.48\pm 0.08$	& $17.38\pm 0.11$	& $16.92\pm 0.10$	\\
%$626\pm 41$ 		& \multicolumn{4}{c}{NIR only} 													& $17.78\pm 0.09$	& $17.41\pm 0.13$	& $17.30\pm 0.14$	\\
%$915\pm 210$ 		& \multicolumn{4}{c}{NIR only} 													& $18.13\pm 0.10$	& $17.96\pm 0.12$	& $17.51\pm 0.12$	\\
%$1434\pm 270$ 		& \multicolumn{4}{c}{NIR only} 													& $18.40\pm 0.09$	& $18.28\pm 0.13$	& $17.80\pm 0.12$	\\
$2331\pm 396$		& 	$>22.2$			& $20.65\pm 0.13$ 	& $19.84\pm 0.06$ 	& $19.43\pm 0.06$ 	& $19.22\pm 0.14$	& $18.85\pm 0.14$ 	& $18.54\pm 0.15$ 	\\
$3165\pm 394$		& 	$>22.2$			& $21.23\pm 0.25$	& $20.27\pm 0.10$ 	& $19.77\pm 0.07$ 	& $19.56\pm 0.23$ 	& $19.26\pm 0.20$ 	& $18.88\pm 0.19$ 	\\
$4209\pm 605$		&  	$>22.5$			& $21.69\pm 0.33$	& $20.66\pm 0.11$ 	& $20.42\pm 0.11$ 	& $20.02\pm 0.25$ 	& $19.64\pm 0.23$ 	& $19.43\pm 0.28$ 	\\	
$7038\pm 1203$		& 	$...$			& $...$				& $21.30\pm 0.10$ 	& $21.08\pm 0.11$ 	& $20.63\pm 0.20$ 	& $20.33\pm 0.20$ 	& $20.12\pm 0.25$ 	\\
$11337\pm 2973$		& 	$...$			& $...$				& $21.93\pm 0.10$ 	& $21.57\pm 0.12$ 	& $21.27\pm 0.18$ 	& $20.97\pm 0.25$ 	& $20.76\pm 0.24$ 	\\
$18358\pm 1916$ 	& 	$...$			& $...$				& $22.58\pm 0.38$ 	& $22.23\pm 0.33$ 	& $21.90\pm 0.35$ 	& $21.61\pm 0.39$ 	& $21.40\pm 0.50$ 	\\
$90787\pm 3616$ 	& 	$>25.1$			& $25.63\pm 0.60$ 	& $24.72\pm 0.28$ 	& $24.38\pm 0.29$ 	& $>22.7$			& $>22.4$			& $>21.9$			\\
$178637\pm 4062$ 	& 	$>25.3$			& $>25.6$			& $25.63\pm 0.45$ 	& $25.29\pm 0.52$ 	& $>22.8$			& $>22.5$  			& $>20.0$			\\ \hdashline
$7453\pm 1617$ 		& 	$>23.3$			& $22.29\pm 0.24$	& $...$ 			& $...$ 			& $...$				& $...$				& $...$				\\
$14747\pm 5527$ 	& 	$>24.4$			& $23.19\pm 0.27$	& $...$ 			& $...$ 			& $...$				& $...$				& $...$				\\
\hline
\multicolumn{8}{c}{GRB 151027B / $T_0 = \text{22:40:40}$ UT (MJD $= 57322.94491$) (21)} \\
$28839\pm 903$ 		& $22.16\pm 0.23$	& $20.41\pm 0.05$ 	& $19.78\pm 0.04$ 	& $19.58\pm 0.04$ 	& $19.47\pm 0.09$ 	& $18.96\pm 0.08$ 	& $19.09\pm 0.16$ 	\\
$30948\pm 1125$ 	& $22.17\pm 0.26$ 	& $20.59\pm 0.07$ 	& $20.02\pm 0.06$ 	& $19.66\pm 0.05$	& $19.48\pm 0.10$ 	& $19.19\pm 0.10$ 	& $19.10\pm 0.20$  	\\
$33291\pm 1129$ 	& $22.69\pm 0.24$ 	& $20.62\pm 0.04$ 	& $20.00\pm 0.04$ 	& $19.67\pm 0.04$ 	& $19.52\pm 0.07$ 	& $19.17\pm 0.08$ 	& $19.10\pm 0.14$	\\
$35634\pm 1130$ 	& $22.45\pm 0.16$ 	& $20.64\pm 0.04$ 	& $20.09\pm 0.04$ 	& $19.77\pm 0.04$ 	& $19.54\pm 0.07$ 	& $19.39\pm 0.08$ 	& $19.24\pm 0.14$ 	\\
$112431\pm 6317$ 	& $>23.6$ 			& $22.05\pm 0.07$  	& $21.53\pm 0.07$ 	& $21.18\pm 0.06$ 	& $21.10\pm 0.15$ 	& $20.73\pm 0.16$ 	& $20.36\pm 0.24$ 	\\
$202273\pm 1606$ 	& $>24.0$ 			& $22.57\pm 0.09$ 	& $22.09\pm 0.09$ 	& $21.85\pm 0.08$ 	& $21.80\pm 0.25$ 	& $21.53\pm 0.29$ 	& $>20.9$			\\ 
$804300\pm 6436$ 	& $>25.8$ 			& $24.75\pm 0.16$ 	& $24.33\pm 0.21$ 	& $23.87\pm 0.23$ 	& $>22.3$ 			& $>21.9$ 			& $>21.5$ 			\\ \hdashline
$31733\pm 2686$		& $22.46\pm 0.16$	& $20.60\pm 0.04$ 	& $19.98\pm 0.04$ 	& $19.66\pm 0.04$ 	& $19.47\pm 0.07$ 	& $19.11\pm 0.07$ 	& $18.95\pm 0.11$ 	\\ \hline
\multicolumn{8}{c}{GRB 151112A / $T_0 =$ 13:44:48 UT (MJD $ = 57338.57278$) (22)} \\
$41030\pm 3474$ 	& $>25.4$ 			& $23.58\pm 0.09$ 	& $22.63\pm 0.08$ 	& $22.23\pm 0.09$ 	& $21.74\pm 0.26$ 	& $21.25\pm 0.22$ 	& $20.59\pm 0.26$ 	\\
$130717\pm 2389$ 	& $>25.7$ 			& $24.56\pm 0.14$ 	& $23.63\pm 0.12$ 	& $23.50\pm 0.20$ 	& $>22.3$ 			& $> 21.8$ 			& $>20.8$ 			\\
$225119\pm 2393$ 	& $>25.2$ 			& $>25.0$ 			& $>24.3$ 			& $>23.7$ 			& $>22.0$ 			& $> 21.6$ 			& $>20.6$			\\ \hline
\end{longtable}
\tablefoot{
\tablefoottext{a}{Mid-time of the GROND exposure in seconds after the detection of the prompt emission ($T_0$).} \tablefoottext{b}{Magnitudes are not corrected for the Galactic foreground extinction
given in Tab. \ref{tab:sample}. Upper limits are 3$\sigma$.} Below the dashed lines we give the
magnitudes of stacked observations. Empty entries ($...$) indicate technical problems, reflections
rings at the position of the source, or when we used the magnitudes of stacked/single observations
instead.}
   \tablebib{
	(1)~\citet{pagani2007}; (2)~\citet{evangelista2008};
    (3)~\citet{schady2008}; (4)~\citet{goldstein2008}; (5)~\citet{perri2009};
    (6)~\citet{krimm2009}; (7)~\citet{ukwatta2009}; (8)~\citet{rowlinson2009};
    (9)~\citet{rowlinson2010}; (10)~\citet{mereghetti2010}; (11)~\citet{marshall2010};
    (12)~\citet{saxton2011}; (13)~\citet{page2012}; (14)~\citet{yershov2012};
    (15)~\citet{ukwatta2013}; (16)~\citet{page2013}; (17)~\citet{racusin2014};
    (18)~\citet{kocevski2014}; (19)~\citet{davanzo2014b}; (20)~\citet{page2014};
    (21)~\citet{ukwatta2015}; (22)~\citet{malesani2015b}
    .}

\endgroup